\documentclass[11pt]{article}
\pdfoutput=1
\usepackage{jheppub}
\usepackage{upgreek}
\usepackage{float, extarrows, tikz-cd}
\usepackage{graphicx}
\usepackage{subfig}
\usepackage{slashed}
\usepackage{tabularx,ragged2e}
\usepackage{longtable}
\usepackage{amssymb}
\usepackage{amsmath,amssymb}
\usepackage{slashed}
\usepackage{caption}
\usepackage{xcolor}
\usepackage{dsfont}
\usepackage{verbatim}
\usepackage{mathtools,xcolor,ytableau,amsfonts,tikz}
\usepackage{physics}
\usepackage{simpler-wick}
\usepackage{microtype}
\usepackage[nobottomtitles*]{titlesec}
\usepackage{color,soul}
\setulcolor{orange}
\newcolumntype{C}{>{\Centering\arraybackslash}X}

\usepackage{array}
\newcolumntype{P}[1]{>{\centering\arraybackslash}p{#1}}

\numberwithin{equation}{section}

\newcommand{\mL}{\mathcal{L}}
\newcommand{\mO}{\mathcal{O}}

\newcommand{\mR}{\mathcal{R}}

\newcommand{\pd}{\partial}

\newcommand{\ellb}{\bar{\ell}}

\newcommand{\bz}{\bar{z}}

\def\<{\langle}
\def\>{\rangle}

\newcommand{\mt}{\tilde{m}}

\author[a,b]{Gabriel Cuomo,}   
\author[a,c]{Sergei Dubovsky,}        
\author[d]{Guzm\'an Hern\'andez-Chifflet,}      
\author[e]{Alexander Monin,}                         
\author[a]{and Shahrzad Zare}

\affiliation[a]{Center for Cosmology and Particle Physics, Department of Physics, New York University, New York, NY 10003, USA}
\affiliation[b]{Department of Physics, Princeton University, Princeton, NJ  08544, USA}
\affiliation[c]{Institute for Advanced Study, Princeton, NJ 08540, USA}
\affiliation[d]{Instituto de F\'isica, Facultad de Ingenier\'ia, Universidad de la Republica, J.H.y Reissig 565, 11300 Montevideo, Uruguay}  
\affiliation[e]{Department of Physics and Astronomy, University of South Carolina, Columbia, SC 29208, USA}

\emailAdd{gc3265@nyu.edu}				
\emailAdd{sd103@nyu.edu}       
\emailAdd{guzmanhc@fing.edu.uy}
\emailAdd{amonin@mailbox.sc.eduk}   
\emailAdd{sz2507@nyu.edu}

\begin{document}

\title{The EFT of Large Spin Mesons}

\abstract{We use effective string theory to study mesons with large spin $J$ in large $N_c$ QCD as rotating open strings. In the first part of this work, we formulate a consistent effective field theory (EFT) for open spinning strings with light quarks. Our EFT provides a consistent treatment of the endpoints' singularities that arise in the massless limit. We obtain results, in a systematic $1/J$ expansion, for the spectrum of the leading and daughter Regge trajectories.  Interestingly, we find that the redshift factor associated with the quarks' acceleration implies that the applicability regime of the EFT is narrower compared to that of static flux tubes. In the second part of this work, we discuss several extensions of phenomenological interests, including mesons with heavy quarks, the quarks' spin and the daughter Regge trajectories associated with the worldsheet axion, a massive string mode identified in lattice simulations of $4d$ flux tubes. 
We compare our predictions with $4d$ QCD spectroscopy data, and suggest potential \emph{stringy} interpretations of the observed mesons. We finally comment on the relation between the EFT spectrum and the Axionic String Ansatz, a recently proposed characterization of the spectrum of Yang-Mills glueballs.}
\maketitle

\section{Introduction}

It is believed that large $N$ confining gauge theories, such as $SU(N)$ Yang-Mills or $QCD_N $, admit a dual formulation in terms of a weakly coupled string theory. The concept of confining strings dates back to the observation that the lightest hadronic resonances organize into families following the scaling relation $M^2 \propto J$. This empirical relation inspired early string models, which ultimately evolved into the modern framework of fundamental string theory aiming to describe all particles and interactions.

Despite decades of research, our understanding of the worldsheet dynamics of confining strings remains incomplete. Nonetheless, recent years have witnessed substantial progress, driven by a synergy of theoretical insights and lattice simulations.\footnote{There has also been recent progress in the bootstrap approach to the QCD flux tube, see e.g.~\cite{EliasMiro:2019kyf,EliasMiro:2021nul,Gaikwad:2023hof}.}
Much of this progress concerns the dynamics of long, static flux tubes in $SU(N)$ Yang-Mills theory at large $N$, both in $d=3$ and $d=4$ spacetime dimensions (see \cite{Athenodorou:2024loq} for a recent account). 

In particular, lattice measurements have confirmed the validity of the effective string theory (EST) description of long strings in terms of $d-2$ Goldstones~\cite{Polchinski:1991ax,Dubovsky:2012sh,Aharony:2013ipa}, providing quantitative values for the basic parameters of the worldsheet theory (such as the string tension). Additionally, Monte-Carlo simulations provide a glimpse on the UV dynamics of the flux tube beyond effective field theory. 

Surprisingly, all observed string states in three-dimensional $SU(N)$ gauge theories are well described by a single Goldstone, with no additional massive string degrees of freedom. In contrast, in four-dimensional Yang-Mills theory, lattice data revealed the existence of a massive pseudoscalar mode on the worldsheet, referred to as the “worldsheet axion”, besides the two Goldstones \cite{Dubovsky:2013gi,Athenodorou:2017cmw}. Both lattice measurements and theoretical arguments support the hypothesis that no additional degrees of freedom exist on the worldsheet of the confining string of $SU(N)$ Yang-Mills theory in $d=3$ and $d=4$.\footnote{Intriguingly, there exist integrable Lorentz-invariant worldsheet S-matrices precisely in terms of a single massless mode in three dimensions, and in terms of two Goldstone modes and a massless worldsheet pseudoscalar in four dimensions~\cite{Dubovsky:2015zey}. While the data show that the Yang-Mills string is only approximately described by such integrable theories, one might speculate that the UV dynamics of the flux tube restores integrability at high energies for certain inclusive observables, potentially reflecting the asymptotic freedom of partons.}

Given these advancements in our understanding of confining strings, it is natural to revisit the calculation of the hadron spectrum in large $N$ confining theories from the dual string perspective. This endeavor is further motivated by both experimental data from QCD and recent lattice simulations of the glueball spectrum in $3d$ and $4d$ Yang-Mills theory~\cite{Athenodorou:2016ebg,Conkey:2019blu,Athenodorou:2021qvs}. In particular, identifying signatures of the $4d$ worldsheet axion in these datasets would be a fascinating development.
\begin{figure}[t!]
  \centering
  \subfloat{\includegraphics[width=0.45\textwidth]{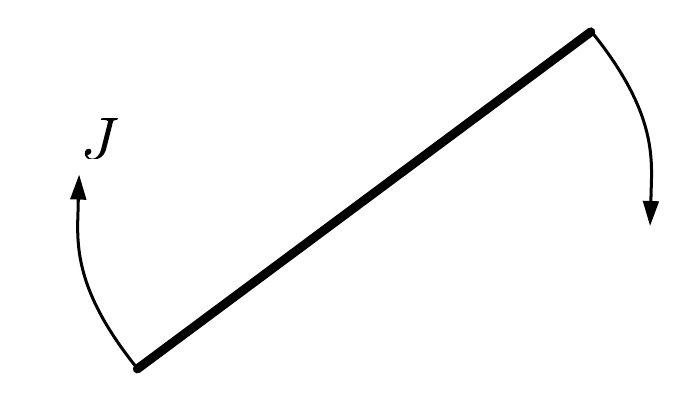}}\hfill
  \subfloat{\includegraphics[width=0.45\textwidth]{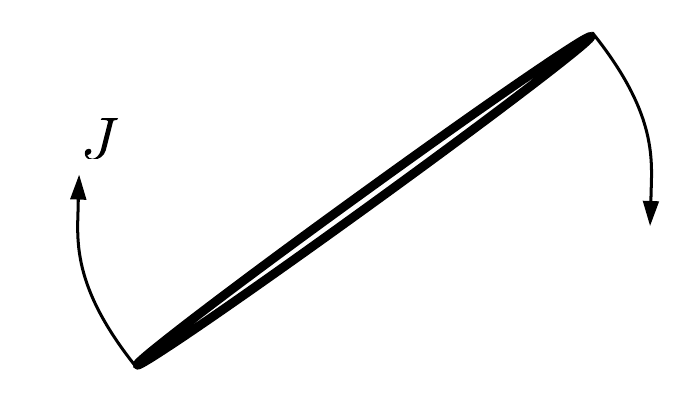}}
  \caption{\small Left: Spin $J$ meson as a rotating open string. Right: Spin $J$ glueball as a rotating closed string.}
  \label{fig:mesonsandglueballs}
\end{figure}

A natural starting point for this analysis is particles with large spin $J$. Mesons and glueballs are described by, respectively, open and closed rotating strings. The string stretches between a quark and an anti-quark for mesons, and it is folded between two cusps for glueballs (see fig. \ref{fig:mesonsandglueballs}). As the angular momentum increases, the string length grows, inducing thus a parametric separation between its size $L \sim \sqrt{J} \bar{\ell}_s$ and the inverse confinement scale $\bar{\ell}_s$. This separation, similarly to long flux tubes, enables the systematic calculation of the mass spectrum of the theory, in the vicinity of the leading Regge trajectory, via a semiclassical expansion in powers of $\bar{\ell}_s^2 / L^2 \sim 1/J$.\footnote{Recall that string splitting eﬀects are negligible for large number of colors.} The primary goals of this work are to provide a systematic formulation of the effective field theory for rotating open strings (in any $d > 2$), and to compare its predictions with QCD meson spectroscopy data.

The effective field theory (EFT) argument outlined above offers a modern explanation for the mass relation $M^2 \sim J/\bar{\ell}_s^2$ obeyed by mesons on the leading Regge trajectory.\footnote{Under some technical assumptions, it may be proven that the relation $M^2 \propto J$ is a universal result for particles on the leading Regge trajectories in large $N$ theories \cite{Caron-Huot:2016icg,Sever:2017ylk}.} The spectrum of daughter Regge trajectories is similarly obtained by quantizing the worldsheet Goldstone modes. However, the systematic calculation of $1/J$ corrections to the spectrum of the leading and daughter Regge trajectories is hindered by a technical challenge: the presence of singularities in the worldsheet metric. These singularities are localized at the cusps for closed strings, and at the string endpoints for mesons composed of massless quarks. Even for mesons made of light, but not necessarily massless quarks, the derivative expansion breaks down near the string endpoints due to their large acceleration.

In the first part of this paper we provide a consistent EFT treatment of the string endpoints' singularities for mesons composed of light quarks. To this aim, we augment the bulk EST action with a suitable boundary action that describes the endpoints. The Wilson coefficients of this boundary action absorb the singular contributions of bulk operators, ensuring that the derivative expansion remains well-posed. Our analysis is largely inspired by \cite{Hellerman:2013kba,Hellerman:2016hnf}, which first addressed this issue focusing on the leading Regge trajectory. Here we extend the framework of those works to the calculation of the spectrum of daughter Regge trajectories. 

A surprising outcome of our analysis is that the effective field theory cutoff is parametrically lower than the naive expectation, $\sim \bar{\ell}_s^{-1}$. This phenomenon arises because the high acceleration of the quarks causes frequencies at the string endpoints to appear significantly redshifted relative to those at the string’s center. Consequently, the derivative expansion at the boundary is formulated in terms of this reduced cutoff, and the expansion for the mass spectrum progresses in powers of $1/\sqrt{J}$ rather than $1/J$. A more detailed summary of the results of Part~\ref{part1} of this paper is given below in Section~\ref{subsec_summary_1}.

In the second part of this paper, we explore several extensions of the effective theory that are of phenomenological interest: the inclusion of quark spin in the EFT, the daughter Regge trajectories associated with the worldsheet axion, and the extension of our approach to mesons containing heavy quarks. We then compare our predictions with QCD spectroscopy data from the Particle Data Group (PDG) \cite{ParticleDataGroup:2024cfk}, focusing on mesons with light quarks. From fits of the leading Regge trajectories, we estimate the leading boundary Wilson coefficient for different quark families. We also suggest potential interpretations of the observed mesons beyond the leading Regge trajectory in terms of states predicted by EST (extrapolating to low values of $J$); fascinatingly, we suggest a candidate for a worldsheet axion excitation, though further experimental data are required to confirm or disprove this interpretation. 

Finally, we close this work commenting on the relation between the effective field theory spectrum and the Axionic String Ansatz (ASA) \cite{Dubovsky:2016cog,Dubovsky:2021cor}, a recently proposed recipe that successfully interprets (and predicts) the quantum numbers of all the glueballs measured on the lattice in $3d$ Yang-Mills theory. A more detailed account of the results of this work's Part \ref{part2} can be found in Section~\ref{subsec_summary_2} below.

An important next step is to extend our approach to glueballs. While the physical picture is similar for mesons and glueballs, with the role of the endpoints replaced by the cusps for glueballs, the closed string setup seems to pose additional technical challenges. In particular, we expect that a systematic treatment of the cusps will require more Wilson coefficients than the case of open string endpoints. We are currently investigating this issue. An important concrete target is the calculation of the energy splittings between particles just above the leading Regge trajectory, which have been measured on the lattice for glueballs with angular momentum up to $J=4$ in $3d$ Yang-Mills theory~\cite{Athenodorou:2016ebg,Conkey:2019blu}. It would also be interesting to identify axion excitations in the recent simulations of the mass spectrum of $4d$ glueballs in $SU(N)$ gauge theory. We plan to report on progress in these directions in the future.

In this work, we focus on classical corrections to the mass spectrum of large $J$ particles. It would also be valuable to analyze quantum corrections to the EFT spectrum of rotating strings. While these corrections are subleading compared to the leading contributions from the string endpoints, they are particularly interesting because they do not depend on new Wilson coefficients. For the leading Regge trajectory, the quantum Regge intercept for mesons was computed in \cite{Hellerman:2013kba}, but performing similar calculations for the daughter Regge trajectories (and for closed strings) does not appear to be entirely straightforward.

Finally, it would be interesting to identify sufficiently \emph{inclusive} observables on the rotating string that are not sensitive to the physics near the string endpoints and are thus unaffected by the redshift of the cutoff described above. Such quantities should allow for more precise EFT predictions, purely in terms of the Wilson coefficients of the bulk EST action that have been measured in simulations of static flux tubes.

\subsection{Summary of Part \ref{part1}}\label{subsec_summary_1}

\hspace{5mm}
Effective String Theory (EST) describes the low energy dynamics of the chromodynamic flux tube when this is much longer than the length scale associated with confinement $\sim\Lambda_{QCD}^{-1}$.  
The EST action is organized as a series in terms of geometric invariants of the worldsheet metric, with the  leading term given by the Nambu-Goto (NG) action.

\begin{figure}[t!]
    \centering
    \includegraphics[width=0.6\textwidth, trim= 2cm 4cm 2cm 4cm]{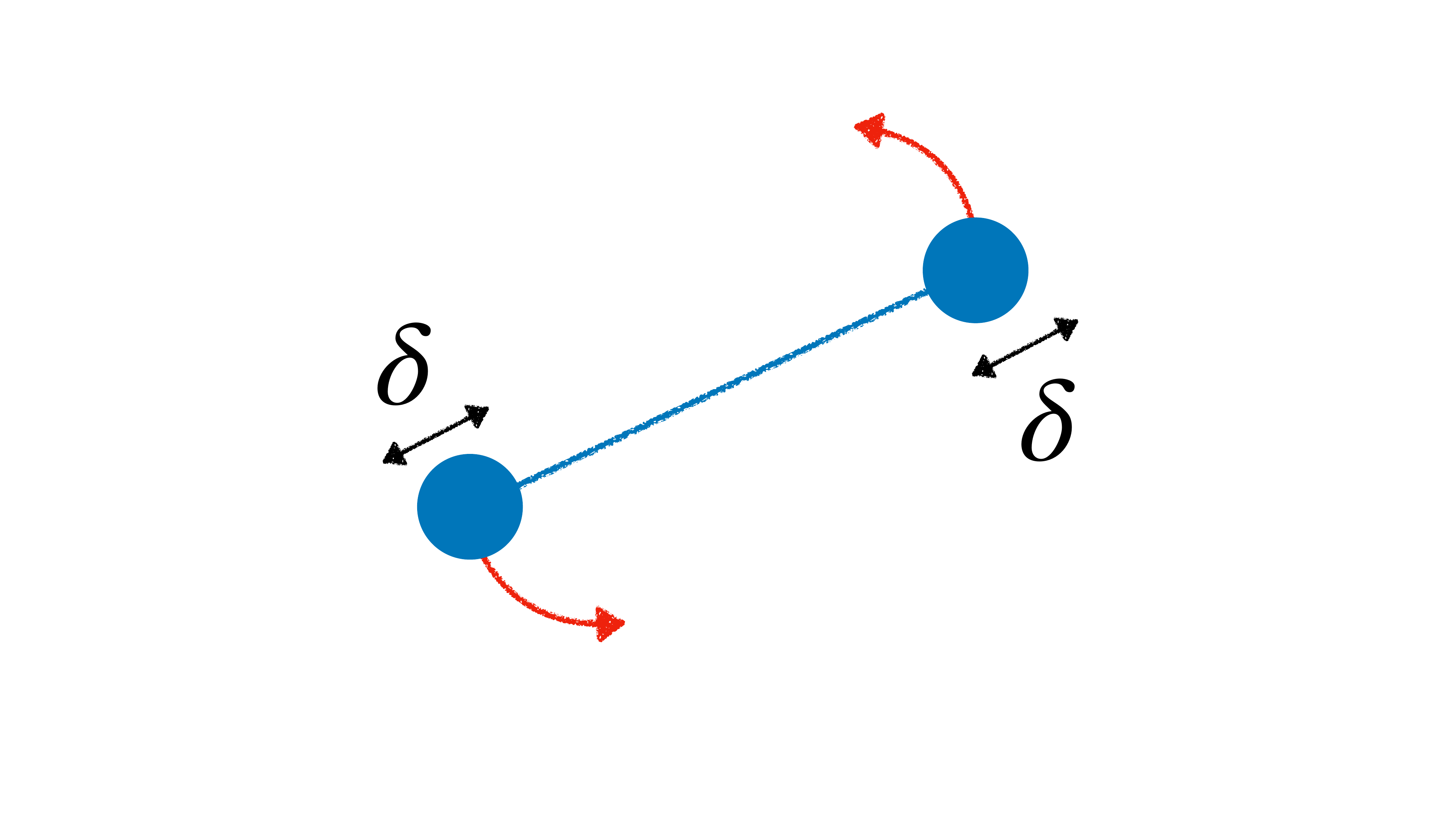}
    \caption{Graphical representation of a meson as a  rotating open string with effective endpoints of size $\delta$. }
    \label{fig:meson}
\end{figure}
In the first part of this work we use EST to study mesons in large $N_c$ QCD-like theories with massless or light quarks in $d=3$ and $d=4$. We think of a meson as composed of a flux tube stretching between a quark and an anti-quark. We ignore string breaking effects, that are suppressed at large $N_c$ for a finite number of flavors.  As it is well known, the length of a classical rotating NG string is proportional to the square root of its angular momentum $J$, see fig.~\ref{fig:meson}. Therefore effective string theory (EST) should allow for the calculation of the spectrum of mesons in a $1/J$ expansion. Famously, this logic predicts the existence of Regge trajectories satisfying $M^2\propto J$. 

To successfully implement this program, we need to address an important technical point: the effective theory close to the string endpoints, which is famously a subtle issue. For instance, when the quarks are taken to be massless, the worldsheet metric at the string endpoints becomes singular (since the projection onto the quark worldline is null), and all higher derivative operators in EST diverge at the endpoints.\footnote{For fundamental strings, for which we only need the Nambu-Goto action, the exact quantization of the theory resolves these divergences.} 
More generally, for light quark masses $m\lesssim\Lambda_{QCD}$ the derivative expansion breaks down close to the endpoints, and all higher derivative operators contribute at the same order in the $1/J$ expansion to the mesons' spectrum--seemingly defying the derivative expansion.

The physical origin of the apparent breakdown of the derivative expansion is that, within EFT, we cannot exactly resolve the positions nor the velocities of the quarks. Rather, we need to supplement the bulk EST with a suitable effective boundary action that describes the endpoints in a multipole expansion. Physically, this is similar to the worldline EFT description for black holes as seen from afar \cite{Goldberger:2007hy}. The Wilson coefficients of the boundary action naturally depend on a cutoff $\delta\sim\Lambda_{QCD}^{-1}$, representing the size of the \emph{effective} quarks. Within this EFT, the contributions from bulk operators that are singular in the limit $\delta\rightarrow 0$ can simply be understood as a renormalization of the boundary coefficient of the endpoints' action. Once these singular contributions are reabsorbed in the boundary Wilson coefficients, higher derivative bulk operators contribute to observables according to the expected EFT counting.

The formulation and study of the large spin mesons EFT for light quarks, including the endpoints action, is the main topic of the first part of this work.  We find that the masses of the particles on the leading and daughter Regge trajectories can be computed in terms of the string tension $1/\ell_s^2\equiv 1/(2\pi\ellb_s^{2})$, and two boundary Wilson coefficients $\mu_1$ and $\mu_2$, up to relative order $\sim J^{-3/2}$. For instance, the mass of the leading Regge trajectory is given by:
\begin{equation}\label{eq_intro_leading_Regge}
M^2=\frac{1}{\ellb_s^2}J+\frac{2\mu_1}{\ellb_s}J^{1/4}-\frac{1}{\ellb_s^2}
+\frac{2\mu_2}{\ellb_s}J^{-1/4}+\frac{3}{4}\mu_1^2J^{-1/2}+\ldots\,.
\end{equation}
Daughter Regge trajectories are associated with excitations of the string and have a gap $\gtrsim 1/L\sim \ellb_s^{-1}/\sqrt{J}$ on top of the leading trajectory.  We provide explicit results for the spectrum in the main text. 

Interestingly, from the study of daughter Regge trajectories we find that we can reliably compute the mass of only those mesons whose gap above the leading Regge trajectory is smaller than $ \Lambda_{QCD}/J^{1/4}$, rather than the expected cutoff $\Lambda_{QCD}$. 
This is due to the redshift of the proper time at the location of the effective quarks with respect to the time coordinate at the center of the string:
\begin{equation}
dt_{quark}\sim J^{-1/4}dt_{bulk}\,,
\end{equation} 
which is a result of the large acceleration.
Therefore, the local cutoff at the endpoints is smaller than in the bulk by a factor of $J^{1/4}$. In other words,  Poincar\'e' invariance implies that the boundary action, is organized in a derivative expansion with cutoff $\Lambda_{b.dry}\sim \Lambda_{QCD}/J^{1/4}$, when written in terms of the bulk time coordinate.

We remark that, once the boundary singularities are renormalized away, higher derivative bulk operators are always at least $O(J^{-2})$ suppressed, and therefore contribute at quite subleading order compared to the boundary Wilson coefficients.

The results of the first part of this work agree with former investigations of large spin mesons' in the literature \cite{Chodos:1973gt,Baker:2000ci,Baker:2002km,Wilczek:2004im,Kruczenski:2004me,Hellerman:2013kba,Hellerman:2016hnf,Sonnenschein:2018aqf}. In particular, Hellerman and Swanson formulated the endpoints EST and derived eq.~\eqref{eq_intro_leading_Regge} working in the Polyakov formalism in \cite{Hellerman:2013kba,Hellerman:2016hnf}. We instead work in the Nambu-Goto formalism, which does not require dealing with constraints and allows obtaining results for the daughter Regge trajectories as well in a straightforward manner. We also note that EFTs with nontrivial space-dependent local cutoffs were recently analyzed in the context of finite density systems \cite{Hellerman:2020eff,Hellerman:2021qzz,Cuomo:2022kio}. Finally, we remark that the fast motion of the quarks makes the boundary EFT that we study in this work very different from the EFTs describing the endpoints or the junctions of static long strings (see e.g.~\cite{Aharony:2010cx,Aharony:2010db,Komargodski:2024swh}).

\subsection{Summary of Part \ref{part2}}\label{subsec_summary_2}

\hspace{5mm}
In the second part of this work we discuss extensions and applications of the EFT developed in the first part that are of phenomenological interest.

First we discuss how to describe the quarks' spin in $d=4$ within EFT.
We find that, generically, states with spin wave-functions that are different from those of the leading order Regge trajectory have a large gap, of the order of the boundary cutoff.  Intuitively, this result can be understood as a consequence of the Thomas' spin precession, which results from the quarks' acceleration; furthermore, interactions with the string renormalize the spins' gap at leading order. We conclude that within EFT, we cannot reliably analyze the energy splitting between different spin states, and the spin variables must be integrated out.

We then analyze the daughter Regge trajectories associated with the QCD worldsheet axion.
Massive worldsheet particles on the rotating strings localize near the endpoints of the string due to a centrifugal force. We argue that as a consequence of the centrifugal barrier, the wave-functions of the corresponding modes are approximately localized near the two endpoints, and the axion's single particle spectrum organizes in a set of approximately doubly degenerate states on top of the leading Regge trajectory (at large $J$). We also point out a puzzle concerning the boundary conditions at the string endpoints, that are needed to compute the mass spectrum of daughter trajectories associated with the worldsheet axion. Contrary to the EFT expectation, based on genericity arguments and symmetries, simulations of open static flux tubes suggest that Neumann boundary conditions provide a more accurate descriptions of the observed data~\cite{Juge:2003ge,Sharifian:2024nco}. If this conclusion were to extend to the rotating string setup, it would suggest that the daughter trajectories of the worldsheet axion may be relatively light at moderate $J$. Due to the physical relevance of this point, in the future it would be interesting to analyze further the axion boundary conditions, perhaps extending the flux tube bootstrap of~\cite{EliasMiro:2019kyf,EliasMiro:2021nul,Gaikwad:2023hof} to include the boundary reflection S-matrix.

Afterwards, we generalize the mesons' EFT to the conceptually simpler case of heavy quarks. We argue that the EFT is properly formulated as a double-scaling limit $J\rightarrow\infty$ with $J/(m^2\ellb_s^2)=\text{fixed}$, where $m$ is the quarks' mass. We find that for the flux tube to be sufficiently long and EST to apply, we need angular momenta $J\gg \sqrt{m\ellb_s}$. We show that the particle's spectrum in the non-relativistic regime, which corresponds to $\sqrt{m\ellb_s}\ll J\ll m^2\ellb_s^2$, can be computed by semiclassically quantizing a simple quantum-mechanical potential model, and we obtain analytical results for the quantum corrections to the leading Regge-trajectory in the heavy quark regime. We finally analyze the quarks' spins, which are within EFT for heavy endpoints, and provide predictions for the gap of the corresponding states. In particular, we predict that in the EFT regime $J\gg \sqrt{m\ellb_s}$, the state with minimal energy among the different spin wavefunctions is the one in which the quarks' spins are aligned with the orbital angular momentum, therefore rigorously establishing the so-called \emph{spin-orbit} inversion scenario~\cite{Isgur:1998kr}.

We then compare our findings with real world mesons spectroscopy data from the Particle Data Group (PDG) \cite{ParticleDataGroup:2024cfk}, focusing on light quark mesons. 
We estimate the string tension and the coefficient $\mu_1$ (for $u$, $d$ and $s$ quarks) fitting the formula~\eqref{eq_intro_leading_Regge}. For daughter Regge trajectories, the scarcity of data points for $J\gtrsim 2$ does not allow for a quantitative comparison. We nonetheless discuss potential interpretations of the observed data in terms of states predicted by EST extrapolating to low values of $J$. To this aim, we discuss how to compute parity and charge conjugation assignments of the particles within EFT. To discard states that do not project onto open strings at $N_c\rightarrow\infty$ (such as glueballs and molecules of two mesons), we compare different isospin channels and identify a small subset of states that may be reliably interpreted as mesons. Intriguingly,  we argue that one of the observed states with $J=2$ might correspond to a worldsheet axion excitation on top of the leading Regge trajectory. Future experimental analyses of $J=2$ and $J=3$ mesons, perhaps along the lines of \cite{Guerrieri:2024jkn}, might prove or disprove this interpretation.

We finally comment on some subtleties that arise when one attempts to extrapolate the EFT Hilbert space to low values of $J$, even when one assumes that the worldsheet does not admit any additional fields besides those included in EFT (as lattice data suggest that is the case in $d=3$).  Intuitively, the subtleties arise because knowing the dimensions and geometry of the phase space at large $J$ near the leading Regge trajectory is not enough to constrain the shape and topology of the phase space at all values of $J$. Our discussion is motivated by a recently proposed recipe to reproduce the spectrum of closed effective strings -- the axionic string Ansatz (ASA) -- that qualitatively matches lattice data for glueballs in $d=3$ all the way down to $J=0$ and disagrees with other quantization prescriptions such as lightcone quantization \cite{Dubovsky:2016cog}. Our discussion illustrates the physical origin of the discrepancies between different quantization procedures in the simpler context of open strings.

\part{The EFT with light quarks}\label{part1}

\section{Lightning review of effective string theory}\label{sec_review}

\subsection{The effective theory of long strings}

\hspace{5mm}
As is well known, both in $d = 3$ and $d = 4$, in the limit where the length of the compact dimension is much larger than the inverse QCD scale, inserting a Polyakov loop in Yang-Mills (YM) theory creates a long, closed, string-like object: the QCD flux tube.

Any string-like object in field theory spontaneously breaks the $d$-dimensional Poincaré group $\text{ISO(d-1,1)}$ down to $ISO(1,1)\times SO(d-2)$, where the first factor represents the $2d$ Poincaré group, and the second one corresponds to rotations in the transverse directions to the string. 

Therefore, the low-energy dynamics of a flux tube of length $L$, much larger than the size of the lightest glueball $m_g^{-1}$, is universally described in terms of the $d-2$-dimensional Goldstones associated with the broken translations.\footnote{Note that no additional Goldstones are required to account for the breaking of boosts and rotations; such a mismatch between the number of broken generators and Goldstones is common for spontaneously broken spacetime symmetries, see e.g. \cite{Low:2001bw,Nicolis:2013sga}. } The corresponding effective theory is usually referred to as \emph{effective string theory} (EST).

Let us review the construction of the EST action in the Nambu-Goto formalism. For clarity, we consider an infinitely long string. 

The action for a string is written in terms of the coordinates $X^\mu$ of the string worldsheet’s embedding into the target space, which are conveniently parametrized as follows in the static gauge:
\begin{equation}\label{eq_static_gauge}
X^\mu=\left(\tau,\sigma,X^i(\tau,\sigma)\right)\,,
\end{equation}
where $\sigma^\alpha = (\tau, \sigma)$ are the worldsheet coordinates, and $X^i$ denotes the transverse string coordinates, which are the $d-2$ dynamical Goldstone fields. Our conventions and some useful geometric identities are summarized in appendix~\ref{app_geometry}.

The action is constrained by the reparametrization invariance of the worldsheet coordinates and at the leading order it is given by the well known Nambu-Goto (NG) term:
\begin{equation}\label{eq_Sbulk}
S_{bulk}=-\frac{1}{\ell_s^2}\int d^2\sigma \sqrt{G}\left[1+O\left(\ellb_s^4 (K^\mu_{\alpha\beta})^4\right)\right]\,,
\end{equation}
where $1/\ell_s^2$ is the string tension and $G=-\det(G_{\alpha\beta})$ is the determinant of the induced metric defined by
\begin{equation}
G_{\alpha\beta}=\pd_\alpha X^\mu\pd_\beta X^\nu\eta_{\mu\nu}\,,
\end{equation}
where $\eta_{\mu\nu}$ is the spacetime metric in mostly minus signature. For future convenience, we also define the (reduced) string length $\ellb_s$ as
\begin{equation}\label{eq_T_vs_ls}
    \frac{1}{\ell_s^2}=\frac{1}{2\pi\ellb_s^2}\,.
\end{equation}
Note that $1/\bar{\ell_s^2}$ coincides with the $\alpha'$ parameter in fundamental string theory; more generally, we expect that the reduced string length parametrically coincides with the mass of the lightest glueball $\ellb_s^{-1}\sim m_g$, and thus sets the cutoff of the theory according to generalized dimensional analysis~\cite{Georgi:1992dw}.

Higher derivative contributions to the action~\eqref{eq_Sbulk} are given in terms of curvature invariants and, because of the Gauss-Codazzi relation,  they can all be written in terms of the second fundamental form of the worldsheet,
\begin{equation}
K^\mu_{\alpha\beta}=\nabla_\alpha\pd_\beta X^\mu\,,
\end{equation}
and derivatives thereof. It turns out that, upon using the equations of motion (EOM) to remove redundant operators and discarding total derivatives, the first nontrivial terms arise at fourth subleading order in derivatives.  There is therefore a high degree of universality in the low energy predictions of EST.  It is also straightforward to include additional, non-Goldstone, fields in the action~\eqref{eq_Sbulk}. These can be analyzed consistently within EFT as long as they are sufficiently light and weakly-coupled.

As is well known from the study of fundamental strings, the NG action is integrable at the classical level.  Integrability is broken by quantum effects away from the critical dimensions $d=26$ and (somewhat surprisingly) $d=3$.  Nonetheless, since the contribution of loops is subleading in the derivative expansion, the low energy dynamics of the flux tube is approximately integrable. As explained in \cite{Dubovsky:2013gi,Dubovsky:2014fma}, this property can be used to improve the convergence of the EFT results for the spectrum of flux tubes wrapping a nontrivial space cycle of length $L$.

\subsection{EST from lattice data}\label{subsec_EST_from_lattice}

\hspace{5mm}
Let us now review what is known about EST in Yang-Mills theory from lattice data. First of all, the value of $\ellb_s$ can be expressed in terms of the mass $m_g$ of the lightest glueball, comparing the mass spectrum with the ground-state energy of a long string. The result in $d=3$ \cite{Athenodorou:2016ebg} and $d=4$ \cite{Athenodorou:2021qvs} reads, in the $N_c\longrightarrow \infty$ extrapolation,
\begin{align}\label{eq_mg_3d}
    m_{g}\vert_{d=3} &
    = 1.646 \ellb_s^{-1}\,,\\
    \label{eq_mg_4d}
    m_{g}\vert_{d=4} &
    = 1.229 \ellb_s^{-1}\,.
\end{align}

In $d=3$ there is a unique higher derivative operator at fourth order in derivatives, which reads:
\begin{equation}
\int d^2\sigma \sqrt{G}\mathcal{R}^2=\int d^2\sigma\sqrt{G}\left[\left(K^\mu_{\alpha\beta} K^\nu_{\gamma\delta}-K^\mu_{\alpha\gamma}K^\nu_{\beta\delta}\right)G^{\alpha\beta}G^{\gamma\delta}\eta_{\mu\nu}\right]^2\,.
\end{equation}
The coefficient of this term was measured by comparing its contribution to the string fluctuation S-matrix with the phase shift extracted via the L\"uscher method from lattice calculations of the spectrum of strings of finite length $L$. Rather surprisingly, it turns out that, using the latest available data \cite{Athenodorou:2011rx,Athenodorou:2016kpd} (which go up to $N_c= 8$), the value of the measured coefficient is zero (within the uncertainties) in dimensional regularization in the minimal subtraction scheme \cite{Dubovsky:2014fma,Chen:2018keo}.

In $d=4$, lattice data surprisingly revealed the existence of a pseudoscalar resonance on the worldsheet \cite{Dubovsky:2013gi}, which we shall denote $a$. While its mass $m_a$ is of the order of the cutoff $\ellb_s^{-1}$, it turns out that its width is quite small: $\Gamma_a/m_a\simeq 0.21 m_a$ for $N_c=3$.  The basic properties of this mode (at least at the qualitative level) can be understood by including a pseudoscalar field in the EST and pretending that its mass $m_a$ is much smaller than the cutoff, as if $a$ were endowed with an approximate shift symmetry broken by effects of order $(m_a\ellb_s)^2$. The action of this field, including the first nontrivial coupling to the string coordinates, reads
\begin{equation}\label{eq_S_pseudoaxion}
S_{a}=\int d^2\sigma\sqrt{G}\left[\frac12(\pd a)^2-\frac{m_a^2}{2} a^2+\frac{Q_a}{32\pi} a K\cdot\tilde{K}\right]\,.
\end{equation}
The first two terms are the free theory ones. Because of parity, the first nontrivial operator to which $a$ can couple is
\begin{equation}\label{eq_K_Kt_def}
K\cdot \tilde{K}=\frac{\varepsilon^{\alpha\beta}}{\sqrt{G}}\pd_\alpha X^\mu\pd_\beta X^\nu\varepsilon_{\mu\nu\rho\sigma}K^\rho_{\gamma\delta}K^\sigma_{\eta\lambda}G^{\delta\lambda}
\frac{\varepsilon^{\gamma\eta}}{\sqrt{G}}\,.
\end{equation}
The integral of $\frac{1}{8\pi}\sqrt{G} K\cdot\tilde{K}$ is a topological invariant that counts the sign-weighted self-intersection number of the string worldsheet \cite{Polyakov:1986cs,Mazur:1986nr}.  Therefore the coupling between $a$ and $K\cdot\tilde{K}$ is analogous to the QCD axion coupling - for this reason $a$ is sometimes referred to as a \emph{worldsheet axion}.  This coupling yields the leading contribution to the width of $a$. The numerically measured values of $m_a$ and $Q_a$ for $N_c=3$ are
\begin{equation}\label{eq_ma_lattice}
m_a= 0.74 \ellb_s^{-1}\,,\qquad
Q_a= 0.37\,;
\end{equation}
these numbers do not appear to change significantly in the large $N_c$ limit \cite{Athenodorou:2017cmw}.
Notice that $m_a$ is slightly below the mass of the lightest glueball, which partially justifies its inclusion within the EFT.\footnote{Notice that eqs.~\eqref{eq_mg_3d} and~\eqref{eq_mg_4d} parametrically agree with the pi-ology expected from generalized dimensional analysis \cite{Georgi:1992dw}, according to which the UV cutoff $\sim m_g$ is expected to be set by $\ellb_s^{-1}$, which is slightly larger that the square root of the tension $1/\ell_s= 1/\left(\sqrt{2\pi}\ellb_s\right)$.}

\section{Endpoint singularities for spinning strings}\label{sec_singularities}

\hspace{5mm}
Before presenting a proper formulation of the endpoints' EFT, the goal of this section is to explain the breakdown of the bulk derivative expansion for spinning strings, which we alluded to in the introduction.  To this aim, it is simplest to consider the explicit calculation of the leading Regge trajectory. We consider the same bulk action as in the previous Section~\eqref{eq_Sbulk}. Here, the spatial worldsheet coordinate runs over a finite range:
\begin{equation}
\sigma \in [-1,1]\,.
\end{equation}
For now, we do not commit to any specific boundary action at the endpoints, that we assume to be identical. In arbitrary $d\geq 3$ spacetime dimensions, the classical profile describing a spinning string is:
\begin{equation}\label{eq_background}
X^\mu=\frac{1}{\omega}\big(\tau, \alpha_f\, \sigma \cos\tau,\alpha_f\,\sigma \sin\tau,\underbrace{0,0,\ldots}_{d-3\text{ times}}\big)\,.
\end{equation}
This is guaranteed to be a solution of the bulk EOMs to all orders in the derivative expansion. This is because eq.~\eqref{eq_background} is the most general ansatz invariant under translations in the transverse directions and the linear combination $H-\omega J$ of time translations and rotations in the first two spatial coordinates.
This profile corresponds to the following background metric:
\begin{equation}\label{eq_metric}
ds^2=\frac{1-\sigma^2\alpha_f^2}{\omega^2}d\tau^2-\frac{d\sigma^2}{\omega^2}\,.
\end{equation}
In eqs.~\eqref{eq_background} and \eqref{eq_metric}, $\omega$ is the angular velocity and we work in units such that $\tau$ and $\sigma$ are dimensionless. The parameter $\alpha_f$ in eq. \eqref{eq_background} is the velocity of the endpoints and depends upon the precise boundary conditions. It determines physical length of the string $L$, which is given by
\begin{equation}\label{eq_string_length}
L=\int d\sigma\sqrt{G_{\sigma\sigma}}=\frac{2\alpha_f}{\omega}\,.
\end{equation}
We see that the string is long for small angular velocity $\omega \ellb_s\ll 1$; as we review below, somewhat counter-intuitively this is precisely the regime that describes a string with large angular momentum. For a pure NG action with free endpoints one finds $\alpha_f=1$; we will see that $\alpha_f\simeq 1$ up to small corrections, even for more general boundary conditions in the large spin limit.

To obtain the mass of the leading Regge trajectory, it is convenient to compute the Lagrangian $\mL(\omega)$ on the background profile \eqref{eq_background}, where $S=\int dt \mL$ is the bulk action \eqref{eq_Sbulk} and $t=\tau/\omega$.  Using Noether's theorem, the Lagrangian is related to the energy $M$ and angular momentum $J$ of the rotating string via\footnote{To obtain the result for $J$, note that a rotation in the $X_1$-$X_2$ plane by a time dependent angle $\delta\phi=\epsilon t$ is equivalent to a shift $\omega\rightarrow\omega+\epsilon$, and that Noether's theorem implies that the variation of the action under a rotation is $\delta S=\int dt J\delta\dot{\phi}$. The expression for the mass is obtained similarly using that the same rotation is equivalent to a time shift $\delta t=t\epsilon/\omega $, and that by Noether's theorem we have $\delta S=\int dt M\delta\dot{t}+\mL\vert^{t=t_f}_{t=t_{i}}$.} \cite{Baker:2000ci,Baker:2002km}
\begin{equation}\label{eq_JE_Noether}
J=\frac{\pd\mL}{\pd\omega}\,,\qquad
M=\omega \frac{\pd\mL}{\pd\omega}-\mL\,.
\end{equation}
From these equations, we obtain the well known result for the pure NG action with free endpoints ($\alpha_f=1$):
\begin{equation}\label{eq_intro_Regge}
\mL =-\frac{\pi }{2\omega\ell_s^2}
\,,\quad
J=\frac{\pi }{2\omega^2\ell_s^2}
\quad \implies\quad M^2=\frac{\pi^2 }{\ell_s^2\omega^2}=\frac{J}{\ellb_s^2}\,.
\end{equation}
As anticipated, the relation between angular velocity and $J$ implies that $\omega$ becomes small at large $J$: $\omega \sim \ellb_s^{-1}/\sqrt{J}$. Therefore, the derivative expansion in the bulk runs in powers of $(\ellb_s/L)^2\sim (\omega \ellb_s)^2$, which is equivalent to the $1/J$ expansion. 

Let us now discuss the expectation for the scaling of subleading corrections to eq.~\eqref{eq_intro_Regge}. According to the naive power counting, the NG term scales as $\sim 1/(\ellb_s\omega)^2\sim J$, which thus provides the loop counting parameter. Therefore, we expect a quantum $J^{0}$ correction to $M^2$ in eq.~\eqref{eq_intro_Regge}. Higher derivative corrections are further suppressed and, as explained in Section~\ref{sec_review}, start at fourth-order in derivatives. Therefore, their contribution to $M$ is expected to be suppressed by a relative $\sim (\ellb_s/L)^4\sim 1/J^2$ factor with respect to the leading order.

As anticipated, this naive counting is incomplete. This is manifested by looking at the metric~\eqref{eq_metric}; setting momentarily $\alpha_f=1$, we see that there is an infinite redshift factor at $\sigma=1$,  and the metric is singular at the string endpoints. This implies that higher derivative curvature invariants also become singular, e.g.:
\begin{equation}\label{eq_K2n}
\left[\ellb_s^2(K_{\alpha\beta}^\mu)^2\right]^{n}=\frac{2^n\ellb_s^{2n}\omega^{2n}}{(1-\sigma^2)^{2n}}\qquad
\text{for }\alpha_f=1\,,
\end{equation}
which diverges at $\sigma=1$.

Within EFT,  this means that we need to avoid going all the way to $\sigma=1$. Instead, we should consider some region around the endpoints and replace it by some effective boundary. It is natural to cutoff the integration where higher derivative operators become of order one, $\ellb_s^2(K_{\alpha\beta}^\mu)^2\sim O(1)$. To this aim, we consider endpoints at $\sigma=\pm(1-z \ellb_s\omega)$ for some $z>0$ that does not scale with $\omega$.  In this way, the contribution of the operators~\eqref{eq_K2n} to the Lagrangian reads 
\begin{equation}\label{eq_higher_der_break}
\begin{split}
\mL &\supset \frac{\ellb_s^{2n-2}}{2\pi}\int_{-1+z\ellb_s\omega}^{1-z\ellb_s\omega} d\sigma\sqrt{G}\left[(K_{\alpha\beta}^\mu)^2\right]^{n}\\[0.5em]
&\simeq \sqrt{\frac{\omega}{\ellb_s}}\frac{2^{\frac{3}{2}-n}  z^{\frac32-2 n}}{\pi  (4 n-3)}
\left[1+\left(z\ellb_s\omega\right)\frac{(4 n-3) (4 n-1)}{4  (4 n-5)}+
O\left(\left(z\ellb_s\omega\right)^2\right)\right]\,.
\end{split}
\end{equation}
As we will see in the next section, a similar result can be obtained in a manifestly gauge invariant approach upon introducing a mass term at the endpoints, which makes the difference $1-\alpha_f$ nonzero.

Contrary to the naive expectation outlined before, eq.~\eqref{eq_higher_der_break} shows that all higher derivative terms contribute at the same order to the leading Regge trajectory. More precisely, each contribution is suppressed only by $\sim(\ellb_s\omega)^{3/2}$ with respect to the NG contribution, with further subleading terms organized in a series in $\ellb_s\omega$ rather than the expected $(\ellb_s\omega)^2$. Similar results are found in the study of other observables such as the energy of daughter Regge trajectories.  Notice also that eq.~\eqref{eq_higher_der_break} does not contain any integer power of $\omega \ellb_s$ - this fact will be important when we will discuss quantum corrections in the next sections.

It seems therefore that we lack an organizing principle for the derivative expansion beyond the leading order.  This is not just because all operators contribute at the same order, but equally dramatically because the expansion in eq.~\eqref{eq_higher_der_break} contains unexpected fractional powers of $(\ellb_s\omega)^2\sim \ellb_s^2/L^2$. It is clear that these issues have to do with the physics of the endpoints, and as such we should expect a more careful treatment and EFT description of the endpoints to resolve the confusion. The formulation of such boundary EFT will be the topic of next two sections.

\section{Effective endpoints: power-counting and results}\label{sec_bdry_action}

\hspace{5mm}
In this section, we discuss the general construction of the endpoint effective action, neglecting bulk higher derivative terms. We will explain how to use this action to absorb the singular contributions from bulk operators in Section~\ref{sec_renormalization}.

Before delving into the technicalities, it is useful to explain the physical picture.  Since the size of the region in which the derivative expansion breaks down is small compared to the string extension~\eqref{eq_string_length}, it should be possible to treat it as an effective endpoint within EFT, whose interaction with the bulk is described in a multipole expansion.  It is however important to note that the worldsheet metric \eqref{eq_metric} varies nontrivially along the string. EFT is applicable only at time and spatial differences much larger than the string length:
\begin{equation}
\Delta\tau^2(1-\alpha_f^2\sigma^2)/\omega^2\gg \ellb_s^2\quad\text{and}\quad
\Delta\sigma^2/\omega^2\gg \ellb_s^2\,.
\end{equation}
Since $(1-\alpha_f^2\sigma^2)\sim\ellb_s\omega$ at the location where we should impose boundary conditions,  we infer that in terms of the physical time $t=\tau/\omega$ at the center of the string, the frequency cutoff at the boundary appears redshifted
\begin{equation}\label{eq_ddt_bdry}
\pd/\pd t\ll \Lambda_{b.dry}\sim \sqrt{\ellb_s\omega}\Lambda_{bulk}=\sqrt{\omega/\ellb_s}\,.
\end{equation}
In other words, Poincar\'e invariance and the large acceleration of the quarks imply that the boundary conditions for the bulk fields admit an expansion in even powers of $\sqrt{\ellb_s \omega}$ (notice that this matches the form of the expansion in eq.~\eqref{eq_higher_der_break}). The cutoff over spatial momenta is not affected by these considerations in the coordinates~\eqref{eq_metric}.

The most important physical consequence of the redshift concerns the calculability of observables within EFT  that are sensitive to the boundary conditions. For such quantities, the derivative expansion breaks down at smaller momenta than in the long string setup. Most notably, this observation implies than the number of daughter Regge trajectories that can be studied within EFT is smaller than the naive expectation, as we will see explicitly in the next sections.

\subsection{Warm-up: massless scalar field on a non-dynamical rotating string}\label{subsec_toy_bc}

\hspace{5mm}
As a warm-up, let us consider an external scalar field on a rigid, i.e. non dynamical, rotating string. We work to quadratic order in the field, with bulk action given by
\begin{equation}\label{eq_S2_toy}
S_{bulk}=\int d\tau \int^{1}_{-1}  d\sigma\sqrt{G}\frac12 (\pd a)^2\,,
\end{equation}
where we called $a$ the scalar field in preparation of the analysis of the $4d$ worldsheet axion, and $(\pd a)^2=G^{\alpha\beta}\pd_\alpha a\pd_\beta a$. For the moment, we neglect higher derivative corrections whose role we will analyze in Section~\ref{sec_renormalization}. The worldsheet metric is given by~\eqref{eq_metric} where we choose $\alpha_f=1-z\ellb_s\omega$ for some $z>0$ which does not scale with $\omega$.

The free bulk EOM reads
\begin{equation}\label{eq_a_massless}
\ddot{a}-\frac{1}{\alpha_f^2}\left(1-\alpha_f^2\sigma ^2\right) a ''+ \sigma  a '=0\,,
\end{equation}
where the dots and the primes respectively stand for time and spatial derivatives.  Upon changing coordinates via
\begin{equation}\label{eq_s_def}
s=\arcsin(\alpha_f\sigma)\,,
\end{equation}
we recast eq.~\eqref{eq_a_massless} into a standard Klein-Gordon equation, whose
general solution is
\begin{equation}
a(\tau,\sigma)=\int\frac{d\epsilon}{2\pi}\left[e^{-i\epsilon\tau +i \epsilon s}f_+(\epsilon)+
e^{-i\epsilon\tau -i \epsilon s}f_-(\epsilon)\right]\,,
\end{equation}
where $f_+(\epsilon)=f_+^*(-\epsilon)$ and $f_-(\epsilon)=f_-^*(-\epsilon)$ denote the Fourier components.

To study the spectrum, we need to supplement eq.~\eqref{eq_a_massless} with the boundary conditions. The simplest way to do so is to write the most general boundary action for $a$ at $\sigma=\pm 1$ in a derivative expansion. We suppose that the boundary conditions preserve the shift-invariance of $a$.  In this case, we can self-consistently write the boundary action in terms of time derivatives of $a$. This is because, as we will show, operators built out of derivatives normal to the boundary are redundant due to the EOM and the leading order Neumann boundary conditions. Working to quadratic order in the field and assuming identical endpoints, we consider the boundary action to be
\begin{equation}\label{eq_toy_bdry_a}
S_{bdry}=-\sum_{\sigma=\pm 1}\frac12\int d\tau\sqrt{\hat{G}_{\tau\tau}}\left[\ellb_s b_a(\hat{\nabla}_t a)^2+\ellb_s^3 d_a(\hat{\nabla}_t^2a)^2+\ldots\right]\,,
\end{equation}
where $b_a$ and $d_a$ are dimensionless Wilson coefficients,  and we defined the endpoints' worldline metric and proper time derivatives as
\begin{equation}
\hat{G}_{\tau\tau}=G_{\tau\tau}\vert_{\sigma=\pm 1}\,,\qquad
\hat{\nabla}_t=\hat{G}_{\tau\tau}^{-1/2}\hat{\nabla}_\tau\,,
\end{equation}
where $\hat{\nabla}_{\tau}$ is the covariant derivative at the endpoints (see appendix~\ref{app_geometry}).  These definitions ensure that the action~\eqref{eq_toy_bdry_a} is invariant under reparametrizations of $\tau$. For simplicity, we neglected  operators built out of the acceleration or higher derivatives of the string coordinates $X$'s, whose power counting is more subtle and will be discussed in the next subsection.

Since we work in terms of dimensionless coordinates, in order to power-count, it is enough to count the powers of the metric. 
Bulk terms (up to the renormalization issues that we will address later) are counted as in the long string background, i.e. using $G_{\alpha\beta}\sim 1/\omega^2$. Therefore, the bulk action~\eqref{eq_Sbulk} is of order $\sqrt{G}G_{\alpha\beta}^{-1}\sim \omega^0$, so $a$ is canonically normalized and we count it as $a\sim O(1)$.
At the boundary, we power-count according to the rules
\begin{equation}\label{eq_power_count_toy}
\hat{G}_{\tau\tau}\sim \frac{\ellb_s}{\omega}\,,\qquad \hat{\nabla}_t\sim \sqrt{\frac{\omega}{\ellb_s}}\,,
\end{equation}
which follow from the redshift factor previously discussed. Therefore, the first term in eq.~\eqref{eq_toy_bdry_a} scales as $\sim \sqrt{\ellb_s\omega}$, the second as $(\ellb_s\omega)^{3/2}$, etc.  We can see this more explicitly by writing the boundary conditions that follow from variation of~\eqref{eq_S2_toy} and~\eqref{eq_toy_bdry_a}:
\begin{equation}\label{eq_toy_bc}
\pd_n a\vert_{\sigma=\pm 1}=b_a\ellb_s\hat{\nabla}_t^{2} a-d_a\ellb_s^3\hat{\nabla}_t^4 a+\ldots\,,
\end{equation}
where $\pd_n=n^\alpha\pd_\alpha$ is the normal derivative (see app.~\ref{app_geometry} for notation). The aforementioned scalings can be made manifest by working in terms of the coordinate $s$ introduced in eq.~\eqref{eq_s_def}
\begin{equation}\label{eq_toy_bc_explicit}
\pm \pd_s a \vert_{\sigma \pm 1}=b_a\frac{\sqrt{\ellb_s\omega}}{\sqrt{2 z}}\left(1+\frac14z\ellb_s\omega+\ldots\right)\ddot{a}-d_a\frac{(\ellb_s\omega)^{3/2}}{(2 z)^{3/2}}\left(1+\ldots\right)\pd_\tau^4 a+\ldots\,,
\end{equation}
where the parentheses arise from the expansion of $\sqrt{\hat{G}_{\tau\tau}}\simeq \sqrt{2\ellb_s z/\omega}\left[1+O\left(\ellb_s\omega\right)\right]$.

It is instructive to use eq.~\eqref{eq_toy_bc} to extract the spectrum of single-particle states.  In terms of the coordinate~\eqref{eq_s_def}, the endpoints are located at
\begin{equation}\label{eq_toy_s_vs_sigma}
\sigma=\pm 1\quad\implies\quad
s=\pm\left[ \frac{\pi}{2}-\sqrt{2z\ellb_s\omega}-\frac{1}{6\sqrt{2}}\left(z\ellb_s\omega\right)^{3/2}+\ldots\right]\,.
\end{equation}
Because of worldsheet parity $s\rightarrow -s$, the solutions satisfying the boundary conditions must take the form
\begin{equation}\label{eq_toy_sol}
a=e^{-i\epsilon\tau}\cos(\epsilon s) \quad\text{or}\quad 
a=e^{-i\epsilon\tau}\sin( \epsilon s) \,.
\end{equation}
Up to the leading order, we can identify the boundaries with $s=\pm\pi/2$ and the boundary conditions~\eqref{eq_toy_bc} with Neumann $\pd_s a=0$. Therefore, we find that the allowed frequencies are
\begin{equation}\label{eq_toy_spectrum}
\epsilon_n \simeq n\in\mathds{N}\,,
\end{equation}
with even (odd) eigenfunction for even (odd) $n$.  In physical units $t=\tau/\omega$, single-particle states have energy $\omega \epsilon_n$. Solving eqs.~\eqref{eq_toy_bc} to the subleading order, we find
\begin{equation}\label{eq_toy_a_disp}
\epsilon_n=n\left\{1-\sqrt{\omega\ellb_s} \frac{\sqrt{2}B_a}{\pi}+\omega\ellb_s\frac{2 B_a^2}{\pi^2}+(\omega\ellb_s)^{3/2}\left[\frac{6D_a+B_a^3}{3\sqrt{2}\pi} n^2+O(n^0)\right]+\ldots\right\}\,,\qquad
n\in\mathds{N}\,,
\end{equation}
where we introduced the coefficients $B_a$ and $D_a$ as follows for future convenience: 
\begin{align}\label{eq_Ba_toy}
B_a=-\frac{ b_a+2 z}{  \sqrt{z}}
\,,\qquad
D_a=\frac{3 d_a+2 z \left(3 b_a^2+6 b_a z+4 z^2\right)}{6 z^{3/2}}
\,,
\end{align}
supposing $B_a\sim D_a\sim O(1)$. In the parenthesis of eq.~\eqref{eq_toy_a_disp}, we neglected a $(\ellb_s\omega)^{3/2}n^0$ term, since this receives contributions from an operator built of higher derivatives of the $X$'s, which we neglected in~\eqref{eq_toy_bdry_a}.\footnote{This contribution arises from $(\nabla^3_t X)^2 (\hat{\nabla}_t a)^2$.}

Eq.~\eqref{eq_toy_a_disp} clearly illustrates the power-counting rules. The first correction to the dispersion relation arises from the first term in eq.~\eqref{eq_toy_bdry_a} and is suppressed by $\sim\sqrt{\omega\ellb_s}$. We also find a calculable $\sim\omega\ellb_s$ term determined by the same coefficient $B_a$ of the $\sim\sqrt{\omega\ellb_s}$ correction. Finally, the second term in eq.~\eqref{eq_toy_bdry_a} determines the $(\omega \ellb_s)^{3/2} n^2$ correction to the dispersion relation. By comparing the terms proportional to $B_a$ and $D_a$ in eq.~\eqref{eq_toy_a_disp} we conclude that the derivative expansion is under control for
\begin{equation}\label{eq_derivative_expansion_toy}
\omega\ellb_s n^2\ll 1 \,.
\end{equation}
The regime specified by~\eqref{eq_derivative_expansion_toy} is smaller than the one that one would have guessed neglecting the redshift factor, namely $\epsilon^2\omega^2\ll \ellb_{s}^{-2}$, i.e.  $n^2(\omega\ellb_s)^2\ll 1$.  As we will see, the derivative expansion assumes a similar structure in the calculation of the energy for the daughter Regge trajectories.

We conclude this section with two important technical remarks for what follows.
\begin{itemize}
\item The expected expansion arises as a consequence of the cutoff on the string length at $(1-\alpha_f)=z \ellb_s\omega$. Since the endpoint metric $\hat{G}_{\tau\tau}$ becomes singular for $\alpha_f\rightarrow 1$, we may think of a positive nonzero $z$ as a regulator of the classical boundary divergences. While working with a fixed $z>0$ is justified on physical grounds, it is also possible to achieve the same results by working directly in the limit $z\rightarrow 0$, working thus with null endpoints. To this aim, we introduce the following \emph{alternative metric} at the endpoints:
\begin{equation}\label{eq_bdry_metric2}
\hat{\gamma}_{\tau\tau}=\ellb_s\left[-\left(\hat{\nabla}_\tau \pd_\tau X^\mu\right)\eta_{\mu\nu} \left(\hat{\nabla}_\tau \pd_\tau X^\nu \right)\right]^{1/2}\,,
\end{equation}
which is the square-root of the endpoints' acceleration and therefore constitutes a singular operator on a static background. However, it is simple to check that $\hat{\gamma}_{\tau\tau}$ scales as $\ellb_s/\omega$ on the rotating background, like the wordline metric:\footnote{Note that the worldline connection $\hat{\Gamma}^\tau_{\tau\tau}=\frac12\hat{G}_{\tau\tau}^{-1}\pd_\tau\hat{G}_{\tau\tau}$ vanishes for any $z$ on the background solution.}
\begin{equation}
\hat{\gamma}_{\tau\tau}=\frac{\ellb_s}{\omega}(1-z\ellb_s\omega)\simeq \frac{1}{z}\hat{G}_{\tau\tau}\,.
\end{equation}
We will see in the next section that $\hat{\gamma}_{\tau\tau}$ and $\hat{G}_{\tau\tau}$ are in fact equivalent and can be used interchangeably around the rotating background also when the string fluctuates. It is also possible to check that all the other operators with the right transformation properties for defining a worldline metric are subleading in $\omega$. Therefore, we may write an action equivalent to~\eqref{eq_toy_bdry_a} as
\begin{equation}\label{eq_toy_bdry_a_2}
S_{bdry}=-\sum_{\sigma=\pm 1}\frac12\int d\tau\sqrt{\hat{\gamma}_{\tau\tau}}\left[\ellb_s \tilde{b}_a(\hat{\nabla}_\tau a)^2\hat{\gamma}_{\tau\tau}^{-1}+\ellb_s^3\tilde{d}_a(\hat{\nabla}_\tau^2a)^2\hat{\gamma}_{\tau\tau}^{-2}+\ldots\right]\,.
\end{equation}
The action~\eqref{eq_toy_bdry_a_2} is manifestly reparametrization invariant. The main advantage of the formulation~\eqref{eq_toy_bdry_a_2} is that it holds for arbitrary $z$ and it allows us to work directly with null endpoints setting $z=0$, thus getting rid of the terms resulting from the expansion of $1-\alpha_f$ in eqs.~\eqref{eq_toy_bc_explicit} and~\eqref{eq_toy_s_vs_sigma}. 

Note that for null boundaries one cannot define a normalized normal vector, therefore $\pd_n a$ in the left-hand side of eq.~\eqref{eq_toy_bc} is not well defined. To write the boundary conditions in a geometric form, we note that in terms of the coordinate~\eqref{eq_s_def} the metric takes the form (for $\alpha_f=1$):
\begin{equation}\label{eq_metric_conformal}
    ds^2=\frac{\cos^2(s)}{\omega^2}\left(d\tau^2-ds^2\right)\,,
\end{equation}
which shows that at the endpoint we have a natural conformal geometric structure. This structure is preserved by diffeomorphism of the form $\tau\pm s\rightarrow f(\tau\pm s)$, under which both $\pd/\pd\tau$ and $\pd/\pd s$ transform as vectors.  
Therefore, for null endpoints we may write the boundary conditions that follow from the action~\eqref{eq_S2_toy}$+$\eqref{eq_toy_bdry_a_2} in geometric form as
\begin{equation}\label{eq_toy_bc_smart}
\pm   \pd_s a \vert_{s\pm\frac{\pi}{2}}=\bar{\ell}_s\tilde{b}_a
\hat{\nabla}_{\tau}\left(\hat{\gamma}_{\tau\tau}^{-1/2}\hat{\nabla}_{\tau}a\right)+
\bar{\ell}_s^3\tilde{d}_a \hat{\nabla}^2_{\tau}\left(\hat{\gamma}_{\tau\tau}^{-3/2}\hat{\nabla}^2_{\tau}a\right)+\ldots\,.
\end{equation}
We see that both sides manifestly transform homogeneously under diffeomorphism of the form $\tau\pm s\rightarrow f(\tau\pm s)$.\footnote{Equivalently, in appendix~\ref{app_geometry} we show that the left-hand side of eq.~\eqref{eq_toy_bc_smart} may be written in terms of a Weyl invariant normal $N^\alpha_\tau$, that makes sense for both timelike and null boundaries.}

It is simple to solve~\eqref{eq_toy_bc_smart} and check that the energy levels following from~\eqref{eq_toy_bdry_a_2} take the form~\eqref{eq_toy_a_disp} with the identifications 
\begin{equation}
B_a=\sqrt{2}\,\tilde{b}_a\,,\qquad
D_a=\sqrt{2}\,\tilde{d}_a\,.
\label{eq_toy_B_D_2}
\end{equation}
which are significantly simpler than~\eqref{eq_Ba_toy} since we work in the limit $z\rightarrow 0$.
\item The boundary conditions~\eqref{eq_toy_bc} are imposed at the string endpoint $\sigma=1$. Since the solutions of the EOMs are functions of $\alpha_f\sigma$, when solving the boundary condition~\eqref{eq_toy_bc} perturbatively in $\omega$ for $z>0$, we further need to expand $(1-\alpha_f)$ using~\eqref{eq_toy_s_vs_sigma}; as a result, the physical parameters $B_a$ and $D_a$ that enter the dispersion relation~\eqref{eq_toy_a_disp} are nontrivial linear combinations of the ones multiplying derivatives of $a$ in eq.~\eqref{eq_toy_bc}. It is possible and technically more convenient to write the boundary conditions directly in terms of the expansion of the field around the singular point $\alpha_f\sigma=1$ of the EOMs.
To this aim, one considers the following (near boundary) expansion of the solutions to eq.~\eqref{eq_a_massless}:\footnote{This expansion is analogous to the mode expansion of bulk fields near the boundary of spacetime in the AdS/CFT.}
\begin{equation}\label{eq_a_near}
\begin{split}
a(\tau,\sigma) &=   \left[1+(1-\alpha_f\sigma)\frac{d^2}{d\tau^2}+
\frac16(1-\alpha_f\sigma)^2
\left(\frac{d^2}{d\tau^2}+\frac{d^4}{d\tau^4}\right)+
O\left((1-\alpha_f\sigma)^3\right)\right]
\beta_1(\tau)
\\ & +
\sqrt{1-\alpha_f\sigma}\left[1+(1-\alpha_f\sigma)
\left(\frac{1}{12}+\frac{1}{3}\frac{d^2}{d\tau^2}\right)
+O\left((1-\alpha_f\sigma)^2\right)\right]\beta_2(\tau)\,,
\end{split}
\end{equation}
where $\beta_1(\tau)$ and $\beta_2(\tau)$ are the boundary \emph{data} which specify the solution. The expansion for $1+\alpha_f\sigma\rightarrow 0$ is identical. The boundary condition~\eqref{eq_toy_bc} is conveniently solved perturbatively in terms of the boundary modes $\beta_1$ and $\beta_2$ as (at both endpoints)
\begin{equation}\label{eq_toy_bc2}
\beta_2=\sqrt{\omega \ellb_s}B_a\ddot{\beta}_1
+(\omega\ellb_s)^{3/2}
\left[D_a \frac{d^4\beta_1}{d\tau^4}+O\left(\ddot{\beta}_1\right)\right]
+O\left(\omega^{5/2}\ellb_s^{5/2}\right)\,.
\end{equation}
Notice that eq.~\eqref{eq_toy_bc2} manifestly depends solely upon the coefficients $B_a$ and $D_a$ (and not on $z$) that determine the dispersion relation~\eqref{eq_toy_a_disp}. This is because the boundary conditions~\eqref{eq_toy_bc2} (at both endpoints) completely determine the profile of the field in the bulk of the string with no reference to the regulator parameter $z$. We will use near boundary expansions of the form~\eqref{eq_a_near} in the next sections to study fluctuations of the string coordinates.
\end{itemize}

\subsection{The general boundary action}\label{subsec_general_Sbdry}

\subsubsection{The construction}

\hspace{5mm}
Let us now discuss the endpoints of a dynamical string.  Namely, we want to write boundary actions that encode the boundary conditions for the $X^\mu(\tau,\sigma)$ fields at $\sigma=\pm 1$ in a consistent derivative expansion. As in the scalar example previously discussed,  the boundary conditions must be obtained from the variation of a Poincar\'e invariant functional made of derivatives of the string coordinates at the endpoints.
First, let us consider the simplest possible endpoint action:
\begin{equation}\label{eq_Sbdry_Massive}
S_{bdry}=-\sum_{\sigma=\pm1} m\int d\tau\sqrt{\hat{G}_{\tau\tau}}\,,
\end{equation}
where, here and for most of this section, we assume identical endpoints.
Eq.~\eqref{eq_Sbdry_Massive} models mass $m$ quarks at the endpoints. More generally, we expect that such a term will always be present in an effective description of the endpoints. In this section, we assume the minimal scenario where the quark masses are comparable or smaller than the Yang-Mills scale, i.e., $m\sim 1/\ellb_s$. We postpone the discussion of heavy quarks $m\gg 1/\ellb_s$ to Section~\ref{sec_heavy_quarks}.

The boundary conditions that follow from \eqref{eq_Sbdry_Massive} and \eqref{eq_Sbulk} read
\begin{equation}\label{eq_bc_Massive}
\frac{1}{\ell_s^2}\sqrt{\hat{G}_{\tau\tau}}\,\pd_n X^\mu=-m\,\hat{G}_{\tau\tau}^{-1/2}\hat{\nabla}_{\tau}\pd_\tau X^\mu\,.
\end{equation}
Eq.~\eqref{eq_bc_Massive} on the ansatz~\eqref{eq_background} imposes 
\begin{equation}\label{eq_alpha_Massive}
\alpha_f=1-\pi m\omega\ellb_s^2+O\left(\ellb_s^2\omega^2\right)\,.
\end{equation}
As announced, for $m\sim 1/\ellb_s$ the quark mass term \emph{shortens} the string similarly to the cutoff prescription adopted in Section~\ref{sec_singularities}. Plugging back in the action, we see that the boundary action \eqref{eq_Sbdry_Massive} scales as $\sqrt{\omega/\ellb_s}$, while the NG term scales as $1/(\ell_s^2\omega^2)$ (recall $X\sim 1/\omega$ from eq.~\eqref{eq_background}). We may therefore think of the quark mass term~\eqref{eq_Sbdry_Massive} as a perturbation of the Neumann boundary conditions $\sqrt{\hat{G}_{\tau\tau}}\,\pd_n X^\mu=0$.  

Additional boundary operators are further suppressed. To see this, notice that leading boundary conditions~\eqref{eq_bc_Massive} imply that, working in a derivative expansion, we can remove redundant operators according to the relations\footnote{For bulk operators it is well known that operators vanishing on the leading EOMs can be eliminated via a field redefinition \cite{Weinberg:1996kr}. A similar statement holds also for boundary operators provided we allow for singular field redefinitions, that can be treated with standard methods within EFT, see e.g. \cite{Garcia-Saenz:2017wzf}. We checked explicitly that different choices of the basis of operators for the boundary action yield the same results for the leading and the daughter Regge trajectories.} 
\begin{equation}\label{eq_redundancy}
\frac{1}{\ell_s^2}\pd_n X^\mu\approx m \hat{\nabla}_t^2 X^\mu\quad
\text{and}\quad \hat{G}_{\tau\tau}\approx 2\pi\ellb_s m\, \hat{\gamma}_{\tau\tau}\,,
\end{equation}
where the second is obtained by squaring eq.~\eqref{eq_bc_Massive} and using $\pd_n X\cdot\pd_n X=-1$. Eq. \eqref{eq_redundancy} implies a somewhat unusual equivalence between the square-root of the quark accelaration and the worldline metric. This is a special property of the rotating background and does not hold for a static string. Using the relations ~\eqref{eq_redundancy} and the discussed geometric arguments, we power-count boundary operators according to the rules
\begin{equation}\label{eq_bc_counting}
X\sim \frac{1}{\omega}\,,\quad
\hat{G}_{\tau\tau}\sim \hat{\gamma}_{\tau\tau}\sim\frac{\ellb_s}{\omega}\,,\quad
\pd_n X\sim 1\,,\quad
\hat{\nabla}_t\sim  \sqrt{\frac{\omega}{\ellb_s}}\,.
\end{equation}
These rules allow us to estimate the relative size of a given operator contribution to an arbitrary observable.

It is now straightforward to write the first correction to the endpoint action~\eqref{eq_bc_Massive}.  The analogy with the massless scalar example presented in eq.~\eqref{eq_toy_bdry_a} is the simplest if we
momentarily choose a basis where according to~\eqref{eq_redundancy}, we systematically get rid of operators with multiple time derivatives $\hat{\nabla}^k_t X$, in favor of those with normal derivatives $\hat{\nabla}^{k-2}_t\pd_n X^\mu$. To the first subleading order, 
\begin{equation}\label{eq_Sbdry1}
S_{bdry}=-m\sum_{\sigma=\pm 1}\int d\tau\sqrt{\hat{G}_{\tau\tau}}
\left[1+\ellb_s^2 \bar{b}_1\left(\hat{\nabla}_t\pd_n X\cdot\hat{\nabla}_t\pd_n X\right)+\ldots\right]
\,,
\end{equation}
where $m\sim 1/\ellb_s$ and $\bar{b}_1\sim O(1)$. We assumed time reversal and neglected terms with more than overall $2$ time derivatives (also accounting for partial integration).  According to the power counting, the leading order scales as $1/\sqrt{\omega\ellb_s}$, while the subleading scales as $\sqrt{\omega\ellb_s}$. In general, the expansion runs in single powers of $\omega\ellb_s$ as expected.

Let us illustrate our discussion by looking at the Lagrangian and the Regge intercept for the theory defined by the NG action and the b.dry action \eqref{eq_Sbdry1}. To this aim, we compute the Lagrangian $\mL$ for arbitrary $\alpha_f$ on the background \eqref{eq_background}
\begin{equation}\label{eq_L_sec3}
\mL =-\frac{1}{\ell_s^2\omega} \left[\alpha_f\sqrt{1-\alpha_f^2}+\arcsin(\alpha_f)\right]
-2m\left[
\sqrt{1-\alpha_f^2}
-\bar{b}_1\frac{\ellb_s^2 \omega^2}{\sqrt{1-\alpha_f^2}}
\right]\,,
\end{equation}
where the first parenthesis arises from the bulk action,  the second one from the boundary terms. We then extremize with respect to $\alpha_f$:
\begin{equation}\label{eq_dalpha_pre}
\frac{\pd \mL}{\pd \alpha_f}=0\quad
\implies
\quad
\alpha_f=1-\pi m\omega\ellb_s^2+ \frac{1}{2}  \left(\pi ^2 \ellb_s^2 m^2-\bar{b}_1\right) \ellb_s^2\omega^2+O\left(\ellb_s^3\omega^3\right)\,.
\end{equation}

Plugging in eq. \eqref{eq_L_sec3}, we find
\begin{equation}\label{eq_L_sec3_2}
\mL=-\frac{1}{4\ellb_s^2\omega} -\frac{4\sqrt{2 \pi } }{3}m^{3/2}\ellb_s \sqrt{\omega}
+\frac{1}{5} \sqrt{\frac{2}{\pi }} \sqrt{m}\ellb_s \left(5 \bar{b}_1+2 \pi ^2 m^2 \ellb_s^2\right)\omega ^{3/2} +\ldots\,.
\end{equation}

Eq. \eqref{eq_L_sec3_2} contains terms at order $\sim\sqrt{\omega}$ and $\omega^{3/2}$ with independent coefficients ($m$ and $\bar{b}_1$), as it was expected from the discussion in Section~\ref{sec_singularities}.  We then obtain the relation between $\omega$ and $J$ from~\eqref{eq_JE_Noether}
\begin{equation}\label{eq_omega_J_1}
\begin{split}
\omega=\frac{1}{2\ellb_s\sqrt{J}}&\left[1-\frac{2(m\ellb_s)^{3/2}}{3J^{3/4}}
+\frac{3 \sqrt{\ellb_s m} \left(5 \bar{b}_1+2 \pi ^2 \ellb_s^2 m^2\right)}{20 \sqrt{\pi }\,J^{5/4}}
+\frac{4(m\ellb_s)^3\pi}{9 J^{3/2}}
+O\left(\frac{1}{J^{7/4}}\right)
\right]\,,
\end{split}
\end{equation}
from which we get the Regge intercept (neglecting quantum corrections):
\begin{equation}\label{eq_leading_Regge_1}
\begin{split}
M^2_{classical}&=\frac{J}{\ellb_s^2}+\frac{8 \sqrt{\pi } \,m^{3/2}}{3 \sqrt{\ellb_s}} J^{1/4}
-\frac{\sqrt{m} \left(5 \bar{b}_1+2 \pi ^2 \ellb_s^2 m^2\right)}{5 \sqrt{\pi } \ellb_s^{3/2}}J^{-1/4}\\
&+\frac{4}{3} \pi  \ellb_s m^3 J^{-1/2}+O\left(J^{-3/4}\right)\,.
\end{split}
\end{equation}
The existence of a $J^{1/4}$ correction from the boundary mass term is a well known result. The existence of a $J^{-1/4}$ term (and in general of $\sim J^{1/4-n/2}$ with $n\in\mathds{N}$ contributions) was previously discussed in \cite{Hellerman:2013kba,Hellerman:2016hnf} in the Polyakov formalism using a seemingly different EFT for the endpoints. We will present an equivalent formulation of the action~\eqref{eq_Sbdry1} which is more closely related to that of those works in the next section. Finally the $J^{-1/2}$ contribution arises when using~\eqref{eq_omega_J_1} to relate $\omega$ and $J$ at subleading orders.

In eq.~\eqref{eq_leading_Regge_1} we neglected quantum corrections. These provide a $\sim 1/J$ relative correction to the result. We will discuss this point further in Section~\ref{sec_sqrt_formula}.

We will soon show that the expansion for daughter Regge trajectories is similar to eq.~\eqref{eq_leading_Regge_1}. The small $\ellb_s\omega$ expansion for the length of the string $L=2\alpha_f/\omega$ takes instead a different form, with the first correction at order $\ellb_s\omega\sim1/\sqrt{J}$,  see~\eqref{eq_dalpha_pre}, therefore seemingly violating the power-counting.  The reason why this is allowed is that the string length, at least as defined in eq.~\eqref{eq_string_length}, is not a physical observable within EFT. Physically, this is because the string length depends upon the precise size of the endpoints, which are fuzzy objects within EFT. Technically, this is because the definition~\eqref{eq_string_length} is not invariant under field redefinitions, which are instead crucial to get rid of redundant operators via~\eqref{eq_redundancy}.  We will indeed see that a different choice for the basis of operators of the boundary action leads to a different value for $1-\alpha_f$ without affecting \emph{on-shell} observables.

\subsubsection{Alternative parametrization and the effective string length}

\hspace{5mm}
In principle, we could now proceed and use the action~\eqref{eq_Sbdry1} to compute the spectrum of the rotating string up to relative order $\sim J^{-3/2}$. However, doing so would require us to deal with some technical complications at intermediate steps of the calculation. This is because the expansion of the operators in eq.~\eqref{eq_Sbdry} yields inverse powers of $1-\alpha_f$, that in turn introduce spurious powers of $\omega$ due to eq.~\eqref{eq_dalpha_pre}. While these spurious powers eventually cancel from the final result, the power counting and the expansion discussed in the previous sections are obscured in intermediate steps.

To remedy this state of affair, in this section we use the redundancies~\eqref{eq_redundancy} to write the boundary action in a different basis of operators,  where the \emph{effective} quarks move at the speed of light $\alpha_f=1$ to all orders in the $1/J$ expansion.  In this basis the power counting discussed in the previous section remains manifest also at intermediate steps. Additionally, the derivation of this simplified approach will also clarify the physical reasons for the validity of the EFT approach.

The discussion proceed in two steps.  First, we note that the relations~\eqref{eq_redundancy} allow to choose a different basis of operators, where $\pd_n X$ does not appear at all, and where we trade $\hat{G}_{\tau\tau}$ for $\hat{\gamma}_{\tau\tau}$. From these observations, we infer that the action~\eqref{eq_Sbdry1} is equivalent to the following 
\begin{equation}\label{eq_Sbdry}
\begin{split}
S_{bdry} =-m\sum_{\sigma=\pm 1}&\left[
b_m \int d\tau\sqrt{\hat{G}_{\tau\tau}}-b_{\gamma}
\int d\tau\sqrt{\hat{\gamma}_{\tau\tau}}\right.
\\+
&
\left. \ellb_s^4b_1\int d\tau\sqrt{\hat{\gamma}_{\tau\tau}}  (\tilde{\nabla}_\tau^3X)^2\hat{\gamma}_{\tau\tau}^{-3} 
+O\left((\ellb_s\omega)^{3/2}\right)
\right]
\,,
\end{split}
\end{equation}
where in the second line the covariant derivative $\tilde{\nabla}_\tau$ is defined using the following connection
\begin{equation}
\tilde{\Gamma}^{\tau}_{\tau\tau}=\frac{1}{2}\hat{\gamma}_{\tau\tau}^{-1}\pd_\tau\hat{\gamma}_{\tau\tau}\,.
\end{equation}
Note that the usual connection $\hat{\Gamma}^{\tau}_{\tau\tau}=\frac{1}{2}\hat{G}_{\tau\tau}^{-1}\pd_\tau\hat{G}_{\tau\tau}$ still appears in the definition~\eqref{eq_bdry_metric2} of~$\hat{\gamma}_{\tau\tau}$. According to the power-counting, the first line of eq.~\eqref{eq_Sbdry} scales as $\sim 1/\sqrt{\ellb_s\omega}$,  while the second as $\sim \sqrt{\ellb_s\omega}$. 

The action~\eqref{eq_Sbdry} is not yet the most convenient for our purposes. Nonetheless, it is useful to discuss some of its features before performing further manipulations. Note in particular that in the first line of~\eqref{eq_Sbdry} we retained an arbitrary linear combination of $\hat{G}_{\tau\tau}$ and $\hat{\gamma}_{\tau\tau}$. However, due to the discussed redundancy only two combinations of the coefficients $b_m$, $b_\gamma$, and $b_1$ enter in physical observables.  We will find below that these combinations are
\begin{align}\label{eq_bm}
\mu_1&=-\sqrt{2} b_{\gamma} m+
\frac{4}{3} \sqrt{\pi } \,b_m^{3/2}m^{3/2}\ellb_s^{1/2}
\,,\\
\mu_2&=-m\frac{b_1 }{\sqrt{2}}+\ellb_s m^2\frac{b_{\gamma}b_m\pi}{2\sqrt{2}}
-\ellb_s^{3/2} m^{5/2}\frac{\pi ^{3/2}}{5} b_m^{5/2} \,. \label{eq_bm2}
\end{align}
Thus $\mu_1$ and $\mu_2$ are the true non-redundant Wilson coefficient: on dimensional grounds we expect $\mu_1\sim \mu_2\sim 1/\ellb_s$ for a theory with light quarks.  Nonetheless, the leading correction to the string length, which is not an on-shell observable as previously observed, is controlled only by $b_m$:
\begin{equation}\label{eq_dalpha}
\begin{split}
\alpha_f&=1-b_m m\omega\ellb_s^2\pi+\left(
b_{\gamma}\frac{\sqrt{b_m \ellb_s^3m^3\pi^3 }}{\sqrt{2}}
-\frac{\pi^2}{2}  b_m^2 \ellb_s^2 m^2\right)\ellb_s^2\omega^2\\
&+\left(
\frac{\pi^2}{4}b_{\gamma}^2\ellb_s^2m^2-b_1\frac{\sqrt{b_m\ellb_s^3m^3\pi^3}}{\sqrt{2}}\right)
\ellb_s^3\omega^3+O\left(\ellb_s^4\omega^4\right)\,.
\end{split}
\end{equation}
This observation has an important physical consequence: by choosing a sufficiently large coefficient $b_m$ while keeping fixed the coefficients $\mu_1$ and $\mu_2$ in eqs.~\eqref{eq_bm} and~\eqref{eq_bm2}, we can effectively cutoff the string before the derivative expansion breaks down without modifying the spectrum. We may therefore think of $b_m $ as a \emph{regulator} for the endpoints singularities.\footnote{As usual, quantum corrections require an additional, different, regulator - such as dimensional regularization.}
In some sense, the freedom of changing the effective string length via a redundant coupling provides a technical explanation for the validity of the EFT approach. 

Retaining a finite  value of~$1-\alpha_f>0$ will be important when discussing bulk operators in Section~\ref{sec_renormalization}, but, as announced at the beginning of this section, to make the power counting manifest in our calculations we would like to work in a scheme in which the endpoints move at the speed of light exactly. Naively, we might try to achieve this by setting $b_m=0$ in eq.~\eqref{eq_Sbdry}, but this is not enough due to the subleading terms in the expansion~\eqref{eq_dalpha}.  Instead, it is convenient to perform a further step and introduce an auxiliary einbein field $e$ to rewrite the first term in eq.~\eqref{eq_Sbdry} as
\begin{equation}
-m\sum_{\sigma=\pm 1}
b_m \int d\tau\sqrt{\hat{G}_{\tau\tau}}\rightarrow-\frac{m}{2}\sum_{\sigma=\pm 1}
 \int d\tau\left(\frac{\dot{X}^2}{e}+b_m^2 e\right)\,,
\end{equation}
while leaving the rest of the action unchanged. Solving for $e$ on its EOMs for $b_m>0$ we recover the previous expression (assuming $e\geq 0$).  If we instead set the redundant coefficient to zero, $b_m= 0$, we find that the EOM for the einbein requires the string endpoints to be precisely null:
\begin{equation}\label{eq_constraint_null}
\dot{X}^2|_{\sigma=\pm1}= 0\,.
\end{equation}
Note that this constraint implies that the modified boundary metric~\eqref{eq_bdry_metric2} can be simply written as
\begin{equation}\label{eq_bdry_metric2_smart}
\hat{\gamma}_{\tau\tau}=\ellb_s\left(-\ddot{X}^\mu\eta_{\mu\nu} \ddot{X}^\nu \right)^{1/2}\,,
\end{equation}
with no reference to a connection.\footnote{To derive eq.~\eqref{eq_bdry_metric2} we used that the connection~$\hat{\Gamma}^\tau_{\tau\tau}$ remains finite in the limit $\dot{X}^2\rightarrow 0$,  as it can be checked solving for~\eqref{eq_constraint_null} explicitly as in eq.~\eqref{eq_sol_alpha} below.}
We therefore conclude that we can work in a scheme where the endpoints' effective action reads
\begin{equation}\label{eq_Sbdry_smart}
\begin{split}
S_{bdry} =m\sum_{\sigma=\pm 1}\left[b_{\gamma}
\int d\tau\sqrt{\hat{\gamma}_{\tau\tau}}
-\ellb_s^4b_1\int d\tau\sqrt{\hat{\gamma}_{\tau\tau}}  (\tilde{\nabla}_\tau^3X)^2\hat{\gamma}_{\tau\tau}^{-3} 
+O\left((\ellb_s\omega)^{3/2}\right)
\right]
\,,
\end{split}
\end{equation}
with the constraint~\eqref{eq_constraint_null} that the endpoints move on lightlike trajectories.  The action~\eqref{eq_Sbdry_smart} is analogous to the one discussed in~\cite{Hellerman:2016hnf}, where the authors worked in Polyakov formalism. Note that eq.~\eqref{eq_Sbdry_smart} does not include operators containing $\pd_s X^\mu$, where $s$ is the natural normal coordinate in the null case according to the discussion around~\eqref{eq_metric_conformal}, since these vanish on the leading order boundary conditions that follow from the Nambu-Goto action.

As an illustration, let us discuss the calculation of the leading Regge trajectory again, using the action~\eqref{eq_Sbdry_smart}. The constraint~\eqref{eq_constraint_null} imposes that on the background profile~\eqref{eq_background} we must set
\begin{equation}
\alpha_f=1\,.
\end{equation}
The Lagrangian is then straightforward to compute. Using~\eqref{eq_bdry_metric2_smart} and noticing that $\tilde{\Gamma}^\tau_{\tau\tau}=0$ on the solution~\eqref{eq_background}, we find
\begin{equation}
    \mL=-\frac{1}{4\ellb_s^2 \omega }
    -\sqrt{2} \mu_1\sqrt{\ellb_s\omega}  -2\sqrt{2}  \mu_2  (\ellb_s  \omega)^{3/2}+O\left(\ellb_s^{5/2}\omega^{5/2}\right)\,,
\end{equation}
where the Wilson coefficients are the ones given in~\eqref{eq_bm} and~\eqref{eq_bm2} in the limit $b_m\rightarrow 0$:
\begin{equation}
\mu_1\stackrel{b_m\rightarrow 0}{=}-\sqrt{2}m b_\gamma\,,\qquad
\mu_2\stackrel{b_m\rightarrow 0}{=}-\frac{1}{\sqrt{2}}m b_1\,.
\end{equation}
We then find $\omega$ as 
\begin{equation}\label{eq_omega_J_exp}
\omega=\frac{1}{2\ellb_s}J^{-1/2}-\frac{\mu_1}{4} J^{-5/4} -\frac{3 \mu_2}{4} J^{-7/4}
+
O\left(J^{-9/4}\right)\,,
\end{equation}
from which we recover the same expansion in eq.~\eqref{eq_leading_Regge_1} (neglecting the Casimir energy)
\begin{equation}\label{eq_Regge}
M_{classical}=\frac{\sqrt{J}}{\ellb_s}+\frac{\mu_1}{J^{1/4}}+\frac{\mu_2}{J^{3/4}}
+\frac{\mu_1^2\ellb_s}{8 }J^{-2}+
O\left(J^{-5/4}\right)\,.
\end{equation}
The same result is obtained using the action~\eqref{eq_Sbdry}, with the Wilson coefficients given by eq.~\eqref{eq_bm} and~\eqref{eq_bm2}.
On physical grounds, we expect that the effective endpoints contribute to the energy with a positive amount, hence $\mu_1>0$; this can indeed be proven rigorously using Regge theory and the analytic S-matrix bootstrap \cite{Sever:2017ylk}.

In the following sections, we will discuss the calculation of the spectrum of the daughter Regge trajectories. We will mostly work with exactly massless endpoints, but we will also comment on the calculation using the action~\eqref{eq_Sbdry}, as this will be important when discussing the renormalization of bulk operators. An important subtlety when working with massive endpoints with the action~\eqref{eq_Sbdry} is that it is important to keep $b_m>0$ at intermediate steps. This is because keeping $b_m>0$ ensures that the first two corrections to $1-\alpha_f$ in eq.~\eqref{eq_dalpha} do not vanish and, as we will see, the expansion into fluctuations of the operators in eq.~\eqref{eq_Sbdry} yields inverse powers of $1-\alpha_f$, which obscure the power-counting as mentioned in the introduction to this section. We checked that the final results for the spectrum are independent of the scheme when expressed in terms of the physical Wilson coefficients~\eqref{eq_bm} and~\eqref{eq_bm2}.

\subsection{Fluctuations}

\subsubsection{Leading order analysis}\label{subsec_LO_fluct}

\hspace{5mm}
We parametrize fluctuations around the background~\eqref{eq_background} with $d-1$ fields $\{\alpha,\,\phi ,\,\rho^i\}$ as
\begin{equation}\label{eq_flucts}
X^\mu=\frac{1}{\omega}\left\{\tau, \left[\alpha_f\, \sigma+\alpha(\tau,\sigma) \right]\cos\left(\tau+\phi(\tau,\sigma)\right),
\left[\alpha_f\, \sigma+\alpha(\tau,\sigma) \right] \sin\left(\tau+\phi(\tau,\sigma)\right),\rho_i(\tau,\sigma)\right\}\,,
\end{equation}
where $i=1,\ldots,d-3$. One might think that $\alpha(\tau,\sigma)$ can be removed by a gauge transformation. This is true everywhere but at $\sigma=\pm 1$, since we are only allowed to consider diffeomorphisms that preserve the position of the boundaries. We will see that $\alpha(\tau,\sigma)$ enters in the bulk action only in total derivative terms, and only the $\alpha(\tau,\pm 1)$ boundary modes are physical. The parametrization~\eqref{eq_flucts} was formerly used in \cite{Sonnenschein:2018aqf}.

Let us consider first the NG action, for arbitrary $\alpha_f$, neglecting the boundary terms. Neglecting total time derivatives, the expansion to quadratic order yields:
\begin{equation}\label{eq_S2}
\begin{split}
S_{bulk}&\simeq \frac{ \alpha_f }{2\omega^2\ell_s^2}\int d^2\sigma\sqrt{1-\alpha_f^2\sigma^2}\left[\frac{(\dot{\rho}_i)^2}{1-\alpha_f^2\sigma^2}-\frac{(\rho_{i}')^2}{\alpha_f^2}
+\frac{\alpha_f^2\sigma^2\dot{\phi}^2}{(1-\alpha_f^2\sigma^2)^2}-\frac{\sigma^2(\phi')^2}{1-\alpha_f^2\sigma^2}\right]\\
&+\frac{1}{\omega^2\ell_s^2}\int d^2\sigma\frac{\pd}{\pd\sigma}
\left[-\sqrt{1-\alpha_f^2\sigma^2}\,\alpha+\frac{\alpha_f\sigma\alpha^2}{2\sqrt{1-\alpha_f^2\sigma^2}}+\frac{\alpha_f^2\sigma^2\dot{\phi}\,\alpha}{\sqrt{1-\alpha_f^2\sigma^2}}\right]
\,.
\end{split}
\end{equation}
As anticipated, $\alpha$ enters only in a total derivative term, and is thus purely a boundary mode.

As formerly noted, the constraint~\eqref{eq_constraint_null} implies $\alpha_f=1$ at leading order in fluctuations; this sets to zero the term linear in $\alpha$ in eq.~\eqref{eq_S2}. At subleading order, we may solve perturbatively for the null constraint in terms of the boundary mode $\alpha$:
\begin{equation}\label{eq_sol_alpha}
    \alpha\vert_{\sigma=\pm 1}=-\sigma\dot{\phi}+\sigma\left(\dot{\phi}^2-\frac12\ddot{\phi}^2-\frac12\dot{\rho}^2\right)+\ldots\,,
\end{equation}
where the term that is quadratic in the fields will be important to obtain the expansion of the boundary action~\eqref{eq_Sbdry_smart} to quadratic order, that we report in the next subsection. Plugging back into eq.~\eqref{eq_S2} we obtain
\begin{equation}\label{eq_S2_2}
\begin{split}
S_{bulk}&\simeq \frac{ 1 }{2\ell_s^2\omega^2}\int d^2\sigma\sqrt{1-\sigma^2}\left[\frac{(\dot{\rho}_i)^2}{1-\sigma^2}-(\rho_i')^2
+\frac{\sigma^2\dot{\phi}^2}{(1-\sigma^2)^2}-\frac{\sigma^2(\phi')^2}{1-\sigma^2}\right]\\
&-\frac{1}{2\ell_s^2\omega^2}\sum_{\pm}\lim_{\sigma\rightarrow\pm 1}\int d\tau
\frac{\sigma^2\dot{\phi}^2}{\sqrt{1-\sigma^2}}\,.
\end{split}
\end{equation}
The boundary term in the second line ensures that the action is finite for configurations such that the angular mode $\phi$ does not vanish at the endpoints.  
Within the standard approach, in which eq.~\eqref{eq_constraint_null} is not imposed as a constraint, the action~\eqref{eq_S2_2} is obtained noticing that there is no kinetic term for $\alpha$ in eq.~\eqref{eq_S2} and we can therefore integrate it out on its EOM.

We are now ready to extract the leading order spectrum from the action~\eqref{eq_S2_2}. The action for the transverse mode(s) $\rho_i$ is identical to that of a massless scalar in eq.~\eqref{eq_S2_toy}, yielding the EOM~\eqref{eq_a_massless} with Neumann boundary conditions at the leading order:
\begin{equation}\label{eq_rho_bc_LO}
\lim_{\sigma\rightarrow\pm 1}\frac{\mp\sqrt{1-\sigma^2}}{\ell_s^2\omega^2}\rho'_i=0\,.
\end{equation}
Proceeding as in Section~\ref{subsec_toy_bc}, we conclude that each of the transverse modes is associated with a spectrum of excitations whose energy levels are integerly spaced in units of the angular velocity:
\begin{equation}\label{eq_rho_disp_LO}
\epsilon_n = n\in\mathds{N}\,.
\end{equation}

At a quantum level, the modes with $n\geq 1$ in eq.~\eqref{eq_rho_disp_LO} are associated with creation and annihilation operators $\hat{a}^\dagger_{\rho_i,n},\,\hat{a}_{\rho_i,n}$, and generate the corresponding Fock space. The $n=0$ mode as usual needs to be treated differently and is associated with translations of the string. Since we can always work in the rest-frame of the particle this mode is irrelevant for our purposes. Also the $n=1$ mode is special. To understand why, it is simpler to specialize momentarily to $d=4$ where there is a single transverse coordinate. The creation operator associated with the $n=1$ mode acts on the vacuum as
\begin{equation}\label{eq_arho1_gapped_NGB}
\hat{a}_{\rho,1}^\dagger|J,J\rangle=|J+1,J\rangle\,,
\end{equation} 
where $|J,J\rangle$ indicates the EFT ground state,  which is a highest weight state of the spin $J$ representation~\cite{Cuomo:2020fsb}. Therefore $\hat{a}_{\rho,1}^\dagger$ creates a non-highest weight state.  The same is true for all the $n=1$ modes in $d\geq 4$.
In more physical terms, the $n=1$ modes create states which can be obtained by applying a rotation to states with a different $J$ on the leading Regge trajectory.
Hence, to study the spectrum of particles in spin-$J$ representations it is enough to focus on highest weight states. Therefore only the $n\geq 2$ modes are relevant for our discussion.

Let us now discuss the angular variable $\phi$. The EOM reads
\begin{equation}\label{eom_phi}
\ddot{\phi}-(1-\sigma^2)\phi''-\left(2-\sigma^2\right)\phi'=0\,.
\end{equation}
The most general solution may be written as a superposition of plane waves of the form (see also \cite{Sonnenschein:2018aqf})
\begin{equation}\label{eq_phi}
\phi=e^{-i\epsilon\tau}
\frac{(1-\sigma ^2)^{3/4} }{\sigma }
\left[c_1 P_{\epsilon -\frac{1}{2}}^{\frac{3}{2}}(\sigma )+c_2 \,Q_{\epsilon -\frac{1}{2}}^{\frac{3}{2}}(\sigma )\right]
\,,
\end{equation}
where $P$ and $Q$ are Legendre functions. For $\alpha_f<1$, the solution is given by eq.~\eqref{eq_phi} replacing $\sigma\rightarrow \alpha_f \sigma$. The overall $1/\sigma$ prefactor is due to the parametrization~\eqref{eq_flucts}, which implies that $\delta X\propto \sigma\phi $ to linear order.  The ratio $c_2/c_1$ and the allowed values of $\epsilon$ in eq.~\eqref{eq_phi} are found imposing the boundary conditions at both endpoints: 
\begin{equation}\label{eq_phi_LO_bc}
\lim_{\sigma\rightarrow\pm 1}\frac{\mp\phi'+\sigma^2\ddot{\phi}}{\omega^2\ell_s^2\sqrt{1-\sigma^2}}=0\,.
\end{equation}
The limit in this expression naively looks singular, but it is not. This is because, as for the transverse fluctuations, the endpoints $\sigma=\pm1$ are singular points of the differential equation~\eqref{eom_phi}. To see that the limit is indeed well defined, it is convenient to consider the near boundary expansion for $\phi$:
\begin{equation}\label{eq_phi_near}
\begin{split}
\phi&=
\beta_1^\phi(\tau)-(1-\sigma)\ddot{\beta}_1^{\phi}(\tau)+
(1-\sigma)^{3/2} \beta_2^\phi(\tau)+O\left((1-\sigma)^2\right)\,,
\end{split}
\end{equation}
where we use a superscript to distinguish the transverse and angular boundary modes.  The expansion for $1+\sigma\rightarrow 0$ is identical. Plugging~\eqref{eq_phi_near} in~\eqref{eq_phi_LO_bc}, we see that the first two terms cancel and in the limit $\sigma\rightarrow \pm 1 $ we are left with a nonsingular result
\begin{equation}\label{eq_phi_LO_bc_near}
\beta_2^\phi=0\,.
\end{equation}
Imposing eq.~\eqref{eq_phi_LO_bc_near} at both endpoints implies that the solutions~\eqref{eq_phi} must satisfy
\begin{equation}\label{eq_phi_disp_LO}
c_2=0\quad \text{and}\quad
\epsilon_n= n \in\mathds{N}\,.
\end{equation}
Eq.~\eqref{eq_phi_disp_LO} yields a Fock space spectrum identical to that of the transverse modes, as expected from the standard results in critical string theory (this calculation is independent of the number of spacetime dimensions). 

As for the transverse coordinates, the $n=0$ and $n=1$ modes are associated, respectively, with rigid rotations and translations; therefore they must be excluded when constructing the spectrum of physical particles at fixed angular momentum. In particular,  highest-weight states in different representations are related via the action of the $0$-mode of $\phi$, which enters the mode decomposition via
\begin{equation}
\phi=
\hat{\phi}_0+\hat{p}_0\tau+\text{wave-modes}\,,
\end{equation}
where $\hat{p}_0\propto \langle J\rangle-\hat{J}$ and $\hat{\phi}_0$ is a $2\pi$-periodic quantum mechanical variable. The canonical commutation relation imply that $\left[\hat{\phi}_0,\hat{J}\right]=i$, therefore the operator $e^{i\hat{\phi}_0}$ increases the angular momentum; e.g. in four dimensions in the same notation of eq.~\eqref{eq_arho1_gapped_NGB} we have
\begin{equation}\label{eq_exp_i_phi0}
e^{i\hat{\phi}_0}|J,J\rangle=|J+1,J+1\rangle\,.
\end{equation}
Note that eqs.~\eqref{eq_arho1_gapped_NGB} and~\eqref{eq_exp_i_phi0} imply that the lowering operator of the $SU(2)$ algebra is given by $\hat{J}_{-}=e^{-i\hat{\phi}_0}\hat{a}_{\rho,1}^\dagger$, as it can be checked via Noether's procedure.\footnote{This is analogous to the role of gapped Goldstones \cite{Nicolis:2012vf} in nonlinear sigma models at finite chemical potential and finite volume \cite{Cuomo:2020fsb}.}

Upon quantizing the system we conclude that daughter Regge trajectories corresponding to multi-particle states created by $\rho_i$ and $\phi$ have a gap over the ground-state~\eqref{eq_Regge} given by
\begin{equation}\label{eq_Fock_gap}
\begin{split}
\delta M&=\omega\sum_i\sum_{k\geq 2}  \,n^{(\rho_i)}_k\epsilon^{(\rho_i)}_{k}+
\omega\sum_{k\geq 2}  \,n^{(\phi)}_k\epsilon^{(\phi)}_{k}\\
&\simeq\omega\sum_{k\geq 2}\left[\sum_i  \,n^{(\rho_i)}_k+  n^{(\phi)}_k\right] k\,,
\end{split}
\end{equation}
where the $n_k$'s denote the occupations numbers and we used superscripts to distinguish between the modes in obvious notation.  We stress again that in eq.~\eqref{eq_Fock_gap} there is no contribution from the $k=0$ and $k=1$ modes, for the reasons explained above.  Using \eqref{eq_omega_J_exp} we recover the well known result $\delta M^2=\ellb_s^{-2} \sum_{k\geq 2}\left[\sum_i  \,n^{(\rho_i)}_k+  n^{(\phi)}_k\right] k$ to leading order in $J$.

Finally, we remark that upon canonically normalizing the bulk fields we infer that nonlinear terms are suppressed in $\omega\ell_s$,  with the loop counting parameter given by $\omega^2\ell_s^2$ as anticipated. We will discuss loop corrections further in Section~\ref{sec_sqrt_formula}.

\subsubsection{Subleading orders: massless endpoints' scheme}
\label{sec_transverse}

\hspace{5mm}
Let us now study the corrections to the energy spectrum that arise from the boundary action~\eqref{eq_Sbdry_smart}. Upon expressing $\alpha$ through~\eqref{eq_sol_alpha} and neglecting total derivatives, the expansion to quadratic order of eq.~\eqref{eq_Sbdry_smart} yields
\begin{equation}\label{eq_Sbdry_smart2}
\begin{split}
S_{bdry} =m\ellb_s\sum_{\sigma=\pm 1}\int d&\tau \Bigg\{\frac{b_{\gamma}}{4\sqrt{\omega\ellb_s}}
\left[
(\dot{\rho}_i)^2+(\ddot{\rho}_i)^2-\frac{1}{2}\dot{\phi}^2+\ddot{\phi}^{\,2}-\frac{1}{2}\dddot{\phi}^{\,2}
\right] \\
&
+\frac{b_1 \sqrt{\ellb_s\omega}}{4}\left[
(\dot{\rho}_i)^2-5(\ddot{\rho}_i)^2+4(\dddot{\rho}_i)^2+\frac32\dot{\phi}^2+4\ddot{\phi}^{\,2}+\frac72\dddot{\phi}^{\,2}-\ddddot{\phi}^{\,2}
\right]\\
&
+O\left((\ellb_s\omega)^{3/2}\right)
\Bigg\}
\,.
\end{split}
\end{equation}
Note that because of the $O(d-3)$ unbroken symmetry acting on the $\rho_i$ there is no quadratic mixing between the transverse modes and $\phi$. We therefore discuss transverse and angular modes separately below.

We start from the transverse modes. From the variation of the boundary action~\eqref{eq_Sbdry_smart2}, we find that the leading order boundary condition~\eqref{eq_rho_bc_LO} is corrected as
\begin{equation}
\begin{split}
\lim_{\sigma\rightarrow\pm 1}\frac{\pm\sqrt{1-\sigma^2}}{\ell_s^2\omega^2}\rho'_i =\frac{m\ellb_s}{2}&\Bigg[\left(\frac{b_{\gamma}}{\sqrt{\omega\ellb_s}}-b_1\sqrt{\omega\ellb_s}\right)\frac{\partial^2\rho_i}{\partial\tau^2}
+\left(\frac{b_{\gamma}}{\sqrt{\omega\ellb_s}}-5b_1\sqrt{\omega\ellb_s}\right)\frac{\partial^4\rho_i}{\partial\tau^4}\\
&
-4b_1\sqrt{\omega\ellb_s}\frac{\partial^6\rho_i}{\partial\tau^6}+O\left((\omega\ellb_s)^{3/2}\frac{\partial^8\rho_i}{\partial\tau^8}\right)
    \Bigg]\,.
    \end{split}
\end{equation}
This condition may be written in a manifestly finite form using the near boundary expansion~\eqref{eq_a_near} of the solution:
\begin{equation}\label{eq_rho_bc_NNLO}
\begin{split}
\beta_2^{\rho}=0&+
\pi  \mu_1 \omega ^{3/2}\ellb_s^{5/2}
\left(\frac{d^2\beta_1^\rho}{d\tau^2}+\frac{d^4\beta_1^\rho}{d\tau^4}\right) \\
&-2 \pi  \mu_2 \omega ^{5/2}\ellb_s^{7/2}
\left[\left(\frac{d^2\beta_1^\rho}{d\tau^2}+\frac{d^4\beta_1^\rho}{d\tau^4}\right)+4
\left(\frac{d^4\beta_1^\rho}{d\tau^4}+\frac{d^6\beta_1^\rho}{d\tau^6}\right)
\right]
+
O\left(\omega^{7/2}\ellb_s^{9/2} \frac{d^8\beta_1^\rho}{d\tau^8}\right)\,,
\end{split}
\end{equation} 
where we used eq.~\eqref{eq_T_vs_ls}.

Imposing the boundary condition~\eqref{eq_rho_bc_NNLO} on plane wave solutions of the form~\eqref{eq_toy_sol}, we find that the eigenfrequencies are corrected as
\begin{equation}\label{eq_rho_spectrum_NNLO}
\epsilon_n=n\left[1+\sqrt{2}(\omega\ellb_s)^{3/2} \mu_1\ellb_s  \left(n^2-1\right)
+2\sqrt{2}(\omega\ellb_s)^{5/2}
\mu_2 \ellb_s  (n^2-1)(4n^2-1)
+\ldots\right]\,,
\end{equation}
where $n\in\mathds{N}$.  Similarly to eq.~\eqref{eq_toy_a_disp}, the derivative expansion breaks down for $n^2\omega\ellb_s\sim 1$. Eq.~\eqref{eq_rho_spectrum_NNLO} neglects  $O\left(n^6(\omega\ellb_s)^{7/2}\right)$ corrections from higher derivative boundary terms. For the moment we are also neglecting $\omega^2\ellb_s^2$ one-loop corrections to the spectrum. Notice that $\epsilon_n$ is unmodified by the higher derivative corrections for $n=0$ and $n=1$, since as explained these are protected modes implementing translations and rotations. 

The calculation of the corrections to the leading order boundary condtions~\eqref{eq_phi_LO_bc} for $\phi$ proceeds analgously. We directly report the result in terms of the boundary modes defined in eq.~\eqref{eq_phi_near}:
\begin{equation}\label{eq_phi_bc_NNLO}
\begin{split}
\beta_2^{\phi}=0&+
\frac{\pi}{3}\mu_1\omega^{3/2}\ellb_s^{5/2}
\left(
\frac{d^2\beta_1^\phi}{d\tau^2}+
2 \frac{d^4\beta_1^\phi}{d\tau^4}+\frac{d^6\beta_1^\phi}{d\tau^6}\right) \\[0.3em]
&-\frac{2\pi}{3}\mu_2\omega^{5/2} \ellb_s^{7/2}
\left[3\left(
\frac{d^2\beta_1^\phi}{d\tau^2}+
2 \frac{d^4\beta_1^\phi}{d\tau^4}+\frac{d^6\beta_1^\phi}{d\tau^6}\right)
+2\left(
\frac{d^4\beta_1^\phi}{d\tau^4}+
2 \frac{d^6\beta_1^\phi}{d\tau^6}+\frac{d^8\beta_1^\phi}{d\tau^8}\right)
\right] \\[0.3em]
&+O\left(\omega^{7/2}\ellb_s^{9/2}\frac{d^{10}\beta_1^\phi}{d\tau^{10}}\right)\,.
\end{split}
\end{equation}
From eq.~\eqref{eq_phi_bc_NNLO} we find the corrections to eq.~\eqref{eq_phi_LO_bc}:
\begin{align}
c_2& =c_1\left[0+
\frac{1}{\sqrt{2}}\mu_1\ellb_s(\omega\ellb_s)^{3/2} n(n^2-1) 
+\sqrt{2}\mu_2\ellb_s(\omega\ellb_s)^{5/2}
n (2 n^2-3)(n^2-1)+\ldots\right]\,,\\[0.4em]
\epsilon_n&=n\left[1-\frac{1}{\sqrt{2}}\mu_1\ellb_s(\omega\ellb_s)^{3/2}   \left(n^2-1\right)-\sqrt{2}\mu_2\ellb_s(\omega\ellb_s)^{5/2}
(2n^2-3)(n^2-1)+\ldots\right]\,,
\label{eq_phi_spectrum_NNLO}
\end{align}
where $n\in\mathds{N}$.  Comments analogous to those below eq.~\eqref{eq_rho_spectrum_NNLO} apply here. In particular, notice again that the result for the $n=0$ and $n=1$ modes is unmodified by higher-derivative corrections due to their protected nature. We also remark that the first correction to the leading order result in eq.~\eqref{eq_phi_spectrum_NNLO} takes the opposite sign compared to the transverse modes~\eqref{eq_rho_spectrum_NNLO}. As formerly commented, we must have $\mu_1>0$, and therefore the daughter trajectories associated with the angular mode are lighter than the ones for the transverse modes.

\subsubsection{Subleading orders: massive endpoints' scheme}

\hspace{5mm} Let us now briefly comment on the calculation using the action~\eqref{eq_Sbdry} in which the endpoints are not constrained by eq.~\eqref{eq_constraint_null}, and the quarks velocity $\alpha_f$ is given in eq.~\eqref{eq_dalpha} (this ensures that in the expansion of the action, the terms linear in the fluctuations vanish). We focus first on the transverse modes $\rho_i$. Expanding~\eqref{eq_Sbdry}, we get
\begin{equation}\label{eq_Sbdry2_rho}
S_{bdry}\supset m\sum_{\sigma=\pm 1}\int d\tau\left\{
\frac{b_m (\dot{\rho}_i)^2}{2\omega \sqrt{1-\alpha_f^2}  }
+ \frac{b_{\gamma } \ellb_s }{4\alpha_f\sqrt{\alpha_f\ellb_s \omega }}( \ddot{\rho}_i)^2
+b_1\ellb_s \sqrt{\frac{\ellb_s \omega}{\alpha_f^5 }}\left[
\left(\frac{d^3\rho_i}{d\tau^3}\right)^2-\frac54(\ddot{\rho}_i)^2\right]
\right\}\,.
\end{equation} 
We see that, differently than eq.~\eqref{eq_Sbdry_smart2}, since $\sqrt{1-\alpha_f^2}\sim \sqrt{\omega\ellb_s}$, the first term in eq.~\eqref{eq_Sbdry2_rho} is proportional to $\omega^{-3/2}$ rather than the expected power $\omega^{-1/2}$.  This might seem to suggest that the leading-order boundary conditions for $\rho$ in eq.~\eqref{eq_rho_bc_LO} is corrected already at order $\sqrt{\omega\ellb_s}$, therefore contradicting the EFT power-counting and the results of the previous section. Luckily, it turns out that upon carefully using the value of $\alpha_f$ in eq.~\eqref{eq_dalpha}, this contribution cancels out upon summing the variation of the bulk and the boundary action, and the final result agrees with the one we found in eq.~\eqref{eq_rho_bc_NNLO}. 

Let us show this up to order $O(\omega)$. We write the boundary term in the variation of the bulk action~\eqref{eq_S2} using the near boundary expansion (which is identical to~\eqref{eq_a_near})
\begin{equation}\label{eq_dSbulk_rho}
\begin{split}
\delta S_{bulk}=
\text{bulk EOMs}+
\sum_{\sigma=\pm 1}\int d\tau&\left[
\frac{ \delta \beta_1^\rho\beta_2^\rho}{\sqrt{2} \omega ^2\ell_s^2}
+ \frac{\sqrt{m  b_m}}{2 \ell_s\omega ^{3/2}}\left(2 \delta \beta^\rho _1 \ddot{\beta}^\rho _1+ \delta \beta _2^\rho\ddot{\beta}^\rho_2\right)
\right.\\
&\left.
+\frac{b_m m}{\sqrt{2}\omega}\left(\delta\beta_2^\rho\ddot{\beta}_1^\rho+\delta\beta_1^\rho\ddot{\beta}_2^\rho\right)
+O\left(\frac{1}{\sqrt{\ellb_s\omega}}\right)
\right]\,,
\end{split}
\end{equation}
where we suppressed the $O(d-3)$ indices and retained terms up to order $1/\omega$. Expanding similarly the variation of eq.~\eqref{eq_Sbdry2_rho} we obtain
\begin{equation}\label{eq_dSbdry_rho}
\delta S_{bdry}=-\sum_{\sigma=\pm 1}\int d\tau\left[
\frac{\sqrt{m b_m}}{\ell_s\omega ^{3/2}} \delta \beta^\rho _1 \ddot{\beta}^\rho _1+\frac{b_m m}{\sqrt{2}\omega}\left(\delta\beta_2^\rho\ddot{\beta}_1^\rho+\delta\beta_1^\rho\ddot{\beta}_2^\rho\right)
+O\left(\frac{1}{\sqrt{\ellb_s\omega}}\right)\right]
,.
\end{equation}
Summing eqs.~\eqref{eq_dSbulk_rho} and~\eqref{eq_dSbdry_rho} we see that the subleading terms proportional to $\delta\beta^\rho_1$ and $\beta^\rho_1$ cancel out and we obtain
\begin{equation}\label{eq_dS_rho}
\delta S_{bulk}+\delta S_{bdry}=\text{bulk EOMs}+
\sum_{\sigma=\pm 1}\int d\tau \left[
\frac{ \delta \beta_1^\rho\beta_2^\rho}{\sqrt{2}\ell_s^2 \omega ^2}
+ \frac{\sqrt{m  b_m}}{2\ell_s \omega ^{3/2}}\delta \beta _2^\rho\ddot{\beta}^\rho_2
+O\left(\frac{1}{\sqrt{\ellb_s\omega}}\right)\right]\,,
\end{equation}
which is compatible with the leading order condition~\eqref{eq_rho_bc_LO} up to order $(\omega\ellb_s)^{3/2}$ as expected:\footnote{Notice that when solving~\eqref{eq_dS_rho} with the ansatz $\beta_2^\rho=F\left(\frac{d}{d\tau}\right)\beta_1^\rho$ we also need to express the variation of $\beta_2^\rho$ in terms of that of $\beta_1^\rho$ as $\delta\beta_2^\rho= F\left(\frac{d}{d\tau}\right)\delta\beta_1^\rho$. }
\begin{equation}
\beta_2^\rho=0+O\left((\omega\ellb_s)^{3/2}\right)\,.
\end{equation}

Similar cancellations occur at subleading order.  The reason for the existence of these seemingly enhanced terms in the boundary action is that $\hat{G}_{\tau\tau}$, and therefore also the boundary connection $\hat{\Gamma}=\frac12\hat{G}_{\tau\tau}^{-1}\pd_\tau\hat{G}_{\tau\tau}$, are not analytic for $\alpha_f\rightarrow 1$. At some intuitive level, such terms arise because the redundant operator in eq.~\eqref{eq_Sbdry} is effectively equivalent to a \emph{cutoff} regulator near the endpoints. While conceptually satisfying, it is often the case that cutoff schemes yield enhanced contributions at intermediate steps in EFT. 

The technical reason why such spurious terms in the expansion do not contribute to observables, is that they always depend on the specific value of the regulator coefficient $b_m$ (and are often singular in the limit $b_m\rightarrow 0$), which determines the leading correction to the string length~\eqref{eq_dalpha}. Based on the arguments presented at the beginning of this section, only the combinations in eq.~\eqref{eq_bm} and \eqref{eq_bm2}, which are regular in the limit $b_m\rightarrow 0$, can contribute to physical observables. Unfortunately, retaining $b_m>0$ in the action~\eqref{eq_Sbdry} is nonetheless important at intermediate steps if we do not impose the condition~\eqref{eq_constraint_null}.

Despite these complications, it is nonetheless straightforward to compute the subleading corrections to the boundary conditions for the transverse modes. By expanding the variation of the bulk and boundary actions as in eqs.~\eqref{eq_dSbulk_rho} and~\eqref{eq_dSbdry_rho}, we eventually recover the previous result~\eqref{eq_rho_bc_NNLO} in terms of the physical Wilson coefficients in eqs.~\eqref{eq_bm} and~\eqref{eq_bm2}.

The analysis of angular fluctuations from the action~\eqref{eq_Sbdry} proceeds similarly to that of the transverse modes. The main difference is that we now need to compute also subleading corrections to the expression for the boundary mode $\alpha$ when we integrate it out. 

It turns out that the cancellations described above are even more dramatic for the boundary EOMs of $\alpha$ and $\phi$ than for the $\rho_i$'s. For instance the expansion of eq.~\eqref{eq_Sbdry} produces the following terms
\begin{align}\nonumber
S_{bdry}\supset  m \sum_{\sigma=\pm 1}\int d\tau &\left[\frac{b_m(\alpha\pm\dot{\phi})^2}{4\sqrt{2}\omega(1-\alpha_f)^{3/2}}+\frac{b_{\gamma}\ellb_s(\dot{\alpha}\pm\ddot{\phi})^2}{8\sqrt{\ellb_s\omega(1-\alpha_f)}}+\frac{b_m(3\alpha^2+8\dot{\alpha}^2\pm14\alpha\dot{\phi}-5\dot{\phi}^2)}{12\omega\sqrt{2(1-\alpha_f)}}
\right.\\[0.3em] &\left.\;\;\;
+\frac{5b_1\ellb_s^{3/2}\sqrt{\omega}(\dot{\alpha}\pm\ddot{\phi})^2
}{4(1-\alpha_f)^2}+O\left(\frac{1}{\sqrt{\omega\ellb_s}}\right)\right]\,,
\end{align}
where the first term scales as $1/\omega^{5/2}$ and the last three terms are all of order $1/\omega^{3/2}$ using $1-\alpha_f\sim\omega\ellb_s$.  Luckily again, the first term vanishes on the leading order solution for $\alpha$, given by $\alpha\vert_{\sigma=\pm1}=\mp\dot\phi$. Upon carefully expanding both the bulk and boundary actions we can solve perturbatively for $\alpha$ and $\beta_2^\phi$ (cf.~\eqref{eq_phi_near}), and we find similar cancellations at subleading order. The expansion for $\alpha$ takes a form similar to eq.~\eqref{eq_dalpha}
\begin{equation}
\alpha\vert_{\sigma=\pm 1}=\mp\left[\dot{\beta}_1^{\phi}-(\omega\ellb_s)^2\alpha_2-(\omega\ellb_s)^3\alpha_3+O\left((\ellb_s\omega)^4\right)\right]\,.
\end{equation}
The precise value of $\alpha_2$ and $\alpha_3$ is scheme-dependent for the same reason as the background string length. We show for illustration the result for $\alpha_2$ obtained from the action~\eqref{eq_Sbdry}
\begin{equation}
\alpha_2=\left(\frac{\pi^2}{2}  b_m^2 \ellb_s^2 m^2-\frac{ b_{\gamma}\sqrt{b_m}}{\sqrt{2}}\pi ^{3/2}  \ellb_s^{3/2} m^{3/2}
\right)\left(\frac{d^5\beta_1^\phi}{d\tau^5}+
2\frac{d^3\beta_1^\phi}{d\tau^3}+\frac{d\beta_1^\phi}{d\tau}\right)\,.
\end{equation}
The expression for $\alpha_3$ is lengthy and we do not report it.  Using this result and the near boundary expansion to the third subleading order, as reported in appendix~\ref{app_near_bdry}, we eventually recover the result~\eqref{eq_phi_bc_NNLO} given in the previous subsection. 

\subsection{Comments on one-loop corrections }\label{sec_sqrt_formula}

\hspace{5mm}
As already commented, the loop counting parameter is $1/\omega^2\sim J$.  A detailed account of one-loop effects is beyond the scope of this work. Here we limit ourselves to some comments.

As well known, the NG action at the classical level is equivalent to the Polyakov action, which consists of $d$ free fields subject to the Virasoro constraints. The Polyakov formulation is convenient for the purposes of this section, since it makes the classical integrability of the NG action manifest.
As emphasized in \cite{Dubovsky:2012sh,Hellerman:2014cba}, at one-loop level, the same equivalence holds provided one adds the so called Polchinski-Strominger (PS) term to the Polyakov action \cite{Polchinski:1991ax}.
In modern language, the PS term is a Wess-Zumino term compensating for the anomalous Weyl transformation properties of the path-integral measure away from $d=26$ \cite{Hellerman:2014cba}. To compute one-loop corrections in Polyakov formalism, one therefore simply needs to sum the result from the free bosons's action (which is simple to obtain), to the correction from a single insertion of the PS term.

For the leading Regge trajectory, the calculation was done in \cite{Hellerman:2013kba}, where it was found  that the result is independent of the number of dimensions and reads\footnote{Ref.~\cite{Baker:2002km} obtained a different result working in the Nambu-Goto formalism; we suspect that the mismatch is due to subtleties in dealing with UV divergences compatibly with Poincar\'e symmetry in the Nambu-Goto formalism. }
\begin{equation}\label{eq_M2_quantum}
\delta M^2_{quantum}=-\frac{1}{\ellb_s^2}+O\left(J^{-3/4}\right)\,.
\end{equation}
The $O(J^{-3/4})$ correction is expected to arise from the deviations from Neumann boundary conditions. We remark that, as well known,  in the covariant formalism one directly computes the mass square rather than the mass.

To perform a similar calculation for excited states is less straightforward unfortunately. This is because the Polchinski-Strominger term corrects the structure of the constraints nontrivially. Nevertheless, there are specific regimes where the free theory result, that coincides with the critical string theory expression:
\begin{equation}\label{eq_Fock_M2_gap}
\delta M^2=\frac{1}{\ellb_s^2}\sum_{k\geq 2}\left[\sum_i  \,n^{(\rho_i)}_k+  n^{(\phi)}_k\right] k\,,
\end{equation}
offers an improvement over the naive expansion~\eqref{eq_Fock_gap}. This is because the formula~\eqref{eq_Fock_M2_gap} is exact in the classical limit, and therefore accounts for infinitely many corrections from nonlinear vertices (which are not suppressed in derivatives) in the NG formalism. In other words,  the classically exact Polyakov result should provide a better approximation than the naive expansion~\eqref{eq_Fock_gap} for large occupation numbers: $n^{(\rho_i)}_k\gg 1$ and/or $n^{(\phi)}_k\gg 1$.

In practice, the only difference between the formula for the NG expansion~\eqref{eq_Fock_gap} and the result of the Polyakov formalism~\eqref{eq_Fock_M2_gap} is that the latter directly yields to $M^2$ rather than $M$.  Therefore, upon accounting for the corrections that we computed in~\eqref{eq_rho_spectrum_NNLO} and~\eqref{eq_phi_spectrum_NNLO}, our final result for the spectrum is written as
\begin{equation}\label{eq_improve_M}
M=M_{Pol}+\delta M_{NG}\,,
\end{equation}
where $M_{Pol}$ is the exact result from the free Polyakov action
\begin{align}\label{eq_M_Pol}
M_{Pol}=\frac{1}{\ellb_s}\sqrt{(J-1)+\sum_{k\geq 2}\left[\sum_i  \,n^{(\rho_i)}_k+  n^{(\phi)}_k\right] k}\,,
\end{align}
and in the second term we include the correction in the usual derivative expansion
\begin{equation}
\begin{split}
\delta M_{NG}&=\left(\frac{\mu_1}{J^{1/4}}+\frac{\mu_2}{J^{3/4}}
-\frac{\mu_1^2\ellb_s}{8  J} +\ldots\right)\\
&+\omega\sum_i\sum_{k\geq 2}  \,n^{(\rho_i)}_k(\epsilon^{(\rho_i)}_{k}-k)+
\omega\sum_{k\geq 2}  \,n^{(\phi)}_k(\epsilon^{(\phi)}_{k}-k)+\ldots\,,
\end{split}
\end{equation}
where we have subtracted the free part of the excitations' energy, which is already included in $M_{Pol}$, and $\omega$ is given in eq.~\eqref{eq_omega_J_exp}. In general, $\delta M_{NG}$ is constructed such that when expanding the square root, $M_{Pol}+\delta M_{NG}$ agrees with the result of a \emph{naive} calculation in NG formalism. 

Let us finally comment that the square-root formula~\eqref{eq_M_Pol} might improve the convergence of the derivative expansion already for low-momentum few-particle states. This is because actual one-loop corrections are expected to be further supressed by factors of $2\pi$ in the denominator with respect to the terms arising from the expansion of~\eqref{eq_M_Pol}. While this is admittedly just optimistic numerology in the absence of one-loop results to compare to, we remark that an analogous phenomenon was analyzed to study long flux tubes in \cite{Dubovsky:2014fma}. In that case,  arranging the derivative expansion similar to~\eqref{eq_improve_M} (using the classical GGRT result for the string spectrum \cite{Goddard:1973qh,Arvis:1983fp}) rather than the naive EFT expansion,  dramatically improves the agreement between lattice results and theoretical predictions.

\subsection{Different endpoints}

\hspace{5mm}
We finally consider mesons made of different quarks, that we model via different boundary conditions at the endpoints. We introduce different Wilson coefficients at the two ends of the string:
\begin{equation}\label{eq_Sbdry(2)} 
S_{bdry} =-\sum_{\sigma=\pm 1}m\left[
 b_{m,\sigma}\int d\tau\sqrt{\hat{G}_{\tau\tau}}-b_{\gamma,\sigma}\int d\tau\sqrt{\hat{\gamma}_{\tau\tau}}+...\right]\,,
\end{equation}
where we work to relative order $\sim \omega^{3/2}$.
Similarly to eq.~\eqref{eq_Sbdry}, the operators multiplied by $ b_{m,\sigma}$ are redundant and may be removed if we work with null endpoints as in eq.~\eqref{eq_constraint_null}. Analogolously to eq.~\eqref{eq_bm}, the physical Wilson coefficients that control the first corrections to the spectrum are given by
\begin{align}\label{eq_bm_different}
\mu_{1,\pm 1}&=-\sqrt{2}m b_{\gamma,\pm 1}+
\frac{4}{3}\sqrt{\ellb_s \pi}\,m^{3/2}
 b_{m,\pm 1}^{3/2}\,.
\end{align}

The calculation of the mass of the leading Regge trajectory and the spectrum of the daughter particles proceeds analogously to the case of identical points. The only difference is that we have to modify the background solution~\eqref{eq_background} to account for the asymmetry of the string:
\begin{equation}\label{eq_background(2)}
X^\mu_{class.}=\frac{1}{\omega}\big(\tau,\ (\alpha_f\sigma+\delta\alpha_{cm}) \cos\tau,\ (\alpha_f\sigma+\delta\alpha_{cm})\sin\tau,\ \underbrace{0,0,\ldots}_{d-3\text{ times}}\big)\,,
\end{equation}
where $\sigma\in(-1,1)$ as before and $\delta\alpha_{cm}$ represents a shift in the center of mass of the string with respect to the symmetric case. Extremizing the classical action we can solve for $\alpha_f$ and $\delta\alpha_{cm}$ perturbatively in $\omega$
\begin{align}\nonumber
    \alpha_f=&1-\pi m\omega\ellb_s^2\frac{b_{m,1}+b_{m,-1}}{2}\\
    &-
    \omega^2\ellb_s^2
     \left[ \frac{\pi^2 m^2\ellb_s^2}{4}(b_{m,1}^2+b_{m,-1}^2)-\frac{\pi^{3/2} (\ellb_s m)^{3/2}}{2 \sqrt{2}}(b_{\gamma,1}\sqrt{b_{m,1}}+b_{\gamma,-1}\sqrt{b_{m,-1}})\right]+\ldots\,,\\ 
     \nonumber
     \delta\alpha_{cm}=&-\pi m\omega\ellb_s^2\frac{b_{m,1}-b_{m,-1}}{2}\\
    &-
    \omega^2\ellb_s^2
     \left[ \frac{\pi^2 m^2\ellb_s^2}{4}(b_{m,1}^2-b_{m,-1}^2)-\frac{\pi^{3/2} (\ellb_s m)^{3/2}}{2 \sqrt{2}}(b_{\gamma,1}\sqrt{b_{m,1}}-b_{\gamma,-1}\sqrt{b_{m,-1}})\right]+\ldots\,.
\end{align}
Proceeding as in Section~\ref{sec_bdry_action}, we arrive at the final result for the mass of the leading Regge trajectory:
\begin{equation}\label{eq_Regge(2)}
M_{classical}=\frac{\sqrt{ J}}{\ellb_s}+\frac{\mu_{1,1}+\mu_{1,-1}}{2J^{1/4}}
+O\left(J^{-3/4}\right)\,.
\end{equation}

To analyze fluctuations we use a parametrization similar to~\eqref{eq_flucts}:
\begin{multline}\label{eq_flucts(2)}
X^\mu=\frac{1}{\omega}\left\{\tau, \left[\alpha_f\, \sigma+\delta\alpha_{cm}+\alpha(\tau,\sigma) \right]\cos\left(\tau+\phi(\tau,\sigma)\right),\right.\\
\left.
\left[\alpha_f\, \sigma+\delta\alpha_{cm}+\alpha(\tau,\sigma) \right] \sin\left(\tau+\phi(\tau,\sigma)\right),\rho_i(\tau,\sigma)\right\}\,.
\end{multline}
The analysis of the bulk and boundary equations proceeds as before. We find that the boundary conditions at the two endpoints of the string depend on the boundary coefficients~\eqref{eq_bm_different} exactly as in the former results~\eqref{eq_rho_bc_NNLO} and~\eqref{eq_phi_bc_NNLO}. We spare the details of the derivation to the reader, and simply provide the results for the spectrum and wave-functions of the single-particle states below.

Let us start from the transverse mode. As explained around~\eqref{eq_toy_sol}, the solution to the bulk equations of motion is given by plane waves of the form
\begin{equation}\label{eq_rho(2)}
\rho_i=e^{-i\epsilon\tau}\left[ \tilde{c}_1\cos(\epsilon \tilde{s})+\tilde{c}_2\sin( \epsilon \tilde{s})\right] \,,
\end{equation}
where $\tilde{s}=\arcsin(\alpha_f\sigma+\delta\alpha_{cm})$, a modified version of eq.~\eqref{eq_s_def} due to the shift in the center of mass. For equal endpoints, worldsheet parity (i.e. charge conjugation) guarantees that the wave-function must be either even or odd under $s\rightarrow-s$, hence setting one between $\tilde{c}_1$ and $\tilde{c}_2$ to zero. For different endpoints instead the boundary conditions break worldsheet parity, and even and odd plane waves mix at subleading order. The ratio $\tilde{c}_2/\tilde{c}_1$ and the allowed values for the energy are found by imposing the boundary conditions at both $\sigma\rightarrow 1$ and $\sigma\rightarrow-1$. We find for $n$ even
\begin{align}
\label{eq_rho_spectrum_ratio_NNLO_even}
\tilde{c}_2& =\tilde{c}_1\left[0-
\frac{\ellb_s\pi}{2\sqrt{2}}(\omega\ellb_s)^{3/2}(\mu_{1,1}-\mu_{1,-1})n(n^2-1) 
+O\left((\omega\ellb_s)^{5/2} \right)\right]\,,\\
\epsilon_n&=n\left[1+\frac{\ellb_s}{\sqrt{2}}(\omega\ellb_s)^{3/2}(\mu_{1,1}+\mu_{1,-1})\left(n^2-1\right)+O\left((\omega\ellb_s)^{5/2} \right)\right]\,.
\label{eq_rho_spectrum_NNLO(2)}
\end{align}
For $n$ odd,  the energy levels are given by the same expression but the ratio should be inverted, which means
\begin{align}
\label{eq_rho_spectrum_ratio_NNLO_odd}
\tilde{c}_2& =\tilde{c}_1\left[0+
\frac{\ellb_s\pi}{2\sqrt{2}}(\omega\ellb_s)^{3/2}(\mu_{1,1}-\mu_{1,-1})n(n^2-1) 
+O\left((\omega\ellb_s)^{5/2} \right)\right]\,.
\end{align}
Note that the mixing between even and odd wavefunctions disappears for $\mu_{1,1}=\mu_{1,-1}$,  as expected, as well as for $n=0,1$. This is because the first two modes correspond to the action of symmetry generators, as explained in Section~\ref{subsec_LO_fluct}. 

For angular fluctuations, the solution to the EOM reads as in eq.~\eqref{eq_phi} with the replacement $\sigma\rightarrow\alpha_f\sigma+\delta\alpha_{cm}/\alpha_f$. Imposing the boundary conditions we find that the coefficients of the wave-functions~\eqref{eq_phi} and the energy levels are given by: 
\begin{align}
c_2& =c_1\left[0+
\frac{\ellb_s}{\sqrt{2}}(\omega\ellb_s)^{3/2}\mu_{1,1}\  n(n^2-1) 
+O\left((\omega\ellb_s)^{5/2} \right)\right]\,,\\[0.4em]
\epsilon_n&=n\left[1-\frac{\ellb_s}{2\sqrt{2}}(\omega\ellb_s)^{3/2}(\mu_{1,1}+\mu_{1,-1})   \left(n^2-1\right)+O\left((\omega\ellb_s)^{5/2} \right)\right]\,.
\end{align}
Note that ratio $c_2/c_1$ depends only on $\mu_{1,1}$ since the Legendre function $Q^{\frac32}_{\epsilon-\frac{1}{2}}(x)$ in general does not have any symmetry property as a function of $x$.

\section{Endpoint renormalization from bulk operators}\label{sec_renormalization}

\hspace{5mm}
Given the construction of the boundary action in the previous section, we are now ready to address the issue raised in Section~\ref{sec_singularities}: the breakdown of the derivative expansion near the string endpoints. In this section we show that higher derivative bulk operators renormalize, already at  the classical level, the boundary action~\eqref{eq_Sbdry}. As a zeroth order check, we observe that the contribution~\eqref{eq_higher_der_break} of an higher derivative operator to the leading Regge trajectory has the same $\omega$-scaling as that of the boundary operators in eq.~\eqref{eq_Sbdry}.  Therefore, we may think that each bulk operator is associated with an infinite tower of boundary operators that cancel its singular contributions.

The physical picture underlying this renormalization procedure is simple.  Within EFT we resolve the location of the effective endpoints only up to a certain uncertainty, parametrized by the boundary cutoff.  By lowering such cutoff we also effectively shorten the bulk string length and therefore modify the contribution of the bulk operators near the boundary. For physical quantities to remain invariant, such contributions must thus be absorbed in a change of the boundary Wilson coefficients. 

In practice we technically implement this strategy by retaining a finite value of the redundant coefficient $b_m$ in eq.~\eqref{eq_Sbdry}; this ensures that the contributions from bulk operators remain finite. This also implies that, when dealing with bulk operators, we cannot work with null endpoints as in the previous sections, and therefore we encounter unphysical powers of $\omega\ellb_s$ in the calculations. It would clearly be desirable to have a more convenient calculational scheme,  more analogous to a mass independent regulator, which would make power-counting manifest at all stages of the calculation. Unfortunately, we were not able to devise such a refined approach.\footnote{For instance, it is simple to check that considering a $D=2+\varepsilon$ dimensional rotating brane and working in dimensional regularization does not regulate the endpoint singularities discussed in Section~\ref{sec_singularities}.  In~\cite{Hellerman:2016hnf} a different scheme, which does not involve introducing a redundant operator, was used to compute the leading Regge trajectory; that scheme does not present any technical advantage compared to ours, and it is unclear to us how to use it in calculations involving fluctuations around the background.}

\subsection{Warm-up: massless scalar field with curvature coupling}

\hspace{5mm}
As in Section~\ref{sec_bdry_action}, it is convenient to first illustrate the renormalization procedure in the toy example of a massless scalar field on a rigid rotating string. We  consider the system defined by the bulk action~\eqref{eq_S2_toy} and the boundary action~\eqref{eq_toy_bdry_a_2}, perturbed by the following coupling to the extrinsic curvature
\begin{equation}\label{eq_S2_toy_K2}
S_{bulk,2}=\ellb_s^2\int d\tau \int^{1}_{-1}  d\sigma\sqrt{G}\frac{c}{4}K^2 (\pd a)^2\,,
\end{equation}
where $c$ is some $O(1)$ coefficient and $K^2=K^\mu_{\alpha\beta}\eta_{\mu\nu}K^\nu_{\gamma\delta}G^{\alpha\gamma}G^{\beta\delta}$ is the extrinsic curvature squared.  As  in Section~\ref{subsec_toy_bc}, the worldsheet metric is given by~\eqref{eq_metric} with $\alpha_f=1-z\ellb_s\omega$. 

We obtained the boundary conditions and the spectrum for the free scalar in eqs.~\eqref{eq_toy_spectrum} and~\eqref{eq_toy_bc2} up to order $O(\omega^{3/2}\ellb_s^{3/2})$. By power-counting,  the operator~\eqref{eq_S2_toy_K2} scales as $\omega^2\ellb_s^2$ and thus it should not affect these results. As formerly explained, because of eq.~\eqref{eq_K2n} $K^2$ however is not suppressed near $\sigma=\pm 1$. This implies that the physical boundary Wilson coefficients in eq.~\eqref{eq_toy_B_D_2} receive a contribution also from the bulk coefficient $c$. Intuitively, this is because we can lower our boundary cutoff increasing $z\rightarrow z+\delta z$ with $\delta z>0$ and reabsorb in the boundary action the contribution of the bulk operators between $\sigma=1$ and $\sigma=1-\delta z \ellb_s\omega/\alpha_f$. Notice that already in the definition~\eqref{eq_toy_B_D_2} the physical Wilson coefficients depend upon the value of $z$. Additionally, the operator in eq.~\eqref{eq_S2_toy_K2} contributes as
\begin{equation}\label{eq_toy_dz_int}
\begin{split}
\delta S_{bdry} &=\int d\tau\int_{1-\delta z\ellb_s\omega/\alpha_f}^1 d\sigma\frac{c\ellb_s^2\omega^2\left[\alpha_f^2\dot{a}^2-(1-\alpha_f^2\sigma^2)(a')^2\right]}{2\alpha_f(1-\sigma^2\alpha_f^2)^{5/2}} +(\sigma\rightarrow-\sigma)\\
&\approx c\frac{\delta z}{z^{5/2}}\sum_{\sigma=\pm 1}\int d\tau\sqrt{\hat{\gamma}_{\tau\tau}}\left[\# \ellb_s\hat{\gamma}_{\tau\tau}^{-1}(\hat{\nabla}_{\tau}a)^2+\#\ellb_s^3
(\hat{\nabla}_\tau^2a)^2\hat{\gamma}_{\tau\tau}^{-2}+\ldots\right]\,,
\end{split}
\end{equation}
where in the second line we Taylor expanded at leading order $a$ near $\sigma=1$ and we used the leading order boundary conditions~\eqref{eq_toy_bc2}. We denoted with $\#$ some $O(1)$ numerical coefficients whose precise value is irrelevant for our discussion. The important message is that we expect a contribution of order $c/z^{3/2}$ for $z\gg 1$ to the coefficients~\eqref{eq_toy_B_D_2}, such that a variation of $z$ can be reabsrobed by a shift of the form~\eqref{eq_toy_dz_int} in the Wilson coefficients of the boundary action~\eqref{eq_toy_bdry_a_2}.

Let us now show how to compute the boundary conditions in the presence of the bulk higher derivative operator~\eqref{eq_S2_toy_K2}. First notice that the bulk EOMs in the presence of the operator~\eqref{eq_S2_toy_K2} can be solved perturbatively in the form
\begin{equation}\label{eq_toy_sol_a1}
a(\tau,\sigma)=a^{(0)}(\tau,\sigma)+\frac{c\ellb_s^2\omega^2}{(1-\alpha_f^2\sigma^2)^2}a^{(1)}(\tau,\sigma)+O\left(\frac{c^2\ellb_s^4\omega^4}{(1-\alpha_f^2\sigma^2)^4}\right)\,,
\end{equation}
where $a^{(0)}$ is the leading order solution~\eqref{eq_toy_sol} and we defined the prefactor of the perturbation such that $a^{(1)}$ is regular for $\alpha_f\sigma\rightarrow\pm 1$. In the case at hand $a^{(1)}$ can be written explicitly as a product of hypergeometric functions and trigonometric functions. For our purposes it suffices to know the expansion of $a^{(1)}$ for $1-\alpha_f\sigma\rightarrow 0$:
\begin{equation}\label{eq_a1_near}
\begin{split}
a^{(1)}(\tau,\sigma) &=   \left[\frac43(1-\alpha_f\sigma)\frac{d^2}{d\tau^2}-\frac83
(1-\alpha_f\sigma)^2\log(1-\alpha_f\sigma)
\left(\frac{d^2}{d\tau^2}+\frac{d^4}{d\tau^4}\right)+
\ldots\right]
\beta_1(\tau)
\\ & +
\sqrt{1-\alpha_f\sigma}\left[\frac13+(1-\alpha_f\sigma)
\left(\frac{11}{12}+\frac{7}{3}\frac{d^2}{d\tau^2}\right)
+\ldots\right]\beta_2(\tau)\,,
\end{split}
\end{equation}
where $\beta_1$ and $\beta_2$ are the boundary modes defined in eq.~\eqref{eq_a_near}. Using eq.~\eqref{eq_a1_near} in eq.~\eqref{eq_toy_sol_a1} we see that the higher derivative correction to the solution is only $\sim\sqrt{\omega\ellb_s}$ suppressed with respect to the leading order for $|\sigma|\simeq  1$. To formally retain control of the expansion we may imagine taking $z\gg 1$. In this way the field at the endpoints admits a double-expansion for $(z\ellb_s\omega)\rightarrow 0$ and $z\gg 1$, whose schematic form up to possible logarithmic corrections is
\begin{equation}\label{eq_double}
a(\tau,\pm 1)=\sum_{n\geq 0}\left[c_n^{(0)}(\tau)+\frac{c}{z^2}c_n^{(1)}(\tau)
+O\left(\frac{c^2}{z^4}\right)\right](z\ellb_s\omega)^{n/2}\,,
\end{equation}
where the $c_n^{(k)}(\tau)$'s are functions of $\beta_1$ and $\beta_2$ as in eqs.~\eqref{eq_a_near} and~\eqref{eq_a1_near}.

To proceed we simply need to compute the variation of the bulk and boundary actions in a double-expansion of the form~\eqref{eq_double}. This procedure is justified since we can always choose $z\gg 1 $, but in practice can be implemented for any value of $z>0$. It is however important to retain $z\neq 0$ to regulate the endpoint singularities. The boundary term in the variation of the bulk action \eqref{eq_S2_toy}$+$\eqref{eq_S2_toy_K2} is
\begin{equation}\label{eq_toy2_dSbulk}
\begin{split}
\delta S_{bulk}+\delta S_{bulk,2}=\text{bulk EOMs}&+
\sum_{\sigma=\pm 1}\int d\tau\left\{\frac{\delta\beta_1\beta_2}{\sqrt{2}}
+\frac{\sqrt{\omega\ellb_s z}}{\sqrt{2}}\left[\delta\beta_2\beta_2+2\delta\beta_1\ddot{\beta}_1\right.
\right.\\ & \left.\left.
+\frac{c}{12 z^4}\left(\delta\beta_2\beta_2-2\delta\beta_1\ddot{\beta}_1\right)+O\left(\frac{c^2}{z^8}\right)\right]+
O\left(z\omega\ellb_s\right)\right\}\,.
\end{split}
\end{equation}
The variation of the boundary action is particularly simple to this order
\begin{equation}\label{eq_toy2_dSbdry}
\delta S_{bdry}=\sum_{\sigma=\pm 1}\int d\tau \,\tilde{b}_a\sqrt{\ellb_s\omega}\delta\beta_1\ddot{\beta}_1+O\left(\ellb_s\omega\right)\,.
\end{equation}
Summing eqs.~\eqref{eq_toy2_dSbulk} and \eqref{eq_toy2_dSbdry} we find the solution
\begin{equation}
\beta_2\simeq\sqrt{\omega \ellb_s}B_a^{(new)}\ddot{\beta}_1\,,
\end{equation}
where $B_a^{(new)}$ can be written in terms of the coefficient defined in~\eqref{eq_toy_B_D_2} as
\begin{equation}\label{eq_toy2_Bnew}
\begin{split}
B_a^{(new)} &=B_a+\frac{c}{6 z^{3/2}}+O\left(\frac{c^2}{z^{7/2}}\right)\\
&=
-\sqrt{2}\, \tilde{b}_a-2 \sqrt{z}+\frac{c}{6 z^{3/2}}+O\left(\frac{c^2}{z^{7/2}}\right)\,.
\end{split}
\end{equation}
As promised, the contribution of the bulk higher derivative operator can be absorbed in a shift of the boundary Wilson coefficient $\tilde{b}_a\rightarrow\tilde{b}_a+c/(\sqrt{2} \,6z^{3/2})$. 

Restoring the subleading orders in eqs.~\eqref{eq_toy2_dSbulk} and~\eqref{eq_toy2_dSbdry}, we find that the boundary conditions take exactly the same form as in eq.~\eqref{eq_toy_bc2}. As in eq.~\eqref{eq_toy2_Bnew}, the only difference is that the coefficient $D_a$ in~\eqref{eq_Ba_toy} is replaced by
\begin{equation}
\begin{split}
D_a^{(new)}&=D_a+\frac{c \left(2 z-4 \sqrt{2z}\, \tilde{b}_a+\bar{b}_a^2\right)}{6 z^{3/2}}+O\left(\frac{c^2}{z^{5/2}}\right)\\
&=\sqrt{2} \,\tilde{d}_a+2 \tilde{b}_a^2 \sqrt{z}+2 \sqrt{2} \,\tilde{b}_a z+\frac{4 z^{3/2}}{3}+
\frac{c \left(2 z-4 \sqrt{2z}\, \tilde{b}_a+\tilde{b}_a^2\right)}{6 z^{3/2}}+O\left(\frac{c^2}{z^{5/2}}\right)\,.
\end{split}
\end{equation}
This is equivalent to a renormalization of the boundary Wilson coefficient $\tilde{d}_a$.

Finally we comment that it is tempting to interpret the $z$-dependence of the coefficients $B_a^{(new)}$ and $D_a^{(new)}$ as a boundary RG flow for the operators multilplied by $\tilde{b}_a$ and $\tilde{d}_a$ in the action~\eqref{eq_toy_bdry_a_2}, with $z$ setting the boundary cutoff.  While conceptually appealing, we did not find any practical use of this viewpoint.  This is because all the operators of the boundary action can be thought as irrelevant deformation of the rotating NG string with free endpoints, and therefore, similarly to the chiral Lagrangian, no interesting RG flow takes place. Note also that bulk and boundary operators always have different scalings in $\omega\ellb_s$, and therefore we do not expect any logarithmic running for the boundary Wilson coefficients at the classical level.

\subsection{Endpoints' renormalization in the dynamical string: an example}\label{subsec_endpoint_ren_example}

\hspace{5mm}
The discussion of boundary conditions for the rotating string coordinates in the presence of bulk higher derivative operators is completely analogous to the example of a massless scalar that we just discussed.  Intuitively, this can be understood using arguments analogous to those presented around eq.~\eqref{eq_toy_dz_int}, as formerly discussed in~\cite{Hellerman:2016hnf}.\footnote{Unfortunately,  it seems hard to turn such physical arguments based on explicitly shortening the integration region of $\sigma$ into concrete calculational schemes.  For instance, in~\cite{Hellerman:2016hnf} the authors use the leading order boundary conditions to rewrite operators integrated along the small strip as in eq.~\eqref{eq_toy_dz_int}, while as we have seen the string length is not invariant under such replacements.}

Let us illustrate this point by considering the NG action perturbed by a nontrivial bulk operator:
\begin{equation}\label{eq_S_k4}
S_{bulk}=-\frac{1}{\ell_s^2}\int d\tau \int^{1}_{-1}  d\sigma\sqrt{G}\left[1-c_4\ellb_s^4 (K^2)^2
\right]\,.
\end{equation}
In the bulk, the operator multiplied by $c_4$ is suppressed by a factor $(\ellb_s\omega)^4$ with respect to the leading order action.

Let us consider first the leading Regge trajectory.  As in the previous section, it is formally advantageous to consider a large value of $z\equiv(1-\alpha_f)/(\ellb_s\omega)$. Because of eq.~\eqref{eq_dalpha}, this is equivalent to taking a large value of the redundant coefficient $b_m$. In this case, we can compute the classical action and extremize over $\alpha_f$ to obtain 
\begin{equation}\label{eq_dalpha_c4}
\begin{split}
\alpha_f =&1-\omega b_m \left[\pi m\ellb_s^2+ \frac{c_4}{b_m^4}\frac{1}{4\pi^3m^3\ellb_s^2}+O\left(\frac{c_4^2}{b_m^8}\right)\right]\\
&+
\omega^2 b_m^2\left[\frac{\pi^2}{2} m^2\ellb_s^4
-\frac{b_{\gamma}}{\sqrt{2} b_m^{3/2}}\pi^{3/2}m^{3/2}\ellb_s^{7/2}
+\frac{c_4}{b_m^4}
\frac{5 \sqrt{2 m b_m} b_{\gamma } -12 \sqrt{\pi \ellb_s} m b_m^2}{16 \pi ^{5/2} \sqrt{\ellb_s} m^3 b_m^2}+O\left(\frac{c_4^2}{b_m^8}\right)
\right]\\
&+O\left(\ellb_s^3\omega^3\right)\,,
\end{split}
\end{equation}
where we work in a double-expansion analogous to eq.~\eqref{eq_double}.  Namely, at each order in $\omega\ellb_s$ we have a series in $c_4/b_m^4$. 

The rest of the calculation proceeds as in Section~\ref{subsec_general_Sbdry}. Eventually we recover the result \eqref{eq_Regge} with the coefficients $\mu_1$ and $\mu_2$ now replaced by 
\begin{align}\nonumber
\mu_1^{(new)} &=
\mu_1
-\frac{c_4}{b_m^{5/2}}\frac{1}{10 \,\pi ^{7/2}  m^{5/2}\ellb_s^{7/2}}
+O\left(\frac{c_4^2}{b_m^{13/2}\ellb_s}\right)\,, \\
\mu_2^{(new)}&=\mu_2+\frac{c_4}{b_m^{3/2}}
\left(
\frac{b_{\gamma}}{8 \sqrt{2} \,\pi ^3 b_m^{3/2} \ellb_s^3 m^2}-\frac{5 \sqrt{\ellb_s m}}{24\, \pi ^{5/2} \ellb_s^3 m^2}\right)
+O\left(\frac{c_4^2}{b_m^{11/2}\ellb_s}\right)\,,\label{eq_Mnew}
\end{align}
where $\mu_1$ and $\mu_2$ were given in eqs.~\eqref{eq_bm} and \eqref{eq_bm2}. As in the previous section, the shifts in eqs.~\eqref{eq_Mnew} can be understood as a renormalization of the coefficients $b_\gamma$ and $b_1$ in eq.~\eqref{eq_Sbdry}.

It is also possible to check that the boundary conditions~\eqref{eq_rho_bc_NNLO} and~\eqref{eq_phi_bc_NNLO} for the fluctuations are unchanged up to the shifts in eq.~\eqref{eq_Mnew}. To see this one proceeds as in the previous section. For instance, the new operator \eqref{eq_S_k4} modifies the solutions to the bulk equations similarly to eq.~\eqref{eq_toy_sol_a1}, e.g. for the transverse modes we have
\begin{equation}
\rho_i(\tau,\sigma)=\rho_i^{(0)}(\tau,\sigma)+c_4\frac{\ellb_s^4\omega^4}{(1-\alpha_f^2\sigma^2)^4}\rho_i^{(1)}(\tau,\sigma)+\ldots\,,
\end{equation}
where the $\rho^{(i)}$ are normalized such that they are regular for $\alpha_f\sigma\rightarrow\pm 1$:
\begin{equation}
\rho^{(i)}(\tau,\sigma)=\sum_{n\geq 0} c^{(i)}_n(\tau)(1-\alpha_f\sigma)^{n/2}\,,
\end{equation}
which again holds up to logarithms. Analogous expressions hold for $\phi$. We provide the explicit results for the near boundary expansion in the presence of the operator $(K^2)^2$ in appendix~\ref{app_near_bdry}.

For $b_m\gg 1$ we impose the boundary conditions at a point $(1-\alpha_f\sigma)=z\omega\ellb_s\gg\omega\ellb_s$. We thus again organize the near boundary expansion for $\rho$ (and $\phi)$ as of a double expansion analogous to eq.~\eqref{eq_double} (neglecting $O(d-3)$ indices):
\begin{equation}
\rho(\tau,1)=\sum_{n\geq 0}\left[c_n^{(0)}(\tau)+\frac{c_4}{z^4}c_n^{(1)}(\tau)
+\ldots\right](z\omega)^n\,.
\end{equation}
We checked that, writing the variation of the bulk and boundary action using such expansions for $\rho$ and $\phi$ as in eqs.~\eqref{eq_toy2_dSbulk} and~\eqref{eq_toy2_dSbdry}, we eventually recover the former results~\eqref{eq_rho_bc_NNLO} and~\eqref{eq_phi_bc_NNLO} for the boundary conditions (and thus for the spectrum, cf.~\eqref{eq_rho_spectrum_NNLO} and~\eqref{eq_phi_spectrum_NNLO}),  with the replacement $\mu_1\rightarrow \mu_1^{(new)}$ and $\mu_2\rightarrow \mu_2^{(new)}$.

\part{Extensions and phenomenology}\label{part2}

\section{Additional degrees of freedom in \texorpdfstring{$4d$}{4d}}

\hspace{5mm}
To apply our results to mesons with light quarks in real world QCD,  it is important to discuss two important extensions of the formalism that we have presented so far: the role of spin at the endpoints and the inclusion of the pseudoscalar resonance on the worldsheet discussed in Section~\ref{subsec_EST_from_lattice}.  In this section we therefore specialize to $d=4$.

\subsection{Endpoints with spin}\label{subsec_spin}

\hspace{5mm}
In QCD, the string endpoints are expected to be endowed with a $2$-dimensional quantum mechanical system describing different spin states of a quark.  As we review in appendix~\ref{app_spin}, it is possible to describe the spin degrees of freedom for a point particle of arbitrary spin $s$ using the coadjoint orbit technique.  This amounts to considering two angles $(\theta_s,\phi_s)$ that parametrize the orientation of the particle spin in the rest frame via
\begin{equation}
\vec{S}_{rest}=s\left(\sin\theta_s\cos\phi_s,\sin\theta_s\sin\phi_s,\cos\theta_s\right)\,.
\end{equation}
To write the spin-component of the Lorentz generator in an arbitrary frame, it is convenient to introduce a bosonic Dirac spinor living on the worldline as
\begin{equation}\label{eq_psi_z_def}
\psi=\frac{1}{\sqrt{2}}\Lambda_u\cdot
\left(\begin{array}{c}
z \\
z \end{array}\right)\quad\text{where}\quad
z=\sqrt{2}\left(\begin{array}{c}
e^{-i\phi_s/2}\cos\frac{\theta_s}{2}\\  
e^{i\phi_s/2}\sin\frac{\theta_s}{2}
\end{array}\right)\,,
\end{equation}
where $\Lambda_u$ is the Dirac boost matrix associated with the four-velocity $u^\mu=\dot{X}^\mu/\sqrt{\dot{X}^2}$ of the particle. The spin components of the Lorentz generators is then written in terms of $\psi$ as
\begin{equation}
S_{\mu\nu}=\frac{i}{4}s\bar{\psi}[\gamma_\mu,\gamma_\nu]\psi\,,
\end{equation}
from which we obtain the Pauli-Lubanski vector: $S^\mu=\frac{1}{2}\varepsilon^{\mu\nu\rho\sigma}u_\nu S_{\rho\sigma}$.  Notice that these satisfy $S_\mu u^\mu=S_{\mu\nu}u^\nu=0$. Finally, in the free limit, the worldline action for the spin degrees of freedom reads:
\begin{equation}\label{eq_free_spin}
S_{spin}=s\int d\tau i\bar{\psi}\dot{\psi}\,.
\end{equation}
It is possible to show that the action~\eqref{eq_free_spin} leads to the expected canonical commutation relations $[S^i_{rest},S^j_{rest}]=i\varepsilon^{ijk}S^k_{rest}$ and the constraint $\vec{S}_{rest}^{\,2}=s(s+1)$. 

Before discussing additional interactions, it is interesting to add the term~\eqref{eq_free_spin} at the string endpoints. Indeed eq.~\eqref{eq_free_spin} already encodes interactions between the spin degrees of freedom, $(\theta_s,\phi_s)$, and the string coordinates via the boost matrix in eq.~\eqref{eq_psi_z_def}. We work in the massive endpoint formalism~\eqref{eq_Sbdry} for simplicity. Expanding~\eqref{eq_free_spin} around the background~\eqref{eq_background} we find
\begin{equation}\label{eq_rotating_free_spin}
\begin{split}
s\int d\tau i\bar{\psi}\dot{\psi} =s\int d\tau
\left(\cos \theta_s \dot{\phi}_s+\cos\theta_s\frac{\Omega}{\omega}\right)\,,
\end{split}
\end{equation}
where we defined a positive frequency as
\begin{equation}
\Omega\equiv\omega\left(\frac{1}{\sqrt{1-\alpha_f^2}}-1\right)\simeq \frac{1}{\ellb_s}\sqrt{\frac{\omega}{ 2\pi b_m m}}\,;
\end{equation}
in the second line we used~\eqref{eq_dalpha} and expanded for $m\omega\ll 1/\ell_s^2$.  From the EOMs of the spin angles, we find that at the classical level the spin moves on trajectories specified by
\begin{equation}\label{eq_Thomas_precession1}
\theta_s=\text{const}.\,,\qquad
\phi_s=-\Omega t+\text{const}.\,
\end{equation}
where $t=\tau/\omega$.
This is just the well known Thomas spin's precession for an accelerated particle.\footnote{Thomas precession is usually stated in terms of the following equation \cite{10.1063/1.528448}
\begin{equation}\label{eq_Thomas_precession2}
\dot{S}^\mu=-u^\mu \dot{u}\cdot S\,.
\end{equation}
This equation can be derived noticing that the fact that $\dot{S}^i_{rest}=0$ in the rest frame for a free particle is equivalent to $\dot{S}^\mu\propto u^\mu$ in covariant form.  Eq.~\eqref{eq_Thomas_precession2} follows then using that $S^\mu u_\mu=0$ implies $\frac{d}{d\tau}(S\cdot u)=0$ (see e.g. \cite{Cognola}).  It is simple to check that eq.~\eqref{eq_Thomas_precession1} is indeed a solution of eq.~\eqref{eq_Thomas_precession2}.} Quantum-mechanically,  since $S^3=s\cos\theta_s$, eq.~\eqref{eq_rotating_free_spin} implies that (in physical units $t=\tau/\omega$)  the Hamiltonian for the spin variables reads
\begin{equation}\label{eq_spin_H}
H=-\Omega S^3\,.
\end{equation}
Therefore different spin states are separated by a gap $\Omega\sim\sqrt{\omega/\ellb_s}$. Recalling eq.~\eqref{eq_ddt_bdry}, we conclude that the splitting between the different spin states is of the order of the strong coupling scale at the boundary.

Being at the cutoff scale, the gap between the lowest spin state and the excited ones is not calculable within EFT.  Notice that indeed eq.~\eqref{eq_Thomas_precession2} depends on the redundant coefficient $b_m$.  Additionally, we show below that the splitting between the spin states receives contributions from infinitely many operators. 

Let us consider couplings between the spin tensor $S_{\mu\nu}$, or equivalently $S^\mu$, and the endpoint worldline. To estimate the size of the interaction terms, it is convenient to write $S^\mu$ as
\begin{equation}\label{eq_Smu_counting}
S^\mu= h u^\mu+\delta S^\mu\,,
\end{equation}
where $h=\vec{u}\cdot\vec{S}_{rest}/u^0\sim O(1)$ is the helicity, $u^\mu\sim1/\sqrt{\omega}$ and $\delta S^\mu\sim O(1)$.  The first term in eq.~\eqref{eq_Smu_counting} is large despite the constraint $S\cdot S=-s^2$ (at the classical level) due to the boost factor. 

Inverting eq.~\eqref{eq_Smu_counting} via $S_{\rho\sigma}=\varepsilon_{\mu\nu\rho\sigma}S^\mu u^\nu$ and recalling $S_{\mu\nu}u^\nu=0$, we identify the following leading interaction term compatibly with parity and time reversal (notice that $S^\mu$ is a pseudovector)
\begin{equation}\label{eq_spin_int}
\int d\tau \sqrt{\hat{\gamma}_{\tau\tau}}f(\mO_s)\quad\text{where}\quad
\mO_s=S_{\mu\nu}\nabla_{\tau}^2X^\mu\nabla^3_{\tau}X^\nu\hat{\gamma}_{\tau\tau}^{-5/2}\,.
\end{equation}
Here $f$ is an arbitrary function.\footnote{For $f(x)=x$, the coupling~\eqref{eq_spin_int} (and its equivalent formulation in eq.~\eqref{eq_heavy_spin_int}) induces the long-range interaction between spin at different endpoints first discussed in \cite{Kogut:1981gm}.} Evaluating~\eqref{eq_spin_int} on the background~\eqref{eq_background} we find 
\begin{equation}
\int d\tau\frac{f(S^3)}{\sqrt{\omega\ellb_s}}\,.
\end{equation}
Therefore the interaction~\eqref{eq_spin_int} contributes to the splitting between different spin states at the same order of the free term.  Since $f$ is an arbitrary function, the splittings between different spin states does not need to scale linearly with $S^3$ (for endpoints in spin $s>1/2$ representations). In practice, EFT breaks down at energies smaller than the ones needed to excite the first nontrivial spin state.

In conclusion, the different spin states of the quarks are separated by (uncalculable) splittings of the order of the cutoff at the endpoints, $\Lambda_{bdry}\sim \sqrt{\omega/\ellb_s}$.  Therefore, within EFT we are forced to integrate out all the spin states but the lowest one.  We conclude that the only effect of spin degrees of freedom at the endpoints is to renormalize the boundary Wilson coefficients in eq.~\eqref{eq_Sbdry1},  which are anyway uncalculable from the low energy viewpoint.

Physically, the degeneracy between the spin states is broken due to the accelerated motion of the endpoints, that implies a nontrivial spin precession.  Due to the sign of the Thomas' force (i.e. the minus sign in eq.~\eqref{eq_spin_H}), this argument suggests that in the lowest energy state the quarks' spin align with the orbital angular momentum of the meson. This possibility is sometimes referred to as spin-orbit inversion \cite{Isgur:1998kr}, in comparison with the usual spin-orbit coupling in atomic physics. We remark however that the interaction terms~\eqref{eq_spin_int} do not have a fixed sign and may therefore affect this conclusion.  In fact, it is also possible that for light quarks the spin states become infinitely heavy and completely decouple for sufficiently large $J$. We will nonetheless argue in favor of the spin orbit inversion phenomenon in Section~\ref{sec_heavy_quarks} for heavy quarks' mesons with large orbital angular momentum.

Finally, we remark that, from the viewpoint of symmetry breaking, it should not come as a surprise that the spin degrees of freedom are gapped. Indeed,  for a point particle, the spin angles $(\theta_s,\phi_s)$ can be thought as Goldstone fields parametrizing the symmetry breaking pattern $SO(3)\rightarrow SO(2)$, where $SO(3)$ is the rotation group in the rest frame of the particle.  The rotating string breaks spontaneously the rotation group even for spinless endpoints, with the string coordinates accounting for its nonlinear realization. Therefore the spin angles at the endpoints provide \emph{redundant} Goldstone fields.  This is not a new phenomenon: it is often the case that in the presence of spontaneously broken spacetime symmetries one can introduce redundant Goldstones, see e.g., \cite{IvanovIHC,Nicolis:2013sga,Delacretaz:2014oxa}. In all the examples of this sort that we are aware of, such additional Goldstones behave similarly to generic matter fields, and acquire a gap.  This is also the fate of the spin states at the string endpoints.

\subsection{Worldsheet axion states on the rotating string}\label{subsec_pseudoaxion}

\hspace{5mm}
As we reviewed in Section~\ref{sec_review}, lattice data revealed the existence of a massive pseudoscalar resonance on the worldsheet. As explained in Section~\ref{subsec_EST_from_lattice}, in EFT we describe it via the action~\eqref{eq_S_pseudoaxion}, \emph{pretending} that its mass is sufficiently small $m_a\ll \ellb_s^{-1}$.  In this section we analyze this massive scalar mode on the rotating string background~\eqref{eq_background}.  The regime of interest to us is
\begin{equation}\label{eq_a_regime}
\omega\ll m_a\ll \ellb_s^{-1}\,.
\end{equation}

Due to the nontrivial dependence on $\sigma$ of the worldsheet metric \eqref{eq_metric}, the energy levels of a free massive particle have a nontrivial dependence on $m_a$ and $\omega$. To see this, consider the free EOM for the worldsheet axion in frequency space $a(\tau,\sigma)=e^{-i\epsilon\tau}a(\sigma)$ (neglecting the higher-derivative coupling to $K\cdot\tilde{K}$):
\begin{equation}\label{eq_a_m_EOM_pre}
-\left(1-\sigma ^2\right) a''(\sigma )+\sigma  a'(\sigma )+
\frac{m_a^2 }{\omega ^2}\left(1-\sigma ^2\right)a(\sigma ) =\epsilon^2 a(\sigma)\,,
\end{equation}
where we set $\alpha_f=1$ for simplicity since we are dealing with light quarks.  Changing coordinates as in eq.~\eqref{eq_s_def}, we recast~\eqref{eq_a_m_EOM_pre} as
\begin{equation}\label{eq_a_m_EOM}
-a''(s)+\frac{m^2_a}{\omega^2}\cos^2(s)a(s)=\epsilon^2 a(s)\,.
\end{equation}
In appendix \ref{app_Mathieu} we discuss the general solution of~\eqref{eq_a_m_EOM} in terms of Mathieu functions. Here we present an approximate analysis in the regime~\eqref{eq_a_regime}, that captures the essential results. 

Eq.~\eqref{eq_a_m_EOM} has the same structure of the Schr\"odinger equation for a particle in a potential $\frac{m_a^2 }{\omega ^2}\cos^2(s)$.  In the limit of interest $m_a\gg \omega$, such potential localizes the field near the endpoints $s=\pm\frac{\pi}{2}$.  To look for approximate solutions,  it is convenient to rewrite $s=\pm(\pi/2-\sqrt{\omega/m_a}\,x)$, so that near each endpoint we obtain the following equation to leading order in $\omega/m_a$
\begin{equation}\label{eq_a_m_EOM_rewritten}
-\frac{1}{2}a''(x)+\frac{x^2}{2} a(x)\simeq\frac{\omega}{2m_a}\epsilon^2 a(x)\quad\text{with}\quad
x>0\,.
\end{equation}
Notice that in the approximation we are working with we obtain two copies of eq.~\eqref{eq_a_m_EOM_rewritten} at the two endpoints. We thus conclude that the particle spectrum of the axion is doubly-degenerate up to corrections exponentially suppressed in $m_a/\omega$, physically associated with the tunneling between the localized states.

Eq.~\eqref{eq_a_m_EOM} is the Schr\"odinger equation for the harmonic oscillator in half-space with energy levels $\frac{\omega}{2m_a}\epsilon^2$.  The allowed values of $\epsilon$ depend on the boundary conditions at $x=0$;\footnote{In the approximation leading to~\eqref{eq_a_m_EOM_rewritten} we simply demand regularity at $x\rightarrow\infty$.} we will discuss below the possibilities for the example at hand. For the moment we simply notice that the eigenfunctions for Dirichlet and Neumann boundary conditions are obtained restricting the usual wave-functions for the harmonic oscillator to, respectively, even and odd quantum numbers. Therefore we have solutions for
\begin{equation}
\frac{\omega}{2m_a}\epsilon^2= \begin{cases}
2k+\frac{1}{2} &\text{for Neumann} \\
2k+\frac{3}{2} &\text{for Dirichlet}
\end{cases}\qquad
\text{with}\quad
k\in\mathds{N}\,,
\end{equation}
where $\mathds{N}$ stands for natural numbers including zero. 
In any case the energy of a single-particle axion mode scales as $\epsilon\sim \sqrt{m_a/\omega}$. This is an important conclusion. In fact, this result shows that the axion states are below cutoff for $m_a\ll \ellb_s^{-1}$ when we account for the redshift factor at the boundary. Indeed eq.~\eqref{eq_ddt_bdry} demands $\omega\epsilon\ll \Lambda_{bdry}\sim\sqrt{\omega/\ellb_s}$.

Let us now briefly discuss the boundary conditions for $a$. In the regime~\eqref{eq_a_regime}, $a$ is approximately invariant under constant shifts. We would like to understand if the string endpoints also preserve such approximate shift symmetry, in which case the Neumann condition would be a good approximation as in Section~\ref{subsec_toy_bc}; conversely, if the (approximate) shift symmetry is explicitly broken at the string endpoints, one expects a Dirichlet condition to provide a better approximation.

One the one hand, even for $m_a=0$, the axion-like coupling in eq.~\eqref{eq_S_pseudoaxion} always breaks the shift invariance of $a$ in the presence of a boundary. This is because $K\cdot\tilde{K}$, despite being a topological invariant, cannot be written as a total derivative of a gauge-invariant invariant quantity, as showed in eq.~\eqref{eq_KKt4d} of the appendix. This suggests that the string endpoints will always break the approximate shift symmetry. Therefore, \emph{a priori} nothing prevents us from considering an arbitrary potential $V(a)$ at the string endpoints, where $V(a)=V(-a)$ due to parity. Under the assumption that $a=0$ is a global minimum of the potential, this would set $a\vert_{\sigma=\pm 1}=0$ up to corrections of order $\sim \sqrt{\omega\ellb_s}$.\footnote{As well known, for Dirichlet boundary conditions it is convenient to write the boundary actions in terms of normal derivatives (see e.g. appendix B of \cite{Cuomo:2021cnb}). In the case at hand, one could consider an action of the form
\begin{equation}\label{eq_Sbdry_a3}
\begin{split}
S_{bdry}=&
-\sum_{\sigma=\pm 1}\int d\tau \,\sigma \,a\pd_s a\\
&-\sum_{\sigma=\pm 1}\int d\tau\sqrt{\hat{\gamma}_{\tau\tau}}\left\{c_1\ellb_s(\pd_s a)^2\hat{\gamma}_{\tau\tau}^{-1}+
c_2\ellb_s^{3}\left[\tilde{\nabla}_\tau(\pd_s a)\right]^2\hat{\gamma}_{\tau\tau}^{-2}+\ldots\right\}\,,
\end{split}
\end{equation}
where $\sigma\pd_sa\vert_{\sigma=\pm1}=N^\alpha_{\tau}\pd_\alpha a$ in terms of the covariant normal defined in eq.~\eqref{eq_covariant_normal}.
The first term in eq.~\eqref{eq_Sbdry_a3} is designed to cancel the boundary term in the variation of the bulk action $\sim\delta a\pd_s a$.  The boundary conditions then arise from the variation with respect to $\pd_s a$ (which is independent of $a$ at the endpoints).  The first term in eq.~\eqref{eq_Sbdry_a3} then scales like $\omega^0$. The terms proportional to $c_1$ and $c_2$ in eq.~\eqref{eq_Sbdry_a3} scale as, respectively, $\sqrt{\omega\ellb_s}$ and $(\omega\ellb_s)^{3/2}$. }

Note that the above theoretical argument suggests that the approximate shift symmetry for $a$ should also be broken at the boundaries of a long static flux tube. This would imply that the gap of the worldsheet axion in a long open flux tube of length $L$ should be larger by a factor $\sim \pi^2/(m_aL^2)$ compared to a closed string due to the Dirichlet boundary condition. Perhaps surprisingly, lattice data suggest instead that the lightest worldsheet axion state in a static open flux tube has approximately the same gap as in a closed flux tube, even for relatively short strings~\cite{Juge:2003ge,Sharifian:2024nco}. This surprising result suggests that the above theoretical argument, hinting at Dirichlet-like boundary conditions for $a$, should be taken with caution. For this reason, we prefer not to commit to a specific choice of boundary conditions.

In summary, the low energy excitations of the worldsheet axion in the rotating string background are localized at distance $(1-|\sigma|)\lesssim \omega/m_a$ from the endpoints.  This implies that at large $J$ the corresponding modes have an approximately doubly degenerate spectrum. The result may be summarized by the formula
\begin{equation}\label{eq_e_pseudoaxion}
\epsilon_n\simeq\sqrt{\frac{m_a}{\omega }}\times\begin{cases}\sqrt{4 \left[\frac{n}{2}\right]+3} & \text{for Dirichlet} \\[0.5em]
\sqrt{4 \left[\frac{n}{2}\right]+1} & \text{for Neumann}
\end{cases}
\quad\text{with}\quad n\in\mathds{N}\,,
\end{equation}
where we reported the result both for Dirichlet and Neumann boundary conditions and $[n/2]$ denotes the integer part of $n/2$ (notice that the lowest energy state corresponds to $n=0$).
In appendix~\ref{app_Mathieu} we provide the exact result for arbitrary values of $m_a/\omega$. Since in practice in QCD $m_a$ is close to the cutoff, we do not consider higher derivative corrections in $\ellb_s m_a$  to this result. Note that the axion states have a gap $\delta M\sim\omega \epsilon\sim \sqrt{m_a \omega}\sim \sqrt{m_a/\ellb_s}J^{-1/4}$. Therefore, for $m_a\sim\ellb_s^{-1}$, the gap of the axion scales as the contribution of an endpoint mass term~\eqref{eq_Sbdry_Massive} to $M$.

\section{Heavy quarks}\label{sec_heavy_quarks}

\hspace{5mm}
In order to understand how EST relates to heavy quarkonium states and what are the different regimes as a function of the quarks' masses, in this section we develop the EFT for mesons with heavy quarks, i.e., with mass $m\gg \ellb_s^{-1}$. The EFT of mesons with heavy quarks is not plagued by the endpoints' singularities that we mentioned in Section~\ref{sec_singularities}, and it is therefore technically straightforward. (In fact, some of the calculations that we present have already appeared in \cite{Sonnenschein:2018aqf}). Nonetheless, the existence of an additional large scale, the quark mass $m$, makes power counting more involved compared to mesons with light quarks. 

We will find that one needs $J\gg \sqrt{\ellb_s m}$ for the flux tube to be much longer than $\ellb_s$, pushing the validity regime of the EFT to higher values of $J$ compared to light quarks. We will also show that the parameter $y\equiv J/(m^2\ellb_s^2)$ controls the EFT spectrum. In the relativistic regime $y\gg 1$, the mass becomes a small perturbation and we recover results similar to those that we discussed so far. In the opposite regime, $y\ll1$, the lowest energy states of the effective theory coincide with the energy levels of two heavy particles in a linear potential.  Notably, in this regime the quarks' spin states are below the cutoff, and can be analyzed within EFT differently than for light quarks (cf. Section~\ref{subsec_spin}).  We will use this to show that in the EFT regime ($J\gg \sqrt{\ellb_s m}$), the quarks' Thomas precession leads to spin-orbit inversion of the energy levels.

\subsection{Double-scaling limit}

\hspace{5mm}
Let us again consider the NG action supplemented with the endpoint mass term~\eqref{eq_Sbdry_Massive}. For a heavy quarkonium meson the quarks velocity $\alpha_f$ does not need to be close to the speed of light for the string to be long. Recalling eq.~\eqref{eq_alpha_Massive}, we may therefore formulate the EFT in the following double-scaling limit
\begin{equation}\label{eq_heavy_double_scaling}
    \ellb_s\omega\rightarrow 0\quad\text{with}\quad
    \tilde{m}\equiv m\omega\ell_s^2=2\pi m\omega\ellb_s^2=\text{fixed}\,,
\end{equation}
which is equivalent to
\begin{equation}\label{eq_heavy_double_scaling2}
J\rightarrow\infty\quad\text{with}\quad y\equiv\frac{J}{m^2\ellb_s^2}=\text{fixed}\,.    
\end{equation}
In other words, we can use EFT to obtain result for the spectrum in a series at large $J$ with coefficients that are functions of $J/(m^2\ellb_s^2)$.

For heavy quarks, the endpoints' redshift factor is controlled by the fixed parameter $m\omega \ell_s^2$ and is not necessarily large. Therefore we can use the bulk power counting also at the endpoints, i.e. $X\sim 1/\omega$ and $G_{\alpha\beta}\sim 1/\omega^2$. Due to the factor of $m$ upfront, in the double-scaling limit~\eqref{eq_heavy_double_scaling}, the quark mass term~\eqref{eq_Sbdry_Massive} scales as the NG bulk action and cannot be thought as a small perturbation of the Neumann boundary conditions.  The scaling of higher derivative boundary terms depends on the Wilson coefficients' dependence on $m$. In the simplest scenario, which is also the one of phenomenological interests in QCD, we think of the endpoints as fundamental particles whose free action is exactly given by~\eqref{eq_Sbdry_Massive}. Note that all higher derivative terms for a free point particle are redundant due to the equations of motion~\cite{Delacretaz:2014oxa}. In this case,  higher derivative boundary terms can be thought as the result of the interactions with the string. Therefore, we expect them to be suppressed simply by powers of $\ellb_s\hat{\nabla}_t$ with respect to the free-particle action.  

In summary, using the equivalence~\eqref{eq_redundancy} between $\ell_s^{-2}\pd_n X$ and $m(\hat{\nabla}_tX)^2$ dictated by the leading order boundary conditions (now being careful about the factors of $m$'s), we conclude that to the first nontrivial subleading order, the boundary action can be written in the same form as eq.~\eqref{eq_Sbdry1} after rescaling $\bar{b}_1\rightarrow \bar{b}_1/(\ellb_s^2 m^2)$:
\begin{equation}\label{eq_heavy_Sbdry}
S_{bdry}=-m\sum_{\sigma=\pm 1}\int d\tau\sqrt{\hat{G}_{\tau\tau}}
\left[1+ \frac{\bar{b}_1}{m^2}\left(\hat{\nabla}_t\pd_n X\cdot\hat{\nabla}_t\pd_n X\right)+\ldots\right]
\,,
\end{equation}
where we used the fact implied by eq.~\eqref{eq_bc_Massive} that $(\hat{\nabla}_t^2X)^2$ is a redundant operator and the last term in eq.~\eqref{eq_heavy_Sbdry} is equivalent to $\sim \ellb_s^4(\hat{\nabla}_t^3X)^2$. As formerly mentioned,  in the double-scaling limit~\eqref{eq_heavy_double_scaling}, the leading term in eq.~\eqref{eq_heavy_Sbdry} contributes at order $m/\omega\sim 1/(\ellb_s\omega)^2$, which is the same as the NG action, while the last term in eq.~\eqref{eq_heavy_Sbdry} is suppressed by a relative factor of $\omega^2/m^2\sim (\ellb_s\omega)^4$, i.e. it is fourth order in derivatives according to the power-counting that we just discussed. 

Let us now illustrate our discussion concretely by calculating the energy of the leading Regge trajectory. Taking the bulk action to be just the NG one\footnote{At fourth order in derivatives, there are also bulk operators that contribute at the same order of $\bar{b}_1/m^2$ in eq.~\eqref{eq_heavy_Sbdry} in the double-scaling limit~\eqref{eq_heavy_double_scaling}; we nonetheless neglect such bulk contributions since our goal here is simply illustrating the structure of the expansion.}, from the boundary conditions we obtain the value of $\alpha_f$ as
\begin{equation}
\alpha_f=\alpha_{f,0}+(\ellb_s\omega)^4\alpha_{f,1}+\ldots\,,
\end{equation}
where 
\begin{align}\label{eq_heavy_alphaf}
\alpha_{f,0}=  \frac12\left(\sqrt{4+\mt^2}-\mt\right)  \,,\qquad\alpha_{f,1}= -\frac{4 \pi ^2 \bar{b}_1}{\mt^2 \sqrt{4+\mt^2}} \,.
\end{align}
Expressing $\mt$ in terms of $\alpha_{f,0}$ and using eq.~\eqref{eq_JE_Noether}, the angular momentum and the mass are conveniently expressed as
\begin{align}\label{eq_heavy_J_general}
J&=
\frac{\alpha_{f,0} \sqrt{1-\alpha_{f,0}^2}+\arcsin(\alpha_{f,0})}{2 \pi  \ellb_s^2 \omega ^2}+
\frac{4 \pi \bar{b}_1 \ellb_s^2 \omega^2  \alpha_{f,0} \left(2+\alpha_{f,0}^2\right) }{\left(1-\alpha_{f,0}^2\right)^{3/2} \left(1+\alpha_{f,0}^2\right)}+O\left(\ellb_s^6\omega^6\right)\,, \\
M&=
\frac{\sqrt{1-\alpha_{f,0}^2}/\alpha_{f,0}+ \arcsin(\alpha_{f,0})}{\pi   \ellb_s^2 \omega }+
\frac{4 \pi   \bar{b}_1 \ellb_s^2 \omega ^3 \alpha_{f,0}}{\left(1-\alpha_{f,0}^2\right)^{3/2} \left(1+\alpha_{f,0}^2\right)}
+O\left(\ellb_s^6\omega^7\right)\,.
\label{eq_heavy_M_general}
\end{align}
Note that when the endpoints have spin, eq.~\eqref{eq_heavy_J_general} is just the orbital contribution to the angular momentum.
From eqs.~\eqref{eq_heavy_alphaf} and~\eqref{eq_heavy_J_general} we can solve for $\omega$ in terms of $J$ (now using $\mt=m\omega \ell_s^2$). We find
\begin{equation}
\ellb_s\omega=\frac{1}{\sqrt{J}}g_1(y)+\frac{\bar{b}_1}{J^{5/2}}g_2(y)+\ldots\,,    
\end{equation}
where we defined two functions $g_1(y)$ and $g_2(y)$. In the two opposite limits these are given by:
\begin{align}
g_1(y)&=\begin{cases}\displaystyle
\frac12
-\frac{\sqrt{\pi }}{3 y^{3/4}}  +\frac{3\pi^{3/2}}{20 y^{5/4}} +\ldots &\text{for }y\gg 2\pi \\[1em]
\displaystyle\frac{y^{1/6}}{(2\pi^2) ^{1/3}}-\frac{7 y^{5/6}}{36 (2 \pi^2)^{2/3}}+\ldots  &\text{for } y\ll 1\,,
\end{cases}\\[1em]
g_2(y)&=\begin{cases}\displaystyle
\frac{3 y^{3/4}}{8 \sqrt{\pi }}+\frac{5\sqrt{\pi }}{32}  y^{1/4}+\ldots &\text{for }y\gg 2\pi \\[1em]
\displaystyle
\frac{2^{4/3} y^{5/6}}{3 \pi ^{4/3}}-\frac{871 y^{13/6}}{19440 \left(2^{1/3}\pi ^{8/3}\right)}+\ldots  &\text{for } y\ll 1\,,
\end{cases}
\end{align}
where $y\gg1$ corresponds to the ultra-relativistic limit $J\gg \ellb_s^2m^2$, while $y\ll 1$ is the non-relativistic (NR) limit $J\ll \ellb_s^2 m^2$.
Plugging back into eq.~\eqref{eq_heavy_alphaf}, we extract $\alpha_{f,0}$
\begin{equation}\label{eq_heavy_alpha0}
 \alpha_{f,0}=
 \begin{cases}
 \displaystyle\left(1-\frac{1}{2} \pi  \sqrt{\frac{1}{y}}+\frac{\pi ^2}{8 y}+\ldots\right)- \frac{\bar{b}_1}{J^2}\left(\frac{3}{8} \sqrt{\pi }y^{1/4}+
 \frac{\pi ^{3/2}}{32 y^{1/4}} +\ldots\right)+\ldots &\text{for }y\gg 1
 \\[1em]
\displaystyle
\left(\frac{y^{1/3}}{2^{2/3} \pi^{1/3}}-\frac{11 y}{72 \pi }+\ldots\right)
-\frac{\bar{b}_1}{J^2}
\left(\frac{2 y}{3 \pi }-\frac{10 2^{2/3} y^{5/3}}{27 \pi ^{5/3}}+\ldots\right)+\ldots
&\text{for } y\ll 1\,.
 \end{cases}
\end{equation}

Finally, using eq.~\eqref{eq_heavy_alpha0} we obtain $M$ from eq.~\eqref{eq_heavy_M_general}. In the relativistic limit, after re-arranging terms , we recover the previous result~\eqref{eq_leading_Regge_1} up to the rescaling $\bar{b}_1\rightarrow\bar{b}_1/(m^2\ellb_s^2)$ that we performed in going from the boundary action for light quarks~\eqref{eq_Sbdry1} to our current conventions in eq.~\eqref{eq_heavy_Sbdry}. Note that although the expansion takes the same form in terms of $J$ for both light and heavy quarks, the quark mass provides the leading contributions to several terms in the large $J$ expansion in the case of heavy endpoints. E.g. the coefficient of the term proportional to $\sim J^{-1/4}$ in eq.~\eqref{eq_leading_Regge_1} is $\sim(m^4+10\bar{b}_1/\ellb_s^4)/m^{3/2}$ (recalling the rescaling of $\bar{b}_1$), and therefore is dominated by the mass contribution for heavy quarks. In other words, for heavy quarks all the $1/J$ (classical) corrections to the relativistic Regge slope are approximately determined only by the quark mass, as expected.

The result takes a qualitatively different form in the NR limit $J\ll (\ellb_s^2 m^2)$. First, notice that in this limit the length of the string~\eqref{eq_string_length} is given by
\begin{equation}\label{eq_heavy_NR}
    L\simeq\frac{2\alpha_{f,0}}{\omega}\simeq\ellb_s\left(\frac{4\pi J^2}{\ellb_s m}\right)^{1/3}\,.
\end{equation}
Therefore, the EFT is reliable only as long as 
\begin{equation}\label{eq_heavy_EFT_regime}
J\gg \sqrt{\ellb_s m}\,.
\end{equation}
Physically, eq.~\eqref{eq_heavy_NR} arises because in this regime the meson can be thought as a bound state of two NR particles in a linear potential. In such a system the Virial theorem relates the potential and the kinetic energy $L/\ell_s^2\sim m\omega^2 L^2$, while the angular momentum is $J\simeq m \omega  L^2$, which results in eq.~\eqref{eq_heavy_NR} and $\omega\sim  \ell_s^{-1}/\sqrt{L m} \sim \ell_s^{-4/3}/(m J)^{1/3}$ as formerly found. In Section~\ref{subsec_NR_limit}, we will discuss the derivation of the NR quantum-mechanical Hamiltonian from EST in more detail. In the NR limit, the mass of a state on the leading Regge trajectory reads
\begin{equation}\label{eq_heavy_Regge_NR}
\begin{split}
 M=2m\left[1
 +\frac{3 }{4(2\pi^2)^{1/3} }\left(\frac{J}{\ellb_s^2 m^2} \right)^{2/3}+\ldots\right] -\frac{2\bar{b}_1 m}{(2\pi^2)^{2/3}J^2} 
\left[\left(\frac{J}{\ellb_s^2 m^2}\right)^{4/3}+\ldots\right] +\ldots\,,
\end{split}
\end{equation}
where we arranged the result in a double-series according to eq.~\eqref{eq_heavy_double_scaling2}. We see that the leading contribution in the NR limit is the rest mass of the quarks. The second term in the first square parenthesis of eq.~\eqref{eq_heavy_Regge_NR} is the first nontrivial contribution and arises both from the string tension and the kinetic energy of the quarks. Finally, the term proportional to $\bar{b}_1$ in eq.~\eqref{eq_heavy_Regge_NR} scales as $\ellb_s^4 m\omega^6L^2$ and is suppressed with respect to the first nontrivial contribution by a factor of $\sim 1/(L^2 m^2)$ in terms of the length~\eqref{eq_heavy_NR}. In the next section, we will discuss the scaling of corrections in the NR regime in more detail. In particular, we will discuss quantum corrections that we neglected so far and are larger than the $\bar{b}_1$ term in eq.~\eqref{eq_heavy_Regge_NR}.

\subsection{String excitations and quantum effects in the non-relativistic limit}\label{subsec_NR_limit}

\hspace{5mm}
The spectrum of fluctuations in the NR regime $J\ll m^2\ellb_s^2$ takes a qualitatively different form compared to the relativistic limit $J\gg m^2\ellb_s^2$. A general analysis was done in \cite{Sonnenschein:2018aqf}; we checked the results and reported some details in appendix~\ref{app_heavy_excitations}. Here, we provide a simple interpretation of the results in the non-relativistic limit.

We find that in the relativistic limit, the spectrum takes the same form as for light quarks, as one would expect. In the NR regime, on the other hand, due to the slow motion, most of the modes behave as in a static string with Dirichlet boundary conditions and therefore have energy $\epsilon\gtrsim \pi/(\omega L)\sim 1/\alpha_{f,0}$ (in units of $\omega$). The only states which make exception are the protected states whose frequencies are fixed to be $\epsilon=0,1$ and do not contribute to the mass spectrum as formerly explained, and the lightest nontrivial state associated with the angular mode, which has a gap $\epsilon_2\simeq\sqrt{3}$.  This mode is physically associated with the stretching of the string and is parametrically lighter than all the others nontrivial states (because $\alpha_{f,0}\ll 1$ in the NR limit).

To illustrate the physics of the NR regime clearly, it is instructive to derive the result for the leading Regge trajectory and the light excitations' gap $\epsilon_2$ in the language of a potential quark model. To this aim, we first restore powers of the speed of light $c$ explicitly and then take the limit $c\rightarrow\infty$ in the leading order action.  In this limit, the Dirichlet-like modes have energy $\pi c/L\rightarrow\infty$ and therefore decouple. Using $\eta_{\mu\nu}=\text{diag}(c^2,-1,-1,\ldots)$ and $X^\mu=(t,\vec{X})$, the leading order action is 
\begin{equation}
\begin{split}
    S&=-\frac{T}{c}\int d^2\sigma\sqrt{G}-mc\sum_{\sigma=\pm 1} \int d\tau\sqrt{\dot{X}^2}\\
    &\rightarrow
   \int dt\left[ -2mc^2+\frac{m}{2}\sum_{\sigma=\pm 1}\dot{\vec{X}}^2-T\int_{-1}^1  d\sigma\sqrt{\pd_\sigma\vec{X}\cdot\pd_\sigma\vec{X}}
+O\left(\frac{1}{c^2}\right)\right]\,,
\end{split}
\end{equation}
where we defined $T=c/\ell_s^{2}$. More precisely, the corrections to the above results are suppressed by inverse powers of the double-scaling parameter $\alpha_{f,0}^2\sim v^2/c^2\sim T^2/(m^2\omega^2 c^2)=\tilde{m}^{-2}$. Equivalently, in terms of the angular momentum, corrections are proportional to inverse powers of $y^{1/3}$, where $y$ is the large double-scaling parameter  in eq.~\eqref{eq_heavy_double_scaling2}; note that after restoring the speed of light, $y^{1/3}$ is given by
\begin{equation}
y^{1/3}=c\left(\frac{m^2}{2\pi J T}\right)^{1/3}=\left(\frac{ m^2\ellb_s^2 c^2}{J}\right)^{1/3}\,.
\end{equation}

As expected in the NR limit, only the boundary fields have a kinetic term. Solving the bulk equations of motion, we therefore arrive at the expected result
\begin{equation}\label{eq_NR_action}
    S=\int dt\left[ -2mc^2+m
    \dot{\vec{X}}_{CM}+\frac{m}{4}\Delta\dot{\vec{ X}}^2
    -T|\Delta\vec{ X}|\right]\,,
\end{equation} 
where we wrote the action in terms of the center of mass coordinates in obvious notation. In the following discussion, we neglect the center of mass variables. 

At large $J$, the system~\eqref{eq_NR_action} can be quantized semiclassically similarly to the rigid rotor or the hydrogen atom (see e.g.~\cite{Monin:2016jmo,Cuomo:2020fsb}). To this aim, we work in spherical coordinates $\Delta\vec{X}=r(\sin\theta\cos\varphi,\sin\theta\sin\varphi,\cos\theta)$ and expand the field around the minimal energy solution of the EOMs with fixed angular momentum $J$:
\begin{equation}\label{eq_heavy_classical_solution}
r=r_0\,,\qquad \theta=\frac{\pi}{2}\,,\qquad
\varphi=\omega t\,,
\end{equation}
where the EOMs and the angular momentum fix $r_0$ and $\omega$ as
\begin{equation}\
    \frac{m}{2}r_0\omega^2=T\,,\quad\frac{m}{2}r^2_0\omega=J\quad\implies\quad
    r_0=\left(\frac{2 J^{2}}{m T}\right)^{\frac13}\,,\quad \omega=\left(\frac{2 T^{2}}{J m}\right)^{\frac13}\,.
\end{equation}
The result~\eqref{eq_heavy_Regge_NR} for the leading Regge trajectory is then obtained by evaluating the classical energy on the solution~\eqref{eq_heavy_classical_solution}:
\begin{equation}\label{eq_heavy_NR_Regge}
\begin{split}
    M_{classical} &=2m c^2+
\frac{3}{4}mr_0^2\omega^2 \left[1+O\left(\frac{1}{c^2}\right)\right]   \\
&   = 2m c^2+
    3\left(\frac{ J^2 T^2}{4 m}\right)^{\frac13}
 \left[1+O\left(\frac{J^{2/3}T^{2/3}}{m^{4/3}c^2}\right)\right]    
    \,.
    \end{split}
\end{equation}
To compute the spectrum of fluctuations, we simply need to quantize the theory at leading order in the $1/J$ expansion. This is accomplished expanding $\Delta\vec{X}$ around the solution~\eqref{eq_heavy_classical_solution}. We obtain the following quadratic action
\begin{equation}\label{eq_NR_L2}
L\simeq \frac{m}{4}(\delta\dot{r}^2-3\omega^2\delta r^2)+\frac{m}{4}r_0^2(\delta\dot{\theta}^2-\omega^2\delta \theta^2)+\frac{m}{4}\left(r_0\delta\dot\varphi+2\omega\delta r\right)^2\,.
\end{equation}
Eq.~\eqref{eq_NR_L2} describes a massless mode and two harmonic oscillators, with frequencies $\omega$ and $\sqrt{3} \omega$. The massless mode and the mass $\omega$ oscillator correspond to symmetry generators (equivalent to the $n=0$ and $n=1$ modes analyzed in Section~\ref{subsec_LO_fluct}) and thus do not create new particles. Therefore, at a quantum level the
nontrivial states are made of $n\geq 0$ \emph{quanta} of the radial oscillator,  each with energy (in units of $\omega$)
\begin{equation}
\epsilon_2=\sqrt{3}\,.
\end{equation}
The corresponding particles have mass $M_{classical}+n\sqrt{3} \omega$.

Finally, let us analyze the structure of quantum corrections. We focus on the leading Regge trajectory~\eqref{eq_heavy_NR_Regge}. To the first subleading order, the corrections arise from the one-loop Casimir energy. This consists of two contributions. The first is due to the heavy Dirichlet-like modes, that at one-loop order simply modify the linear potential by the Casimir contribution of $(d-2)$ massless scalar modes:
\begin{equation}
V(|\Delta\vec{X}|)\supset\sum_{k=1}^{d-2}\sum_{n=1}^\infty
 \frac{\pi cn }{|\Delta\vec{X}|}\left[1+O\left(\frac{1}{c^2}\right)\right]
 =-c\frac{\pi (d-2) }{12 |\Delta\vec{X}|}+O\left(\frac{1}{c}\right)\,.
\end{equation}
This term is suppressed by $\ellb_s^2/L^2\sim c/(TL^2)$ with respect to the leading order potential $TL$; note the explicit factor of $c$ due to the relation $2\pi T/c=\ellb_s^{-2}$. The other contribution is simply due to the oscillators in eq.~\eqref{eq_NR_L2} and trivially reads
\begin{equation}
\frac{1+\sqrt{3}}{2}\omega\left[1+O\left(\frac{1}{c^2}\right)\right]\,.
\end{equation}
Summing up, we arrive at
\begin{equation}\label{eq_heavy_NR_Regge_1loop}
\begin{split}
\delta M^{(1-loop)}&=-c\frac{\pi(d-2) }{12r_0}+\frac{1+\sqrt{3}}{2}\omega+
O\left(\frac{1}{c}\right)\\
&=-\left(\frac{m T}{2 J^2}\right)^{1/3}\left[c\frac{\pi  (d-2) }{12 }
-\left(1+\sqrt{3}\right)\left( \frac{J T}{2  m^{2}}\right)^{1/3}+O
\left(\frac{J^{2/3}T^{2/3}}{m^{4/3} c}\right)\right]\,,
\end{split}
\end{equation}
which is suppressed by a factor $y^{1/3}/J$ compared to the kinetic term in eq.~\eqref{eq_heavy_NR_Regge}.  As formerly noted, the expansion is well posed only as long as $y^{1/3}/J\ll1 $, i.e. $J^2\gg mc^{3/2}\sqrt{T}\sim m c\ellb_s$.

The structure of higher derivative and loop corrections is analogous to the one-loop result~\eqref{eq_heavy_NR_Regge_1loop}. At each order, the most important contribution arises from integrating out the heavy modes, which yield to contributions to the potential starting at relative order $(\ellb_s/L)^{2n}\sim y^{n/3}/J^n$ (due to the Dirichlet-like modes Casimir energy) with additional terms suppressed in $y$. Bulk higher derivative terms result into corrections of order $\sim y^{k/3}/J^n$ with $k<n$, while quantum corrections from the fluctuations of the point-like particles are always suppressed by $1/J^n$. Boundary higher derivative terms are further suppressed in $y$.

\subsection{Quarks' spin and spin-orbit inversion}\label{subsec_heavy_quark_spin}

\hspace{5mm}
Let us analyze the quarks' spin states for heavy quarks.  The kinetic term for the spin is again given by eq.~\eqref{eq_free_spin}, and yields the Hamiltonian~\eqref{eq_spin_H}. The main difference between light and heavy quarks is that in the latter case the gap $\Omega$ is not large and, in the NR limit $J\ll m^2\ellb_s^2 c^2$, is given by
\begin{equation}\label{eq_heavy_H_Thomas_spin}
\Omega\simeq \frac{\omega}{2}\alpha_{f,0}^2= \frac{c J^{1/3} /\ellb_s}{(4 \pi) ^{4/3} (m c \, \ellb_s)^{5/3}} \,,
\end{equation}
where we keep track of the speed of light $c$ as in the previous section.  

All the other couplings between the string coordinates and the spin variables involve higher derivative terms and therefore are subleading compared to the kinetic term contribution. The most important nontrivial contribution arises at first order in derivatives from an operator similar to eq.~\eqref{eq_spin_int}, explicitly given by
\begin{equation}\label{eq_heavy_spin_int}
\frac{m \ellb_s^{3}}{c^4}\int d\tau\sqrt{\hat{G}_{\tau\tau}}\ S_{\mu\nu}\hat{\nabla}^2_tX^\mu\hat{\nabla}^3_tX^\nu\sim
\frac{1}{m \ellb_s c^2}\int d\tau\sqrt{\hat{G}_{\tau\tau}}\ S_{\mu\nu}\pd_nX^\mu\hat{\nabla}_t\pd_n X^\nu\,,
\end{equation}
where we used eq.~\eqref{eq_redundancy} keeping track of the factors of $m$ (and $c$) as commented above eq.~\eqref{eq_heavy_Sbdry}.\footnote{The power of $m$ is also justified on physical grounds if we interpret eq.~\eqref{eq_heavy_spin_int} as a generalized version of the Pauli interaction term $\vec{B}\cdot\vec{S}/m$, with the magnetic field identified as $\vec{B}\sim\ellb_s^{-1}\pd_n\vec{X}\wedge\pd_n\dot{\vec{X}}$ \cite{Kogut:1981gm}.} Evaluating the operator~\eqref{eq_heavy_spin_int} on the classical background, we find that its contribution to the gap between the spin states in the NR limit is proportional to
\begin{equation}\label{eq_heavy_spinH_int}
\delta\Omega\propto\frac{\omega}{m \ellb_s c}\sim \Omega\times\left(\frac{m\ellb_s c}{J^2}\right)^{1/3}\,,
\end{equation}
which is always smaller than eq.~\eqref{eq_heavy_H_Thomas_spin} in the regime of applicability of EFT~\eqref{eq_heavy_EFT_regime}. 

In conclusion, the leading contribution to the energy splitting between the different spin states within EFT arises from the \emph{free} Thomas precession effect and is given by eq.~\eqref{eq_heavy_H_Thomas_spin}. This implies that in the state that minimizes the energy the quarks' spin are aligned with the orbital angular momentum, therefore realizing the spin-orbit inversion scenario alluded in Section~\ref{subsec_spin}. Note however that this prediction is reliable only for $J\gg\sqrt{m c\ellb_s}$, as the extrapolation of the interaction contribution in eq.~\eqref{eq_heavy_spinH_int} makes clear.

\section{A \emph{stringy} view on meson spectroscopy}

\hspace{5mm}
In this section, we compare the predictions of EST with real world data, both for light and heavy quarks' mesons. Unfortunately, for most mesons' families we barely have data beyond $J=2$, making a quantitative comparison impossible. Nonetheless, motivated by the success of similar descriptions for glueballs in $3d$ Yang-Mills \cite{Dubovsky:2016cog,Dubovsky:2021cor}, we will attempt at a qualitative description of the observed spectrum in terms of $stringy$ states, i.e. as particles predicted by EST. 

\subsection{Discrete quantum numbers}
\label{subsec_PC}

\hspace{5mm}
Let us consider a meson made of a string connecting a quark and anti-quark. Besides its mass and spin $J$, we can measure its quantum numbers under parity ($P$) and charge conjugation ($C$). Therefore, in order to compare our analysis to the experimental data, we need to discuss the assignments under $P$ and $C$ for the daughter Regge trajectories (including the ones associated with the worldsheet axion) in $d=4$.

Parity acts in the standard way $X^i\rightarrow -X^i$. In terms of the string fluctuations~\eqref{eq_flucts} and the worldsheet axion, therefore
\begin{equation}\label{eq_P}
P\phi(\tau,\sigma)P=\phi(\tau,\sigma)+\pi\,,\qquad
P\rho(\tau,\sigma)P=-\rho(\tau,\sigma)\,,\qquad
Pa(\tau,\sigma)P=-a(\tau,\sigma)\,.
\end{equation}

Charge conjugation turns the quark at the endpoint into an antiquark, and viceversa.  When the quarks at the endpoints have different flavors, we obtain a meson with opposite charge. Therefore, since $C$ is a conserved quantum number in QCD, mesons with different endpoints are not eigenstates of $C$ and yield two identical towers of states with opposite charges. 

When the quark and the anti-quark have the same flavor, charge conjugation is equivalent to reversing the string orientation and performing a rotation by a factor of $\pi$, such that the background~\eqref{eq_background} remains invariant. Therefore, working at linear order in the fluctuations, we find
\begin{equation}\label{eq_C}
C\phi (\tau,\sigma)C=\phi(\tau,-\sigma)+\pi\,,\qquad
C\rho(\tau,\sigma)C=\rho(\tau,-\sigma)\,,\qquad
Ca(\tau,\sigma)C=a(\tau,-\sigma)\,,
\end{equation} 
which is equivalent to worldsheet parity up to the nonlinear action on the zero-mode of $\phi$.

Note that for long strings worldsheet parity has a different interpretation. Namely, the action $\sigma\rightarrow-\sigma$ corresponds to spacetime $CP$ for long strings, and the worldsheet axion flips sign under that, differently than in eq.~\eqref{eq_C}.

Consider the string ground-state with angular momentum $J$. This has quantum numbers $(P_J,C_J)$, which are arbitrary within EFT.  Because of eq.~\eqref{eq_exp_i_phi0}, since the zero mode of $\phi$ shifts by $\pi$ under both $P$ and $C$, we conclude that the next state in the Regge trajectory has opposite quantum numbers $(P_{J+1},C_{J+1})=(-P_J,-C_J)$. More in general we must have
\begin{equation}\label{general_parity}
(P_{J+\delta J},C_{J+\delta J})=(-1)^{\delta J}(P_J,C_J)\,.
\end{equation}
For the same reasons, this relation holds also for all the daughter Regge trajectories. For heavy quarks, this relation reflects the fact that, in a NR potential model, the wave-function of the quarks is even/odd for even/odd orbital angular momentum.

It is also straightforward to compute the quantum numbers of the daughter Regge trajectories in terms of the leading one.  For parity, this follows directly from eq.~\eqref{eq_P}: daughter Regge trajectories obtained by acting on the vacuum with an even (odd) number of creation operators associated with $\rho$ and $a$, have the same (opposite) parity compared to the state in the leading Regge trajectory with the same angular momentum.

For mesons with identical quarks at the endpoints, eq.~\eqref{eq_C} implies that a single-particle state on the string is $C$ even (odd) if its wavefunction $e^{-i\epsilon \tau}f_\epsilon(\sigma)$ is even (odd) under $\sigma\rightarrow-\sigma$. From our results eqs.~\eqref{eq_phi} and \eqref{eq_phi_disp_LO} for $ \phi$, we conclude that angular modes with $n$ even are $C$-even, while those with $n$ odd are $C$-odd. A similar result is found for $\rho$ and $a$ (see Section~\ref{subsec_toy_bc} and app.~\ref{app_Mathieu}). The $C$ assignment of a daughter Regge trajectory is that of the leading Regge trajectory times the product of the $C$-charges of all the single particles.

In table~\ref{table_PC}, we summarize the quantum numbers of the single-modes created by the string fluctuations and the worldsheet axion. Notice that the lowest energy states are expected to be the daughter Regge trajectories associated with a single $n=2$ mode of $\phi$ or $\rho$, while the lightest axion mode has $n=0$, but this is expected to be almost degenerate with the $n=1$ axionic state for light quarks.

\begin{table}[t]
    \begin{center}
        \begin{tabular}{ |c | c | c | c | }
        \hline
              & $\phi$ & $\rho$ & $a$ \\      \hline 
            $n$ even  & $ P$ $C$ & $ (-P)$ $C$ & $ (-P)$ $C$ \\
          $n$ odd  & $ P$ $(-C)$ & $ (-P)$ $(-C)$ & $ (-P)$ $(-C)$ \\
        \hline 
        \end{tabular}  
        \caption{$P$ and $C$ assignments for the modes of the string fluctuations $\phi$ and $\rho$, and the worldsheet axion $a$. The notation compares the quantum numbers $P$ and $C$ of the leading Regge trajectory with those of the daughter trajectory obtained by exciting the corresponding mode on the string.} \label{table_PC}
    \end{center}
    \end{table}

Let us conclude this section discussing the action of $P$ and $C$ on the different spin states (assuming these are not infinitely heavy) for mesons made of spin one-half quarks. It is convenient to think of a meson as a string of orbital angular momentum $\ell$ tensored with the Hilbert space of the spins. Therefore for each string state (either in the leading or in a daughter Regge trajectory), we obtain a \emph{spin-family} of four states. The Hilbert space of the two spins decomposes into an $s=1$ triplet representation with symmetric wave-function and an $s=0$ singlet with an antisymmetric wavefunction. The tensor product of the triplet states with a string state with $(P,C)$ charges yields three states with the same charges under parity and charge conjugation: $(P_{J},C_{J})=(P,C)$ with $J=\ell-1,\ell,\ell+1$. The state obtained considering the product of the string state and the spin singlet instead has the opposite value of $C$ due to the antisymmetric wavefunction: $(P_{J},C_{J})=(P,-C)$ with $J=\ell$. Note again that the charge conjugation assignment makes sense only for mesons with identical endpoints.

\subsection{Light mesons}
\label{subsec_light_mesons}

\hspace{5mm}
In this section, we discuss mesons made of light quarks ($u,\ d$ and $s$). These are conveniently grouped according to their isospin $I=0,1/2,1$. We recall the reader that the $u$ and $d$ quarks are approximately degenerate and, in the approximation in which they are approximately massless,
form a doublet under isospin $U=\frac{1}{\sqrt{2}}(u,d)$.  In the following, all the data are obtained from the current PDG listings~\cite{Workman:2022ynf}; we provide summary tables in appendix~\ref{meson_table}.

\begin{figure}[t!]
    \centering
    \includegraphics[width=0.5\textwidth]{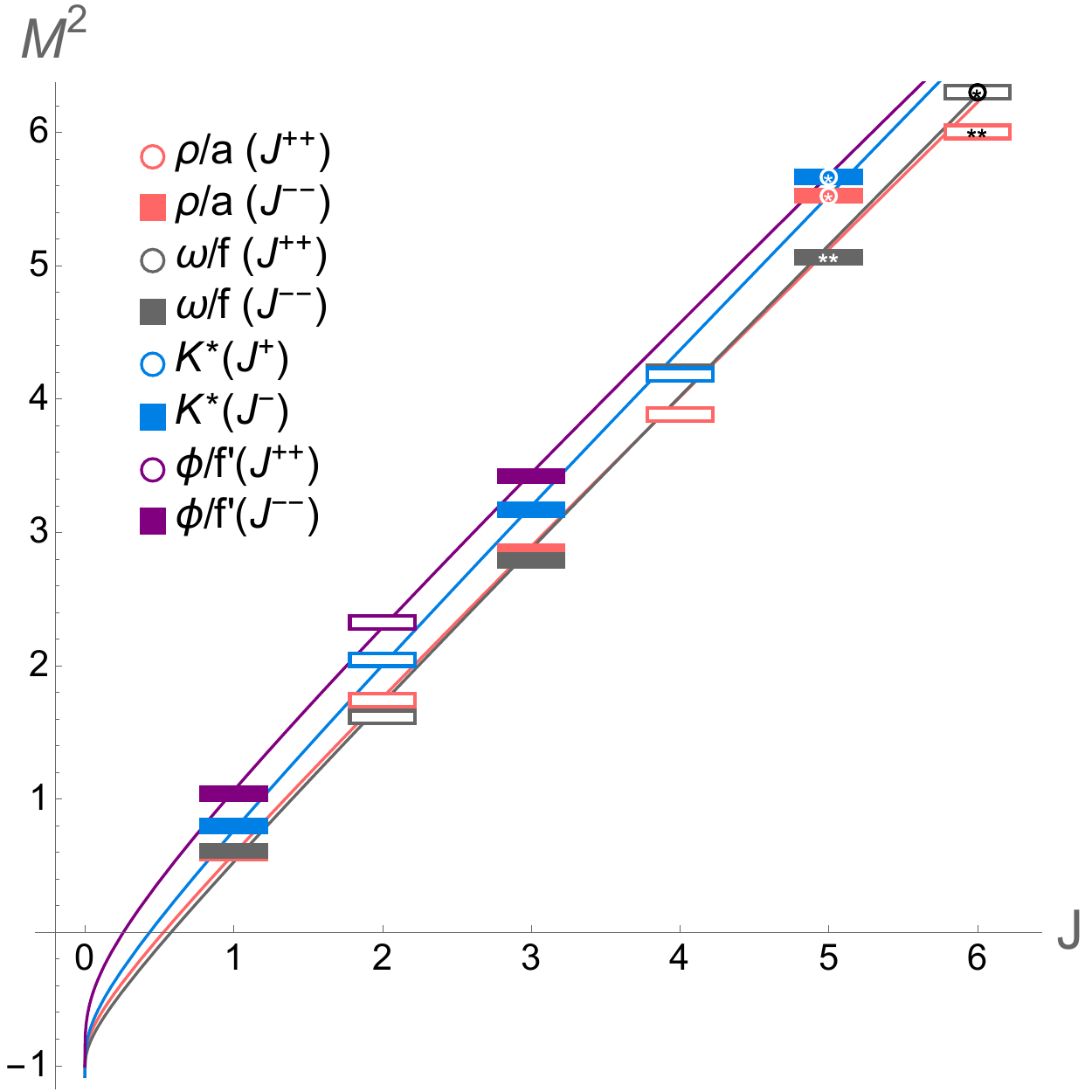}
    \caption{\small \textbf{Leading Regge trajectories:} the plot shows the mass of the lightest particles for each value of $J\geq 1$ for all possible light quark compositions. The $\omega/f$ and $\phi/f'$ trajectory have $I=0$ and (to a very good approximation) correspond, respectively, to $\bar{U}U$ and $\bar{s}s$ states. The $K$ trajectory has $I=1/2$, corresponding to $\bar{s}U$, and the $I=1$ trajectory is made of $\rho/a$ particles, with quark wave-function $\bar{U}\vec{\sigma}U$. The legend details the discrete quantum numbers of the particles. Data points with one star are listed in the summary table of PDG but they are not yet firmly established, while data with two stars are considered poorly established and are listed in the further states section of PDG. The continuous lines are the result of the best fit using the model~\eqref{eq_M2_leading_Regge_fit}. }
    \label{fig:leading_Regge}
\end{figure}

Let us begin by discussing the leading Regge trajectories, that we plot in fig.~\ref{fig:leading_Regge}. Note that we plot two different trajectories for $I=0$, where the $\omega/f$ particles are the ground states for $\bar{U}U$ mesons, and the $\phi/f'$ particles correspond instead to $\bar{s}s$ mesons.\footnote{In principle $\bar{U}U$ and $\bar{s}s$ mesons may mix with each other, but the mixing angle is numerically small.} Parity and charge conjugation quantum numbers, that we indicate with the standard notation $J^{PC}$ ($C$ is measured only for $I=0,1$), follow the predicted pattern~\eqref{general_parity}. We fitted each trajectory (boldly using all the data, including unconfirmed ones, down to $J=1$) with the formula
\begin{equation}\label{eq_M2_leading_Regge_fit}
    M^2_{leading}=\frac{1}{\ellb_s^2}(J-1)+\frac{2\mu_1}{\ellb_s} J^{1/4}\,,
\end{equation}
that is obtained truncating eq.~\eqref{eq_Regge} to the first nontrivial correction and including the quantum predicted Regge intercept~\eqref{eq_M2_quantum}. The fits are shown in fig.~\ref{fig:leading_Regge} and the results for the coefficients are reported in table~\ref{tab_Regge}. 

We remark that the data in fig.~\ref{fig:leading_Regge} include some particles that are not yet firmly established, in particular all the ones with $J>4$. Also because of this, we did not attempt at a systematic analysis of the errors in the fit, but we expect the results in table~\ref{tab_Regge} to be subject to significant uncertainties. Nonetheless, we find the consistency of the results obtained from different families encouraging, as we discuss more in detail below.

\begin{table}[t]
\centering
\begin{tabular}{|c|c|c|}
        \hline 
         &  $\ellb_s^{-2}$ ($\text{GeV}^2$) & $2\mu_1/\ellb_s$ ($\text{GeV}^2$)
         \tabularnewline
        \hline
         $\omega/f$ &  $1.09$ & $0.52$ \tabularnewline
         \hline
         $\rho/a$ &  $1.06$ & $0.59$  \tabularnewline
         \hline
         $K$ & $1.10$ & $0.76$  \tabularnewline
         \hline
         $\phi/f'$ & $1.02$ & $1.06$  \tabularnewline
       \hline
        \end{tabular}
        \caption{Results of the fit of the leading Regge trajectories using eq.~\eqref{eq_M2_leading_Regge_fit}.}
        \label{tab_Regge}
\end{table}

The Regge slope is approximately the same $\ellb_s^{-2}\approx 1.06 \,\text{GeV}^2$ for all families. This is remarkable, especially since for the $\phi/f'$ trajectory we only have three data points. The value of $\mu_1$ is very similar for the $\omega/f$ and the $\rho/a$ trajectories, whose curves almost overlap in plot~\ref{fig:leading_Regge}. These mesons indeed differ only in the quarks' flavor content and one expects them to be degenerate at large $N_c$. The $K$ and $\phi/f'$ trajectories have increasingly larger values of $\mu_1$, as expected since the $s$ quark is heavier than $u$ and $d$.\footnote{If we use $\mu_1$ to naively determine the quark mass from eq.~\eqref{eq_leading_Regge_1}, we obtain $m_u\simeq m_d \approx 240\,\text{MeV}$ from the $\rho/a$ and the $\omega/f$ trajectories, and $m_s\approx 370\,\text{MeV}$ from the $\phi/f'$ trajectory. These values are somewhat smaller than the constituent quark masses quoted in PDG: $m_u\simeq m_d \approx 340\,\text{MeV}$ and $m_s\approx 486 \,\text{MeV}$. This is not suprising since, as we have explained, the coefficient $\mu_1$ provides a reliable determination of the quark masses only for heavy quarks.} Note also that the prediction for different endpoints~\eqref{eq_Regge(2)} implies (in obvious notation)
\begin{equation}
\mu_1\vert_{K}=\frac{\mu_1\vert_{\rho/a,\omega/f}+\mu_1\vert_{\phi/f'}}{2}\,,
\end{equation}
which is in good agreement with the results of the fit in table~\ref{tab_Regge}.

Let us now consider states beyond the leading Regge trajectory. Due to the low values of $J$, typically $J=0,1,2$, we cannot expect a quantitative agreement with the EST predictions. Nonetheless, we find it instructive to analyze the data from a $stringy$ perspective, and discuss their potential interpretations as EST states based on their quantum numbers.

\begin{figure}[t!]
  \centering
  \subfloat[I=0]{\includegraphics[width=0.43\textwidth]{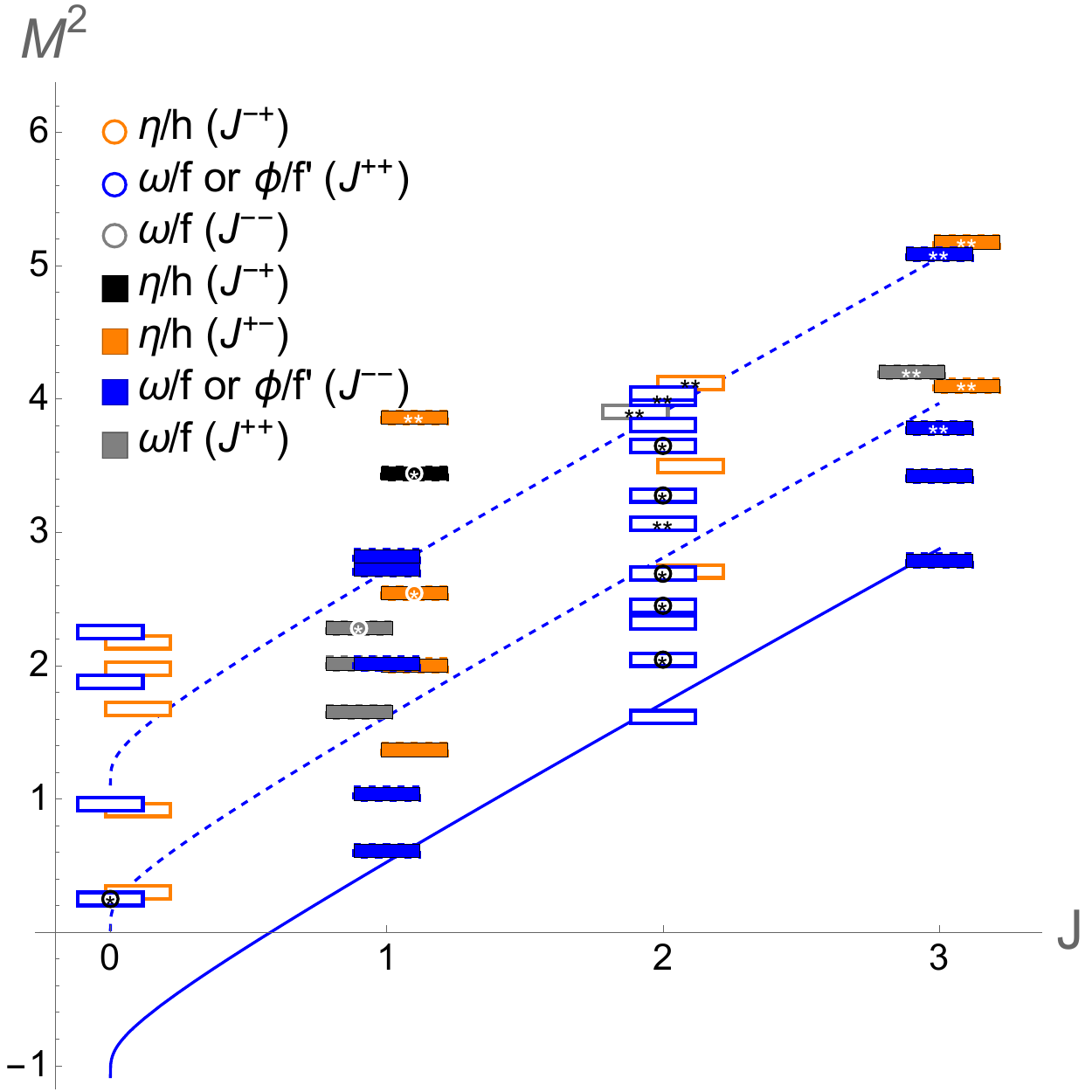}}\\
   \subfloat[I=1/2]{\includegraphics[width=0.43\textwidth]{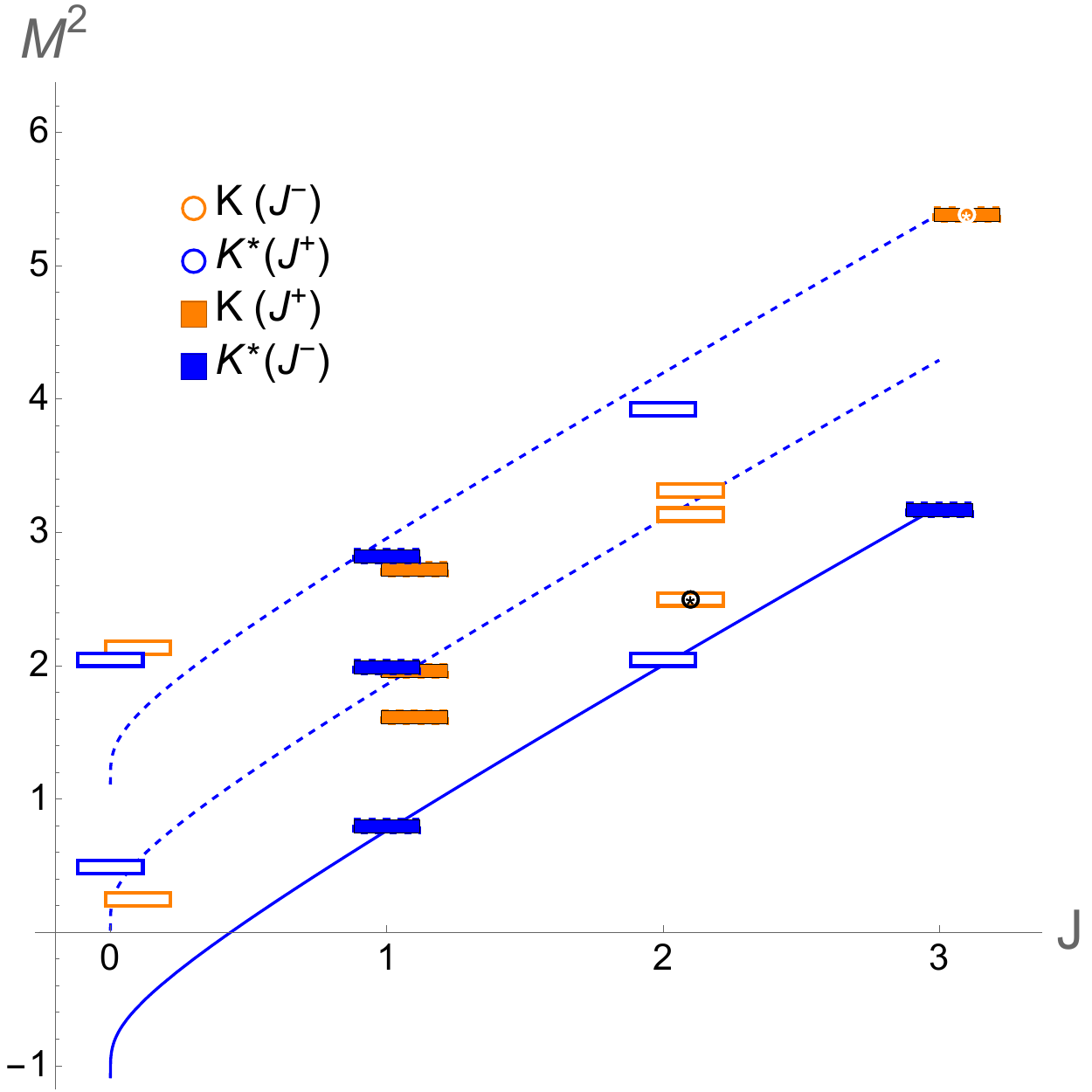}}\hfill
    \subfloat[I=1]{\includegraphics[width=0.43\textwidth]{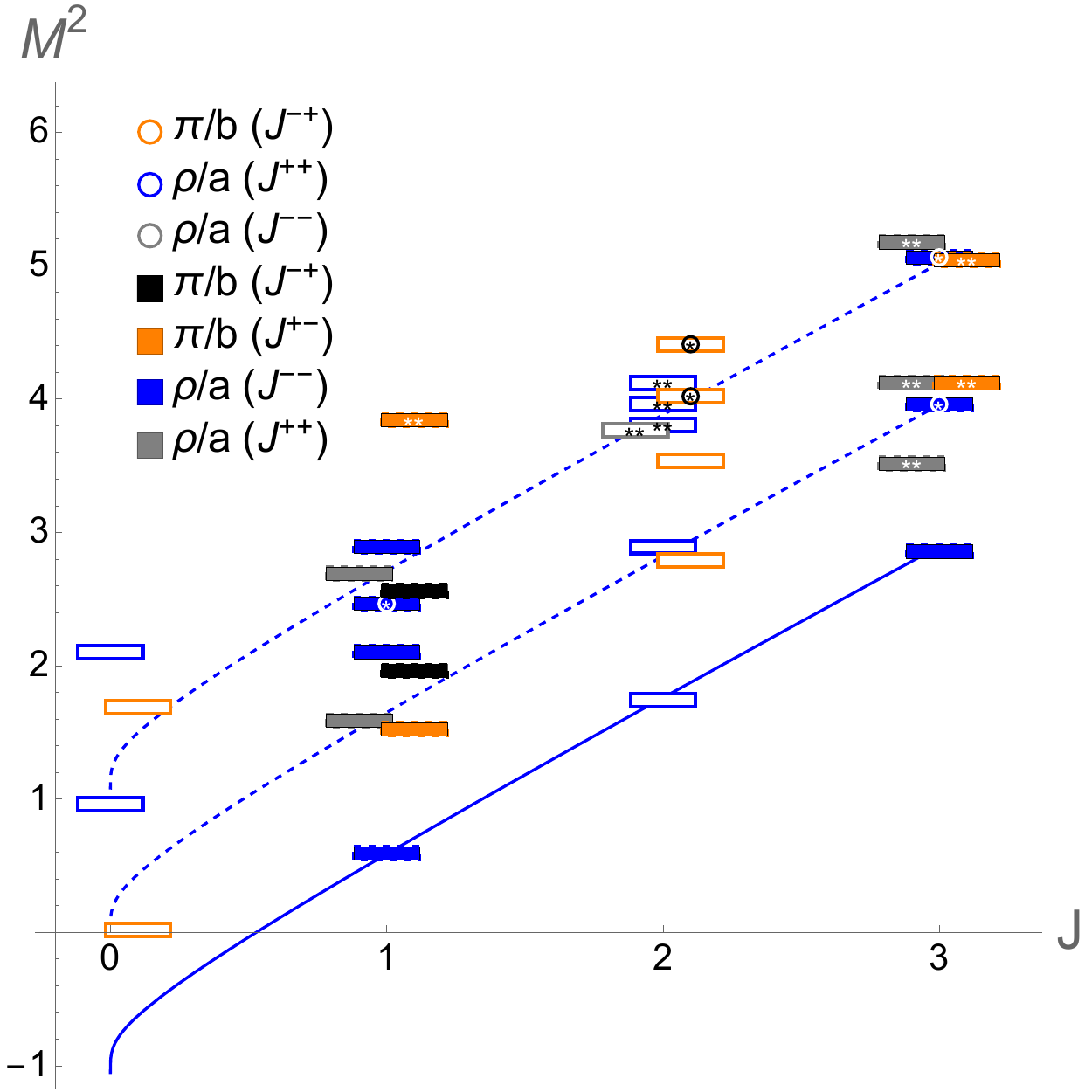}}
  \caption{\small \textbf{Mesons' spectrum for light quarks:} The three plots show the masses and the quantum numbers for mesons with isospon $I=0$, $I=1/2$ and $I=1$. As in fig.~\ref{fig:leading_Regge}, the legend details the discrete quantum numbers of the particles, data points with one star are listed in the summary table of PDG but they are not yet firmly established, while data with two stars are considered poorly established and are listed in the further states section of PDG. The solid lines are the fits for the leading Regge trajectories, and the dashed lines represent the curves $M^2_n(J)=M^2_{leading}(J)+n \ellb_s^{-2}$ with $n=1,2$. }
  \label{fig:mesons_all}
\end{figure}

In figs.~\ref{fig:mesons_all} we plot the mass square as a function of $J$ for most observed established mesons\footnote{We do not show the $\pi(1800)$, $f_0(1710)$, $f_0(2020)$, $\phi(2170)$, $f_2(2300)$, and $f_2(2340)$, which appear around the third daughter Regge trajectory and are thus much heavier than the others particles shown in figs.~\ref{fig:mesons_all}.}, as well as some unestablished ones, that are made of light quarks ($u,\ d$ and $s$) and whose quantum numbers are known. The plot stops at $J=3$, since for higher values of $J$ the only confirmed particles are those on the leading Regge trajectories; we also excluded some unconfirmed very heavy states.

A naive extrapolation of the EST predictions would suggest that the lightest daughter Regge trajectories should be found around  $2\ellb_s^{-2}$ above the leading Regge trajectory in the absence of an axion. This prediction spectacularly fails in all plots: in the $I=1/2,1$ plots, we see that the distance between the leading Regge trajectory and the other particles with the same spin is roughly $\ellb_s^{-2}$, while for $I=0$ it is sometimes even smaller. We also observe many unconfirmed states as well as sometimes partially overlapping resonances (we did not report the widths for simplicity); finally, we might be missing states because we do not consider particles listed in PDG whose quantum numbers are unknown. It appears therefore very hard to provide a rationale for all the observed data points given these features and uncertainties. 

To make progress, we focus on a subset of data which is more promising from our perspective. In experiments, it is hard to distinguish mesons with light quarks from other kinds of particles, such as molecular bound states of mesons, tetraquarks, glueballs, and mixtures thereof. To identify states which are continuously connected to meson particles in the limit $N_c\rightarrow\infty$ we therefore compare different isospin channels. Since at $N_c=\infty$ the meson spectrum is expected to take qualitatively the same form for all possible endpoints, we consider (confirmed) states that appear once, with the same quantum numbers (or just the same $P$ and $J$ when comparing to $I=1/2$) and roughly similar mass, in the $I=1/2$ and $I=1$ plots, and appear twice in the $I=0$ listing (twice to account for both $\bar{U}U$ and $\bar{s} 
s$ mesons). 

\begin{figure}[t!]
  \centering
  \subfloat{\includegraphics[width=0.5\textwidth]{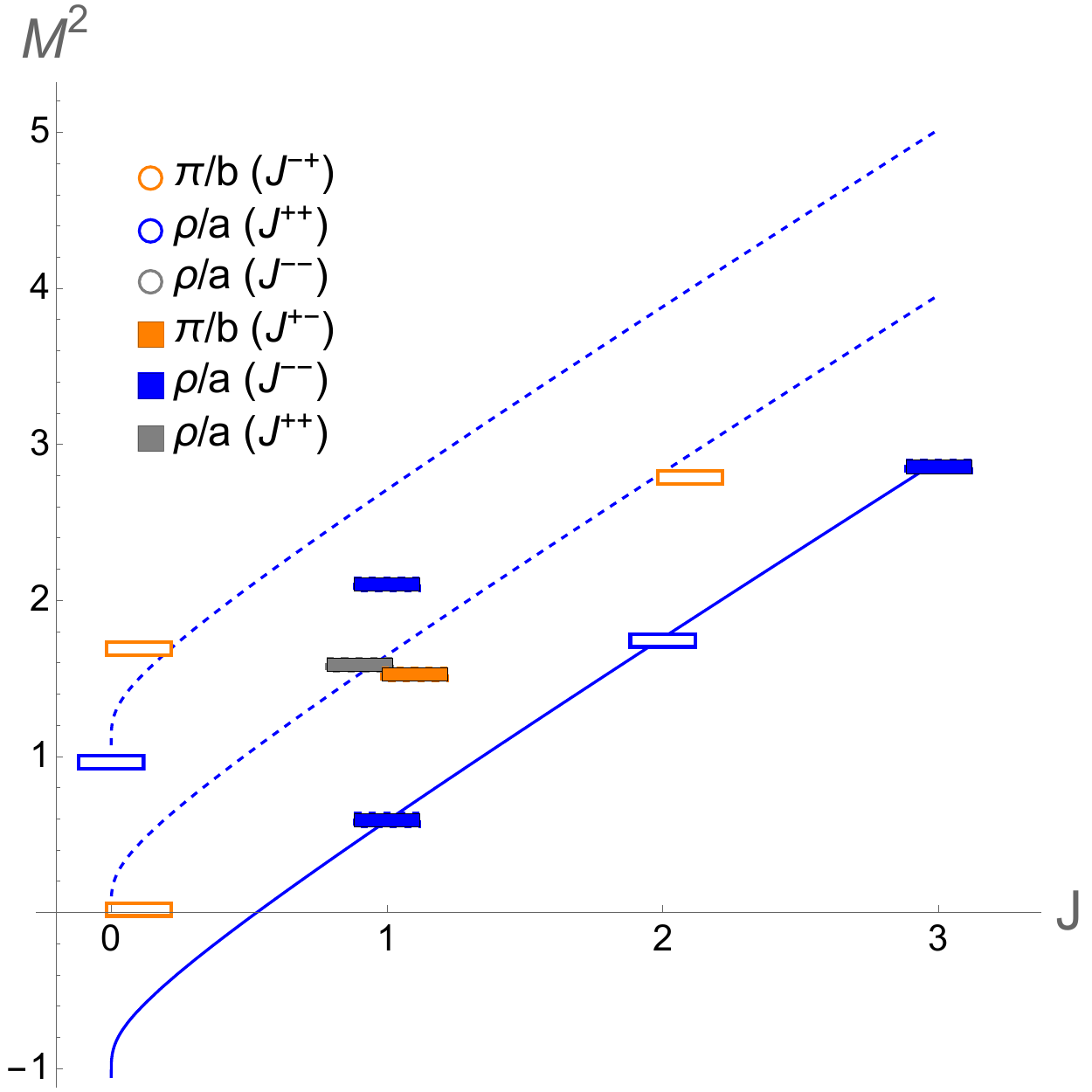}}
  \caption{\small \textbf{Reliable mesons' states:} the plot shows a subset of $I=1$ states that appear also in the other isospin  channels (twice at $I=0$) and that therefore we expect to be continuously connected to the \emph{true} mesons' spectrum at $N_c\rightarrow\infty$. }
  \label{fig:mesons_reliable}
\end{figure}

Proceeding in this way, we conservatively identify a set of $10$ states that we believe should admit an interpretation as proper meson particles. We show them in fig.~\ref{fig:mesons_reliable}, focusing for concreteness on $I=1$. Note that for $J\geq 3$ only the particles on the leading Regge trajectories showed before can be considered reliable meson candidates according to this rationale. The $J=0$ and $J=1$ candidates selected with this rationale agree with the states found in the large $N_c$ lattice study in~\cite{Bali:2013kia}, that measured masses and quantum numbers of the lightest mesons with small angular momentum.

Among the reliable mesons we still observe several particles at $\sim\ellb_s^{-2}$ distance from the leading Regge trajectory. (However, all the $I=0$ particles with mass smaller than $M^2_{leading}+\ellb_s^{-2}$, besides the $\omega/f$ and $\phi/f'$ leading Regge trajectory, have no analogues at $I=1/2$ and $I=1$, suggesting that these should be more properly interpreted as molecular bound states of two lighter mesons). To interpret these unexpectedly light states, the most natural scenario is to suppose that, for some reason, we observe states with different spin wavefunctions in the low $J$ part of the spectrum, as if the meson behaved similarly to the NR heavy quarks' regime. This is because the orbital angular momentum on the leading Regge trajectory is $\ell=J-1$, and therefore at $J=1$ the string would classically be static. The two lightest states are the $\rho$ at $J=1$ and the pion at $J=0$, and, as well known~\cite{Amsler:2018zkm}, these can be interpreted, respectively, as having a symmetric $s=1$ and an antisymmetric $s=0$ spin wavefunction for the quarks, which are connected by a static flux tube with $\ell=0$ orbital angular momentum. The pion is much lighter than the $\rho$ due to (approximate) chiral symmetry breaking. Similarly, the grey and orange states at $J=1$, and the empty blue point at $J=0$, have the right quantum numbers to arrange in a spin quartet together with the $J=2$ $a_2$ particle in the leading Regge trajectory. Note that these four particles are approximately mass degenerate, in agreement with the assumption that the spins' energy splittings are small.

At this point we might expect to see also the states corresponding to different spin wave-functions of the $J=3$ particle on the leading Regge trajectory. However, at $J=2$ there is only a candidate for the antisymmetric state, which has $2^{-+}$ quantum numbers, but no $2^{--}$ candidate for the symmetric state. The \emph{disappearence} of this spin state at $J=2$ seems a robust conclusion and is a well known puzzle in mesons' phenomenology (see e.g.~\cite{Abreu:2020wio} and refs. therein). Indeed in the complete plots~\ref{fig:mesons_all} we see that at $I=1$ the only particle with the right quantum numbers to be interpreted as the missing spin state is poorly established and is quite heavier than the other potential spin state; similarly, at $I=0$ we have only one (poorly established) particle, rather than two as required, with $2^{--}$ quantum numbers. Additionally, even if we insist in interpreting the heaviest $2^{-+}$ and $1^{--}$ particles as spin partners of the $3^{--}$ $\rho_3$, we do not have an understanding of the heaviest $0^{+-}$ state in plot~\ref{fig:mesons_reliable}.

EST suggests also a potentially different interpretation of the data. We may speculate that between $J=2$ and $J=3$ a transition occurs, from a regime in which the meson behaves as if the quarks were heavy to one that follows more closely the expectations from the light quarks' EFT that we described in the first part of this work. According to this scenario, for $J\geq 3$ all the nontrivial spin states, i.e., those with $J<\ell+1$, acquire a large (infinite?) gap above the leading Regge trajectory. Note that indeed the energy of the states in the spin family of the leading Regge trajectory increases with $J$ (at $J=0$ the $\pi$ is much lighter than the $\rho$ obviously, while at $J=1$ the four states in the $a_2$ spin family are approximately degenerate). The disappearence of spin states with $J\geq 2$ would also be compatible with the spectrum being analytic for $J\geq 2$, as expected from general arguments. According to this scenario, the heaviest $2^{-+}$ and $1^{--}$ particles in fig.~\ref{fig:mesons_reliable} are naturally interpreted as some anomalously light string excitations, with the heaviest $0^{+-}$ state obtained from the $1^{--}$ one modifying the spin wave-function. In particular, the $1^{--}$ particle could be a $n=2$ $\phi$ single particle state, which is indeed the expected lightest excitation in the heavy quark regime. Appealingly, the $2^{-+}$ particle one has instead the right quantum numbers to be interpreted as a worldsheet axion $n=0$ mode or a $\rho$ $n=2$ state. In particular, the extrapolation of our formulas suggests that the gap $\delta M^2\approx 2/\ellb_s^2$ of the $n=2$ $\rho$ state is larger than the one of the first axion daughter trajectory at low values of $J$. Indeed, given the lattice measured value of the axion mass~\eqref{eq_ma_lattice}, we can estimate the mass of the associated daughter trajectories using the formula $\delta M^2_n\simeq 2\sqrt{\ell}\omega\epsilon_n/\ellb_s$ and eqs.~\eqref{eq_e_pseudoaxion_app_Dir} and~\eqref{eq_e_pseudoaxion_app_Neu} of the appendix. A naive extrapolation to $\ell=J-1=1$ gives $\delta M^2_0\approx 1/\ellb_s^2$ and $\delta M^2_1\approx 1.3/\ellb_s^2 $ for the first two axion energy levels  assuming Neumann boundary conditions, and $\delta M^2_0\approx 1.6/\ellb_s^2 $ and $\delta M^2_1\approx 2.25/\ellb_s^2 $ assuming Dirichlet boundary conditions. The prediction $\delta M^2_0\approx 1/\ellb_s^2 $ with Neumann boundary conditions is in remarkable agreement with the experimental measure $\delta M^2\simeq 0.99/\ellb_s^2$. Notice, however, that if we were to take the numerical predictions with Neumann boundary conditions seriously, we would also expect a slightly heavier $2^{--}$ partner axion particle, which is nowhere to be seen as commented above. We therefore do not take this numerical agreement at low values of $J$ as an indication that Neumann boundary conditions are preferred for the axion state.

\begin{figure}[t!]
  \centering
  \subfloat{\includegraphics[width=0.45\textwidth]{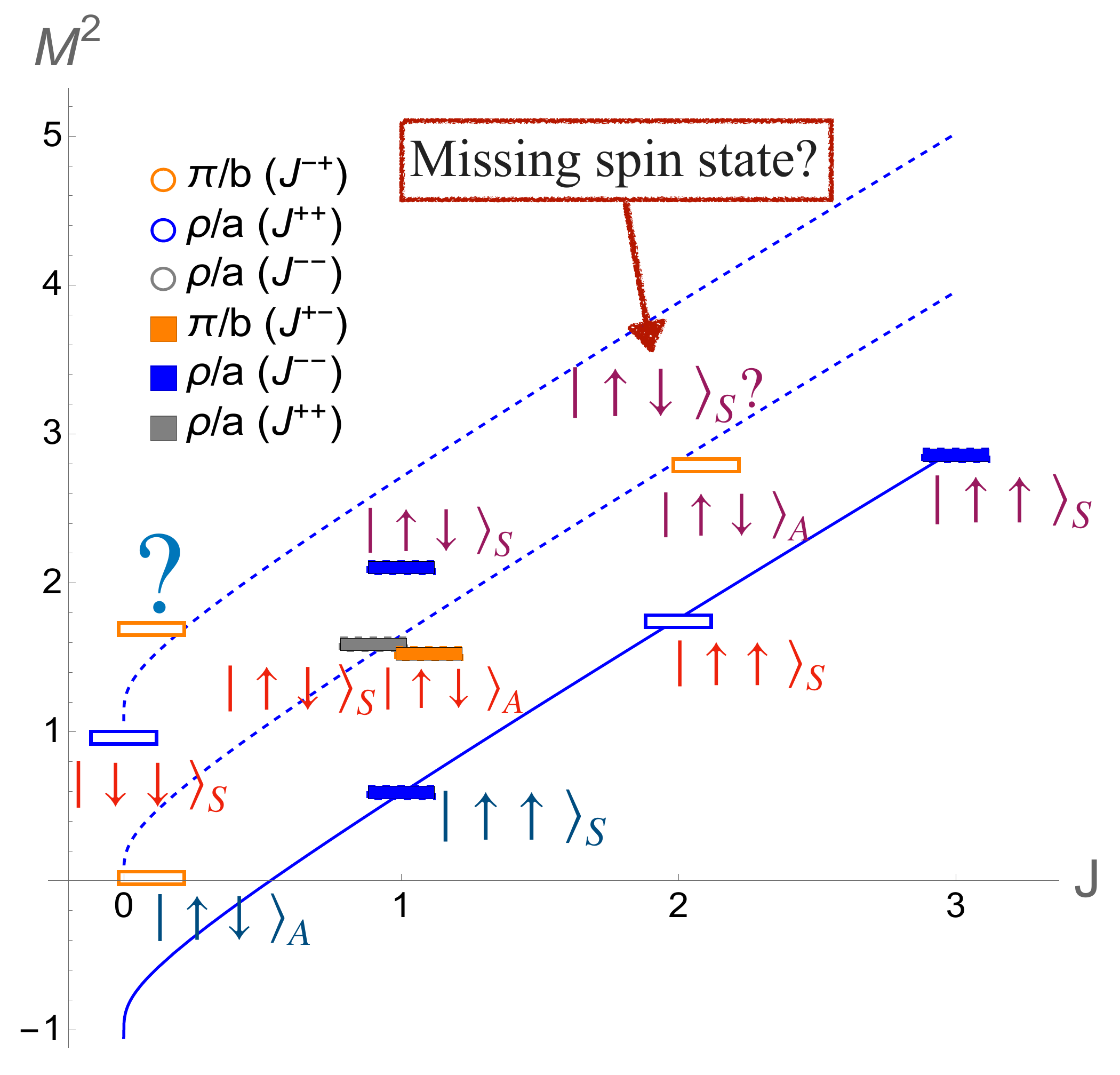}}\hfill
  \subfloat{\includegraphics[width=0.45\textwidth]{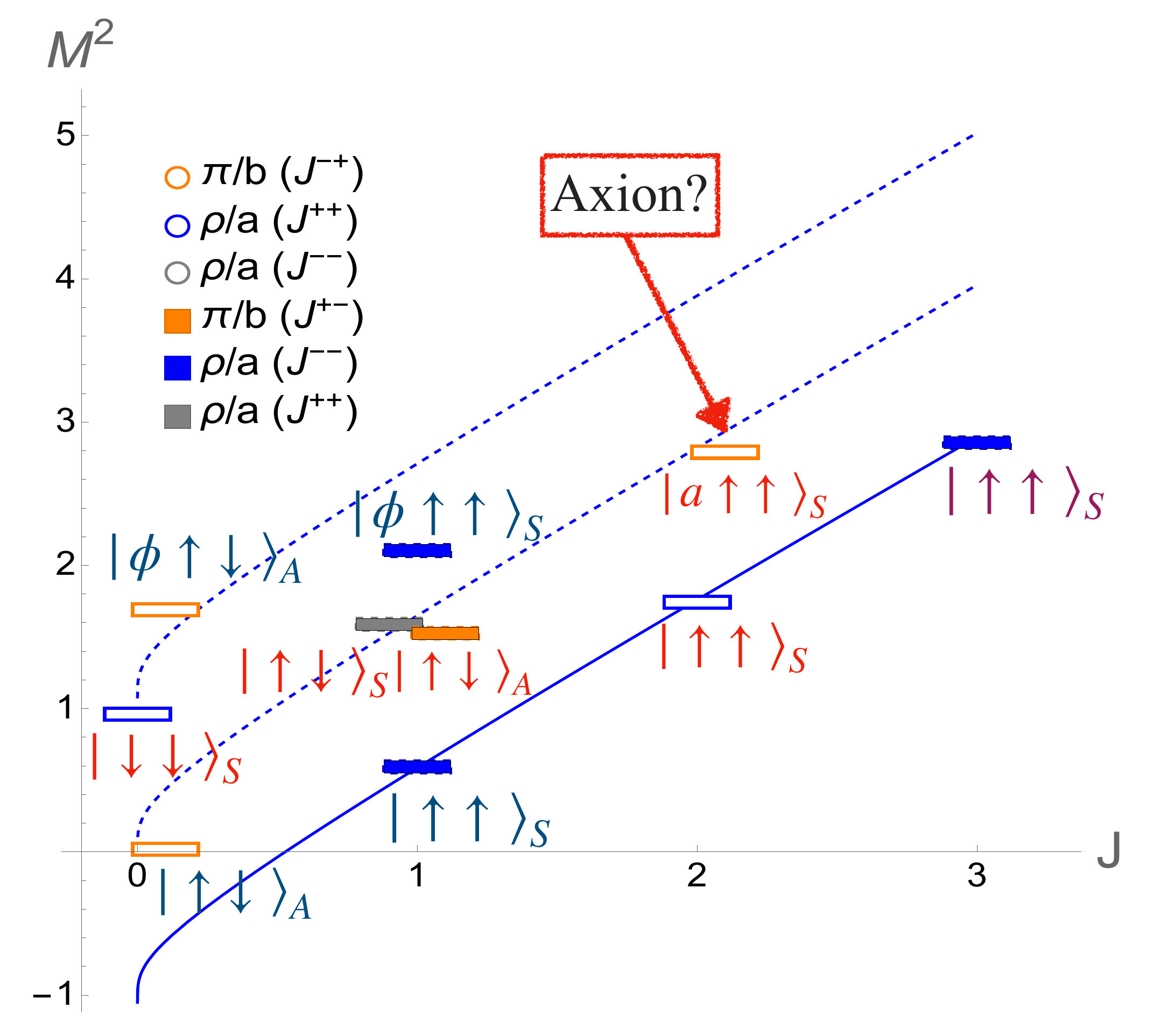}}
  \caption{\small \textbf{Potential interpretations of the states:} Graphic representations of the two scenarios described in the main text. The colored kets label the states, the subscript denoting whether the spin wave-function is symmetric or antisymmetric. Kets with the same color correspond to the same leading Regge trajectory or excitations thereof.}
  \label{fig:mesons_interpretation}
\end{figure}

The two potential interpretations of the data that we described are schematically represented in fig.~\ref{fig:mesons_interpretation}. Clearly, at this stage none of the described scenarios is particularly compelling, and it is well possible that the right interpretation of the data requires more refined ideas. Nonetheless, our analysis provides motivation for experiments to look for more mesons, in all isospin channels, at $J=3$. In particular, if the $2^{-+}$ state in fig.~\ref{fig:mesons_reliable} indeed corresponds to a wordlsheet axion mode, it would be natural to expect at $3^{+-}$ meson, again due to a axion excitation, somewhere in between $\ellb_s^{-2}$ and $2\ellb_s^{-2}$ above the leading Regge trajectory. Alternatively, if the states with different spin wavefunctions do not completely disappear and only slightly increase their gap from $J=1$ to $J=2$ (after all, we are at low values of $J$), we should expect two approximately degenerate particles with quantum numbers $3^{+-}$ and $3^{++}$ - and the absence of the $2^{--}$ particle would then need to be explained by some other mechanism. 

Before concluding, we would like to make a remark about the so-called hybrid mesons. These are the two $\pi/b$ particles and the candidate $\eta/h$ mesons with $1^{-+}$ quantum numbers, denoted by plain black markers in plot~\ref{fig:mesons_all}. These are called hybrid because their quantum numbers cannot be reproduced in a potential quark model, and are therefore usually interpreted as mesons in which the glue binding the quarks is excited \cite{Amsler:2018zkm}. From our viewpoint, these should  therefore be string excitations of the leading Regge trajectory. Consider for instance the lightest $1^{-+}$ $\pi/b$ in the $I=1$ plot in fig.~\ref{fig:mesons_all}. In the interpretation in which the lightest $2^{-+}$ $\pi/b$ is an axion excitation, then the lightest hybrid $1^{-+}$ $\pi/b$ and one of the nearby $1^{--}$ $\rho/a$ could be interpreted as different spin wavefunctions of the $J=2$ worldsheet axion. However, as noted also in \cite{Meyer:2015eta}, for the interpretation of hybrid mesons as string excitations to be reliable one would expect iso-partner states at $I=0$ and $I=1/2$ of the two hybrid $\pi/b$'s; instead, the only potential candidate at $I=0$ is much heavier than the two $I=1$ hybrids, while there is at most one particle candidate for an isopartner at $I=1/2$ (the heaviest $K$ $1^+$ state). Therefore, at this stage, pending further experimental data, the interpretation of the currently observed hybrid mesons remains puzzling.

Let us conclude by remarking that the discussion in this section is only a first attempt at mesons' phenomenology from a $stringy$ viewpoint. More data and more ideas are undoubtedly needed to improve our understanding of QCD from the point of view of the confining string.

\subsection{Comments on other mesons}

We close this section with some brief comments on other mesons' families.

We did not attempt a quantitative analysis of the heavy charmonium and bottomium states due to the reduced validity window of EST (and the scarcity of data for $J\gtrsim 1$). All confirmed states in heavy quarkonium are believed to admit an interpretation in terms of the quantization of a phenomenological potential model of quarks moving in a central potential (qualitatively, and suprisingly often also quantitatively; see e.g.\cite{Chen:2016spr,Amsler:2018zkm}). Since, as explained in Section~\ref{subsec_NR_limit}, for heavy quarks EST reduces to a NR potential model, and the quantum numbers of the bound states do not depend on the details of the potential (as long as it is spherically symmetric), we do not have anything to add to the existing literature on the subject. In other words, all observed heavy mesons admit a reasonable qualitative interpretation in terms of states of the QM potential model~\eqref{eq_NR_action}, including the different spin wavefunctions. The states corresponding to string excitations are expected to be quite heavy (with a $\sim\pi/\ellb_s$ gap on top of the leading Regge trajectory for the low values of $J$ that we can access in experiments) and presumably have not yet been observed.\footnote{On the other hand, assuming Neumann boundary conditions, as suggested by simulations of open static flux tubes~\cite{Juge:2003ge,Sharifian:2024nco}, the lightest daughter trajectory that is not captured by the quark model might be the one associated with a single worldsheet axion state on the string, whose gap is simply given by its mass~\eqref{eq_ma_lattice} and is thus not enhanced by a factor of $\pi$, for sufficiently low values of $J$.}

It might be easier to identify candidates for string excitations and axion states in heavy-light mesons. The application of EST to these particles is a straightforward extension of our work. Unfortunately, very few confirmed particles are listed in PDG, for instance, for both charmed and bottom particles we have data only for $J<2$ after excluding the leading Regge trajectory and unconfirmed states.

In the future, additional data for heavy-heavy and heavy-light mesons might provide a testing ground for our predictions. In particular, it would be interesting to confront data at $J\gtrsim\sqrt{m\ellb_s}$ for heavy quarkonium with the prediction of spin-orbit inversion and the result for the leading Regge trajectory in Section~\ref{sec_heavy_quarks}.

\section{Beyond EFT: inequivalent quantizations and glueballs in \texorpdfstring{$3d$}{3d}}

\hspace{5mm}
Let us now briefly consider what happens when we try to extend our results beyond the vicinity of the leading order Regge trajectory, where we expect the large $J$ expansion to be valid. The discussion will not focus on whether the expansion is under control but rather on the qualitative features of the leading order spectrum, such as energy degeneracies, when extended beyond this regime. As we will review, our motivation for extrapolating the EFT predictions is the study of the glueball spectrum in $d=3,4$ Yang-Mills theory. Indeed, lattice simulations show that the string transverse coordinate qualitatively describes all the observed flux tube states, up to very high energies, also for short strings in $3d$. Similarly, the string coordinates and the worldsheet axion describe all observed flux tube states in $4d$. This suggests that the string coordinates can be thought as the only \emph{fundamental worldsheet fields} in $3d$, and similarly in $4d$ including the axion. In the context of the rotating string, these observations suggest that it should be possible to, at least qualitatively, describe all the glueballs measured on the lattice in $3d$ and $4d$ Yang-Mills theory in terms of states of the rotating string Hilbert space that we described in this work.

Perhaps surprisingly, it turns out that, even after specifying the field content of the theory, the extrapolation of the Hilbert space of the theory to low values of $J$ is not unambiguous beyond the leading Regge trajectory. We will see that the natural semiclassical quantization procedure that emerges when considering the large $J$ effective field theory expansion differs, even when considering critical strings, with respect to what one would obtain from lightcone quantization. Indeed, this fact was already observed in \cite{Dubovsky:2016cog}, where a recipe for quantization of closed effective strings at large $J$---the axionic string Ansatz (ASA)---was used to describe qualitative features of the lattice glueball spectrum all the way to $J\longrightarrow0$. In this section, we identify the physical origin of this discrepancy in the simpler context of open strings---we plan to investigate this issue further in the context of closed strings in a future publication.

We will start with a brief recap of how the closed string spectrum of \cite{Dubovsky:2016cog,Dubovsky:2021cor} is obtained, where we will consider only $d=3$ for now. There, one starts with the leading order effective string theory description at large $J$ for open strings, which is in essence the leading order part of the theory described in this work, although there it was obtained by working with the Polyakov action and solving the Virasoro constraints. From the semiclassical quantization of fluctuations around the classical rotating rod solution at a given $J$, one may construct a Hilbert space of states which we will label $\mathcal{H}_{\mbox{\scriptsize{open,effective}}}$ and within this Hilbert space, one can find the energy degeneracies---the levels---at leading order in $J$. The resulting Hilbert space is spanned by vectors of the form
\begin{equation}
	\ket{\Psi}_{\mbox{\scriptsize{open,effective}}} = \prod_{n\geq 2}\left(a^{\dagger}_n\right)^{m_n}\ket{J}
\end{equation}
where $a_n$'s have the standard commutation relations and are the only oscillators that survive after imposing the Virasoro constraints on top of the fluctuations on the rotating rod background. The $a_n$ can be schematically thought as the annihilation operators for the fluctuations of the angular coordinate $\phi$ in Polyakov formalism. The $a_n$ are built in such a way that they don't change the value of $J$. At leading order in $J$, all states with the same level $N$
\begin{equation}
	N = J + \sum_{n\geq 2}n a^{\dagger}_na_n
\end{equation}
are mass degenerate: $M^2=N/\ellb_s^2$.\footnote{In the present discussion, we ignore all ordering issues, as they are irrelevant for the key point we are trying to make.}

As explained in \cite{Dubovsky:2016cog}, within the ASA the closed string spectrum is postulated to be, at each value of the level $N$, the tensor product of left-moving and right-moving Hilbert spaces
\begin{equation}
	\mathcal{H}_{\mbox{\scriptsize{closed,ASA}}} = \sum_{N} \mathcal{H}_{L}(N)\otimes \mathcal{H}_{R}(N)
\end{equation}     	
where in analogy with what happens in the critical case, the left- and right-moving sectors are postulated to be identical to the open string Hilbert space, which is taken to be the one obtained from the large $J$ effective string description just described
\begin{equation}
	\mathcal{H}_{L} = \mathcal{H}_{R} = \mathcal{H}_{\mbox{\scriptsize{open,effective}}}\,.
\end{equation}
The closed string quantization obtained from this prescription was found in~\cite{Dubovsky:2016cog} to qualitatively match the features of the $d=3$ glueball spectrum. Furthermore, a number of predictions made in~\cite{Dubovsky:2016cog} were later confirmed by~\cite{Conkey:2019blu}. 

Interestingly, as discussed in \cite{Dubovsky:2021cor}, one may also use the closed string large $J$ semiclassical quantization directly to construct the ``effective closed string'' spectrum. The result is that at $d=3$ the Hilbert space $\mathcal{H}_{closed,effective}$ is spanned by vectors of the form 
\begin{equation}
	\ket{\Psi}_{closed,effective} = \prod_{n\geq 2}\left(a^{\dagger}_n\right)^{m_n}\prod_{n'\geq 2}\left(a^{\dagger}_{n'}\right)^{m_{n'}}\ket{J}
	\label{keteffectiveclosed}
\end{equation}
where again the $a_{n},\,a_{n'}$ are oscillators that do not alter $J$. The level is given by
\begin{equation}
	N = J + \sum_{n\geq 2}n a^{\dagger}_na_n + \sum_{n'\geq 2}n'a^{\dagger}_na_{n'}
\end{equation} 

As explained in \cite{Dubovsky:2021cor}, there is crucially the extra subtlety that, due to the imposition of reparametrization invariance, the states in $\mathcal{H}_{closed,effective}$ are those of the form of \eqref{keteffectiveclosed} subject to the extra condition that $N$ must be even. This, in particular, restricts us to states of even $J$ at the leading Regge trajectory.

Interestingly, the two prescriptions to build the closed strings quantizations,  $\mathcal{H}_{closed,ASA}$ and $\mathcal{H}_{closed,effective}$ do not match beyond the leading Regge trajectory when extrapolated to low values of $J$. This is seen from the fact that they give different degeneracies: at a given value of $J$ and $N$ the number of states in $\mathcal{H}_{closed,effective}$ is larger than the number of states in $\mathcal{H}_{closed,ASA}$. For instance we see that at $J=1$ and $N=2$, $\mathcal{H}_{closed,ASA}$ is empty, while $\mathcal{H}_{closed,effective}$ yields $2$ states \cite{Dubovsky:2021cor}. As discussed in \cite{Dubovsky:2021cor}, the two prescriptions match close to the leading Regge trajectory for large $J$. It is still surprising however that the ``closed effective string'' prediction so heavily over-predicts the number of states at higher levels.

Furthermore, both prescriptions also differ from what is obtained from the lightcone quantization. To see this, it is simpler to work in an arbitrary number of dimensions $d>3$. We also consider for simplicity only states with zero spatial momentum. Then the Hilbert space is spanned by vectors of the form
\begin{equation}
	\ket{\Psi}_{open,lightcone} = \prod_{I,n>0}\left(a^{I\dagger}_n\right)^{m_{I,n}}\ket{0}
\end{equation}
for open strings and 
\begin{equation}
	\ket{\Psi}_{closed,lightcone} = \prod_{I,n>0}\left(a^{I\dagger}_n\right)^{m_{I,n}}\prod_{I',n'>0}\left(a^{I'\dagger}_{n'}\right)^{m_{I',n'}}\ket{0}
\end{equation}
for closed strings. Here the label $I = 1\dots d-2$ denotes the fact that the $\alpha^I$ oscillators generate fluctuations in the $d-2$ directions transverse to the lightcone plane. The corresponding masses (squared) are obtained from the zero mode of the Virasoro constraints and yield (again we ignore constant terms that arise from ordering)
\begin{equation}
	M^2_{open} \sim N = \sum_{n,I} n a^{I\dagger}_{n}a^{I}_n
\end{equation}
and
\begin{equation}
	M^2_{closed} \sim \sum_{n,I} n a^{I\dagger}_{n}a^{I}_n + \sum_{n,I} n a'^{I\dagger}_{n}a'^{I}_n 
\end{equation}	
where the $\sim$ symbol denotes up to some dimensionful constant factor and the closed spectrum is subject to the level-matching condition that imposes the total worldsheet spatial momentum to be zero:
\begin{equation}
\sum_{n,I} n a^{I\dagger}_{n}a^{I}_n = \sum_{n,I} n a'^{I\dagger}_{n}a'^{I}_n 
\end{equation} 
As usual in lightcone quantization, the states obtained with this procedure come out in $O(d-2)$ representations, and should then be arranged into irreps. of the massive little group $O(d-1)$. As is well known, such a rearrangement fails for $N=J=1$ in open strings, and for $N=1$ and $J=2$ in closed strings (away from the critical dimension $d=26$ where it is possible to bypass this requirement by making the corresponding states massless), but it can be done for all other states.\footnote{In $d=3$, we are interested in the spectrum that is obtained by analytic continuation of the results of the lightcone quantization procedure outlined in the main text for $d>3$ |(see also \cite{Dubovsky:2016cog}). This spectrum is different than the one that is obtained applying lightcone quantization directly in $d=3$, whose Hilbert space consists of particles with fractional angular momentum and anyonic statistics \cite{Mezincescu:2010yp}.}

It can be seen that the $d=3$ spectrum obtained in this way differs from both the ``ASA'' and ``effective string'' spectra described above. For instance, besides the aforementioned issue at $N=1$, at $N=2$ lightcone quantization predicts two scalars and a $J=4$ particle at $N=4$; ``ASA" instead predicts three scalars, two $J=2$ particles, and one $J=4$; finally, the extrapolation of closed effective string theory yields the same particles as ``ASA", plus two $J=1$ particles and two additional scalars. In general, the full situation is summarized by the following schematic inequalities
\begin{equation}
\left(\begin{array}{c}
\text{closed}\\
\text{lightcone}
\end{array} \right) = \left(\begin{array}{c}
\text{open}\\
\text{lightcone}
\end{array} \right)^2 < \left(\begin{array}{c}
\text{open}\\
\text{effective}
\end{array} \right)^2 < \left(\begin{array}{c}
\text{closed}\\
\text{effective}
\end{array} \right)\,.
\end{equation}
The first equality denotes the fact that the tensor product procedure outlined above, which accounts for level matching (this is what is meant from ``squaring'' in this picture), applied to the open string lightcone spectrum yields the closed string lightcone spectrum (indeed, this fact was the inspiration for performing this procedure within the ASA). The first inequality describes the fact that open string lightcone quantization yields a smaller number of states at fixed $N$ and $J$ than ASA, which is just the open effective string theory spectrum squared. The second inequality further indicates that the closed effective string theory spectrum predicts an even larger number of particles.

Let us attempt to identify the physical origin of how these discrepancies come about. In order to do this let us consider the situation in $d=4$, where the analysis is simpler, although the key subtlety we will point out is also present in the $d=3$ case we have described. The usefulness of considering the $d=4$ case is that here one can start by gauge fixing the lightcone gauge, and then decide whether to perform lightcone quantization or to expand around a fixed angular momentum background. This is because in $d=4$, it is possible to take a background where the string rotates in the plane transverse to the lightcone coordinates, which is not possible in $d=3$.

To illustrate he point, we will consider only open strings. If $2,3$ denote the two transverse coordinates to the lightcone plane and we define (the $\pm$ symbol here should not be confused with the $\pm$ associated to lightcone coordinates)
\begin{equation}
a^{\pm}_n = \left(a^2_n \pm i a^3_n\right)/\sqrt{2}\,,
\end{equation}  
then in lightcone gauge the zero mode of the Virasoro constraints imply that the number operator defined above is given by
\begin{equation}
N = \sum_n n a^{+\dagger}_na^+_n + \sum_n n a^{-\dagger}_na^-_n. 
\end{equation} 
The angular momentum operator for rotation in the $(2,3)$ plane is given by
\begin{equation}
J = \sum_n  a^{+\dagger}_na^+_n - \sum_n a^{-\dagger}_na^-_n. 
\end{equation} 
 
We will now see how different inequivalent quantizations arise when the theory is quantized around the $a^{\pm} =0$ background versus when one quantizes around a rotating rod background with non zero $J$. This subtlety is already captured when one considers just the two $a^+$ oscillators with lower frequency, so in order to show our point we may restrict ourselves to the simpler quantum system with two oscillators and two commuting charges, the ``Hamiltonian''
\begin{equation}
H = \omega_1 a_1^{\dagger}a_1 + \omega_2 a_2^{\dagger}a_2\,,
\end{equation}
 and the ``angular momentum'' 
\begin{equation}
J = a_1^{\dagger}a_1 + a_2^{\dagger}a_2\,,
\label{simpleAngularMomentum}
\end{equation}
where the $a_1$ and $a_2$ oscillators should be thought of as the two $a^+$ oscillators with lowest frequency and the $\omega_i$ represent the frequencies in $N$ of these oscillators (their values of $n$ in formula for $N$ above). The standard quantization -- that is, around the trivial $a_1 = a_2 = 0$ background -- obviously yields states of the form
\begin{equation}
\ket{n,m} = \frac{(a_1^{\dagger})^n}{\sqrt{n!}} \frac{(a_2^{\dagger})^m}{\sqrt{m!}}\ket{0}\,.
\end{equation}
Then if we take the case $\omega_2 > \omega_1$, at fixed $J$ the states with smallest energy are of the form $\ket{J,0}$. This would be the analogue of our leading Regge trajectory. At fixed $J$, there is a finite tower of states on top, of the form $\ket{J-1,1},\ket{J-2,2}\dots \ket{0,J}$.

Let us now consider, in this simple system, the analogue of our effective field theory quantization around the non-zero $J$ background. In order to do this, we will define a new set of canonical variables $\{\delta,J,\alpha,\alpha^{\dagger}\}$ by the canonical transformation
\begin{align}
a_1 &= e^{-i\omega_1(t+\delta)}\left(J_0^{1/2} + \beta(J) \right)\,,\\
a_2 &= e^{-i\omega_1(t+\delta)}\alpha\,,
\end{align}
where $J_0$ is the background value of the angular momentum and $\alpha$ and $\beta$ describe the fluctuations around this background. In order to work at fixed angular momentum, the $\beta$ fluctuation is written in terms of our new canonical variable $J$ by plugging these expressions into \eqref{simpleAngularMomentum} and solving for $\beta$. In these terms, the ``Hamiltonian'' becomes 
\begin{equation}
	H = \omega_1 J + (\omega_2-\omega_1)\alpha^{\dagger}\alpha
\end{equation} 
while the symplectic form yields
\begin{align}
	\omega &= ida_1^{\dagger}\wedge da_1 + ida_2^{\dagger}\wedge da_2\nonumber\\
	&= d\delta\wedge dJ + id\alpha^{\dagger}\wedge d\alpha\,.
\end{align}

When we quantize this canonical system we realize that the eigenvalues of the operator $J$ are integer because this is the conjugate variable to $\delta$, which is $2\pi$ periodic. Furthermore, since $\delta$ does not appear in the Hamiltonian, $J$ is conserved, and so the Hilbert space can be split into the direct sum of sectors with constant integer $J$. The key difference with the trivial quantization in terms of $a_1$ and $a_2$ outlined above is that now for each value of $J$ we have an \emph{infinite} tower of $\alpha$ oscillators, so that at fixed $J$ there are in this quantization many more states than in the standard one. Namely, we can consider states of the form
\begin{equation}
   |J_0,n\rangle= \frac{(\alpha^\dagger)^n}{\sqrt{n!}}|J_0,0\rangle\,,
\end{equation}
for every integer $n\geq 0$, where the leading Regge trajectory state is defined by
\begin{equation}
    J|J_0,0\rangle=J_0|J_0,0\rangle\,,\qquad
    \alpha|J_0,0\rangle=0\,.
\end{equation}
The existence of this infinite tower of states is precisely the reason why the ``effective string'' quantizations outlined above have many more states at fixed $J$ and $N$ than the lightcone quantization.

How can it be, however, that such a simple system---two harmonic oscillators---has two inequivalent quantizations? The reason is that the phase spaces spanned by the canonical variables $\{a_1,a^{\dagger}_1,a_2,a^{\dagger}_2,\}$ and $\{\delta,J,\alpha,\alpha^{\dagger}\}$ are globally inequivalent, given that the first is merely $\mathbb{R}^4$, whereas the second one is actually $\mathbb{R}^3\times S^1$.  In other words, the canonical transformation that relates $\{a_1,a^{\dagger}_1,a_2,a^{\dagger}_2,\}$ and $\{\delta,J,\alpha,\alpha^{\dagger}\}$ is singular at the origin. Given the full theory, it is always possible to identify the correct set of global coordinates for the phase space, but if one starts from the EFT spectrum, that only describes the Hilbert space locally near the leading Regge trajectory, one is faced with a choice when extrapolating to low values of $J$.  Surprisingly, our discussion suggests that the right choice for QCD strings results in a different phase space structure compared to the natural choice in the context of fundamental strings.

\section*{Acknowledgements}

We thank O. Aharony, V. Gorbenko,  S. Hellerman, Z. Komargodski, and R. Rattazzi for useful discussions.  During this work GC, SD and SZ were supported by the Simons Foundation grant 994296 (Simons Collaboration on Confinement and QCD Strings) and by the BSF grant 2018068. SD is also supported in part by the NSF grant PHY-2210349,  and by the IBM Einstein Fellow Fund at the IAS. During this work GHC was supported by CSIC I+D grant number 583 and ANII grant number FCE\_1\_2023\_1\_175902. The work of AM is supported in part by the National Science Foundation under Award No. 2310243.

\appendix

\part*{Appendix}
\addcontentsline{toc}{part}{Appendix}

\section{Geometric identities and notation}\label{app_geometry}

\subsection{Bulk geometry}

\hspace{5mm}
We denote  the string coordinates as $X^\mu$ and assume the ambient space to be flat: $\eta_{\mu\nu}=\eta_{\mu\nu}=\text{diag}(1,-1,-1,\ldots)$. The worldsheet coordinates are denoted by $\sigma^\alpha$ and the induced metric then takes the form
\begin{equation}
G_{\alpha\beta}=\pd_\alpha X^\mu\pd_\beta X^\nu \eta_{\mu\nu}\,.
\end{equation}
The associated connection is denoted $\Gamma^\alpha_{\beta\gamma}$ and $\nabla_\alpha$ is the associated covariant derivative. We also define $G=-\det(G_{\alpha\beta})$.

We define the second fundamental form and the worldsheet curvature via
\begin{equation}
K_{\alpha\beta}^\mu=\nabla_\alpha \pd_\beta X^\mu\,,\qquad
\mR^\alpha_{\;\beta\gamma\delta}=\pd_\gamma \Gamma^\alpha_{\beta\delta}-
\pd_\delta \Gamma^\alpha_{\beta\gamma}+
\Gamma^\alpha_{\gamma\eta}\Gamma^\eta_{\beta\delta}-
\Gamma^\alpha_{\delta\eta}\Gamma^\eta_{\gamma\beta}\,.
\end{equation}
The second fundamental form satisfies:
\begin{equation}
K_{\alpha\beta}^\mu \,\eta_{\mu\nu}\,\pd_\gamma X^\nu=0\,.
\end{equation}
The Riemann tensor and the second fundamental form are related by the Gauss-Codazzi formula:
\begin{equation}\label{eq_Gauss_Codazzi}
\mR_{\alpha\beta\gamma\delta}=\left(
K_{\alpha\gamma}^\mu K_{\beta\delta}^\nu-K_{\alpha\delta}^\mu K_{\beta\gamma}^\nu\right)\eta_{\mu\nu}\,.
\end{equation}

Sometimes it is useful to choose a basis of $d-2$ (spacelike) normal vectors $\{n_A^\mu\}$.\footnote{In $d=3$ there is a unique normal vector and it can be written as $n^\mu=\frac{1}{2}\varepsilon_{\mu\nu\rho}\pd_\alpha  X^\mu\pd_\beta X^\nu\varepsilon^{\alpha\beta}/\sqrt{G}$, where we normalized it such that $n^\mu n_\mu=-1$. } Normalizing them such that $n_A^\mu n_B^\nu\eta_{\mu\nu}=-\delta_{AB}$,  we can decompose the bulk metric as
\begin{equation}
\eta^{\mu\nu}=\pd_\alpha X^\mu\pd_\beta X^\nu G^{\alpha\beta}-n_A^\mu n_B^\nu\delta^{AB}\,.
\end{equation}
There are clearly many equivalent choices for the normals. Indeed we can change the basis of the normal bundle with an arbitrary local orthogonal transformation of the normals: $n^\mu_A\rightarrow R_A^{\,\,B}n_B^\mu$ where $R\in SO(d-2)$.  We may thus define a connection on the normal bundle~\cite{Aharony:2013ipa}:
\begin{equation}
A_{AB\,,\alpha}=n_A^\mu\pd_\alpha n_B^\nu \eta_{\mu\nu}=-A_{BA\,,\alpha}\,.
\end{equation}
Under a local change of basis, parametrized by $R\in SO(d-2)$, $A_{AB\,,\alpha}$ transforms as (omitting the $SO(d-2)$ indices):
\begin{equation}
A_{\alpha}\rightarrow R A_{\alpha}R^T-R\pd_\alpha R^T\,.
\end{equation}
The connection can be used to define covariant derivatives of the normal vectors, that are parallel to the string and are related to the second fundamental form as
\begin{equation}
D_{\alpha}n_A^\mu\equiv\pd_\alpha  n_A^\mu-
 A_{AB,\,\alpha}\delta^{BC}n_C^\mu
=-\pd_\beta X^\mu G^{\beta\gamma} K_{\alpha\gamma}^\nu n_A^\rho\eta_{\nu\rho}
\,.
\end{equation}

Finally, we notice that in $d=4$ the connection on the normal bundle is Abelian:
\begin{equation}
A_\alpha=\frac12 \varepsilon^{AB}n_A^\mu\pd_\alpha n_B^\nu \eta_{\mu\nu}\qquad
(d=4)\,.
\end{equation}
The topological invariant introduced in eq.~\eqref{eq_K_Kt_def} is nothing but the field strength of this Abelian connection
\begin{equation}\label{eq_KKt4d}
\sqrt{G}K\cdot\tilde{K}=\varepsilon^{\alpha\beta}F_{\alpha\beta}\quad\text{where}\quad
F_{\alpha\beta}=\pd_\alpha A_\beta-\pd_\beta A_\alpha\,.
\end{equation}

\subsection{Boundary geometry}

\hspace{5mm}
We denote the endpoint coordinate with $\tau$ and $\sigma^\alpha(\tau)$ denotes the boundary wordline on the worldsheet. The induced metric is
\begin{equation}
\hat{G}_{\tau\tau}=\pd_\tau  \sigma^\alpha \pd_\tau \sigma^\beta G_{\alpha\beta}=\pd_\tau X^\mu\pd_\tau X^\nu \eta_{\mu\nu}\,.
\end{equation}

Let us first discuss the case of a time-like boundary $\hat{G}_{\tau\tau}>0$. In this case we call $\hat{\Gamma}^\tau_{\tau\tau}=\frac{1}{2}
\hat{G}_{\tau\tau}^{-1}\pd_\tau\hat{G}_{\tau\tau}$ the worldline connection and $\hat{\nabla}_\tau$ is the associated covariant derivative. We can decompose the inverse worldsheet metric at the boundary as
\begin{equation}\label{eq_normal_dec}
G^{\alpha\beta}=\pd_\tau\sigma^\alpha\pd_\tau\sigma^\beta\hat{G}^{-1}_{\tau\tau}-\hat{n}^\alpha\hat{n}^\beta\,,
\end{equation}
where $\pd_\tau\sigma^\alpha$ is the vector parallel to the boundary, and $\hat{n}^\alpha$ denotes instead the normal. The normal can be written explicitly as
\begin{equation}\label{eq_normal}
\hat{n}_\alpha=\frac{\sqrt{G}}{\sqrt{\hat{G}_{\tau\tau}}}\varepsilon_{\alpha\beta}\pd_\tau\sigma^\beta\,\quad\implies\quad
\hat{n}_\alpha\hat{n}^\alpha=-1\,.
\end{equation}
From \eqref{eq_normal} it follows that
\begin{equation}
(\hat{n}^\alpha\pd_\alpha X^\mu)^2=\hat{n}^\alpha\pd_\alpha X^\mu\hat{n}^\beta\pd_\beta X^\nu \eta_{\mu\nu}=-1\,.
\end{equation}
The second fundamental form associated with the embedding of the endpoint in the ambient space coincides with the acceleration $\hat{\nabla}_\tau\pd_\tau X^\mu$. We can also consider the second fundamental form associated with the embedding of the endpoint in the worldsheet,
\begin{equation}
\hat{k}^\alpha_{\tau\tau}=\hat{\nabla}_\tau\pd_\tau \sigma^\alpha\,.
\end{equation}
It turns out however that $\hat{k}^\alpha_{\tau\tau}$ is not an independent object, since it satisfies:
\begin{equation}
\hat{k}^\alpha_{\tau\tau}=-\hat{n}^\alpha\left(
\hat{\nabla}_\tau\pd_\tau X^\mu \eta_{\mu\nu}\hat{n}^\beta\pd_\beta X^\nu\right)\,.
\end{equation}

Let us now discuss a null boundary: $\hat{G}_{\tau\tau}=0$. For a null worldline, we may define a reparametrization invariant metric as~\cite{1985rwrc.conf...33B}:
\begin{equation}
    \hat{\gamma}_{\tau\tau}=\ellb_s\left(-\ddot{X}^\mu\eta_{\mu\nu} \ddot{X}^\nu \right)^{1/2}\,,
\end{equation}
that allows to define the following natural connection
\begin{equation}
    \tilde{\Gamma}^{\tau}_{\tau\tau}=\frac{1}{2}\hat{\gamma}_{\tau\tau}^{-1}\pd_\tau\hat{\gamma}_{\tau\tau}\,.
\end{equation}
The main difference compared to the timelike case is that it is not possible to define a normal which is invariant under reparametrizations of the boundary coordinate $\tau$ as in eq.~\eqref{eq_normal}. We may however define a \emph{covariant normal} as
\begin{equation}\label{eq_covariant_normal}
N^\alpha_\tau=\lim_{\sigma^\alpha\rightarrow\sigma^\alpha(\tau)}
\sqrt{G}\,G^{\alpha\beta}\varepsilon_{\beta\gamma}\pd_\tau\sigma^\gamma\,,
\end{equation}
where the limit is well defined under the assumption that the boundary admits a natural conformal structure as in eq.~\eqref{eq_metric_conformal}; note indeed that the definition~\eqref{eq_covariant_normal} is Weyl invariant. It is simple to check that the left-hand side of eq.~\eqref{eq_toy_bc_smart} coincides with $N^\alpha_\tau\pd_\alpha a$. Note that the definition~\eqref{eq_covariant_normal} makes sense for both timelike and null boundaries, and reduces to $n^\alpha\sqrt{\hat{G}_{\tau\tau}}$ in the former case.

\section{Near boundary expansion with the \texorpdfstring{$(K^2)^2$}{K4} term}\label{app_near_bdry}

\hspace{5mm}
As explained in Section~\ref{subsec_endpoint_ren_example}, the solutions of the quadratic EOMs deriving from eq.~\eqref{eq_S_k4} admit the following expansion:
\begin{align}
\rho_i(\tau,\sigma)&=\rho_i^{(0)}(\tau,\sigma)+c_4\frac{\ellb_s^4\omega^4}{(1-\alpha_f^2\sigma^2)^4}\rho_i^{(1)}(\tau,\sigma)+\ldots\,,\\
\phi(\tau,\sigma)&=\phi^{(0)}(\tau,\sigma)+c_4\frac{\ellb_s^4\omega^4}{(1-\alpha_f^2\sigma^2)^4}\phi^{(1)}(\tau,\sigma)+\ldots\,,
\end{align}
where both the $\{\rho_i^{(0)},\,\phi^{(0)}\}$ and the $\{\rho_i^{(1)},\,\phi^{(1)}\}$ are regular for $\alpha_f\sigma\rightarrow\pm 1$. 

We used the following expansions for $1\mp\alpha_f\sigma\equiv \delta\rightarrow 0$ of the leading order solutions:
\begin{align}
\rho_i^{(0)}=&  \left[1+\delta\frac{d^2}{d\tau^2}+
\frac{\delta^2}{6}
\left(\frac{d^2}{d\tau^2}+\frac{d^4}{d\tau^4}\right)
+\frac{\delta^3}{90}\left(4\frac{d^2}{d\tau^2}+5\frac{d^4}{d\tau^4}+\frac{d^6}{d\tau^6}\right)
+O\left(\delta^4\right)
\right]\beta_1^\rho(\tau) \nonumber
\\ \nonumber 
+ &
\delta^{1/2}\left[1+\delta
\left(\frac{1}{12}+\frac{1}{3}\frac{d^2}{d\tau^2}\right)
+\delta^2\left(\frac{3}{160}+\frac{1}{12}\frac{d^2}{d\tau^2}+\frac{1}{30}\frac{d^4}{d\tau^4}\right)
\right. \\ & \phantom{\delta^{1/2}[} \left. 
+\delta^3\left(\frac{5}{896}+\frac{37}{1440}\frac{d^2}{d\tau^2}+\frac{1}{72}\frac{d^4}{d\tau^4}+\frac{1}{630}\frac{d^6}{d\tau^6}\right)
+O\left(\delta^4\right)\right]\beta_2^\rho(\tau)\,,\\
\nonumber
\phi^{(0)}=&
\left[1-\delta\frac{d^2}{d\tau^2}
-\delta^2\left(\frac32\frac{d^2}{d\tau^2}+\frac12\frac{d^4}{d\tau^4}\right)
-\delta^3
\left(\frac{14}{9}\frac{d^2}{d\tau^2}+\frac{11}{18}\frac{d^4}{d\tau^4}+\frac{1}{18}\frac{d^6}{d\tau^6}\right)
\right. \\  \nonumber & \left.\phantom{[}
-\delta^4\left(\frac{47}{30}\frac{d^2}{d\tau^2}+\frac{229}{360}\frac{d^4}{d\tau^4}+\frac{13}{180}\frac{d^6}{d\tau^6}+\frac{1}{360}\frac{d^8}{d\tau^8}
\right)+O\left(\delta^5\right)
\right]\beta_1^\phi(\tau)
\\ \nonumber +&
\delta^{3/2}\left[1+\delta\left(\frac{21}{20}
+\frac{1}{5}\frac{d^2}{d\tau^2}\right)+
\delta^2\left(\frac{237}{224}
+\frac{33}{140}\frac{d^2}{d\tau^2}+\frac{1}{70}\frac{d^4}{d\tau^4}\right)
\right. \\   & \phantom{\delta^{3/2}[} \left.
+\delta^3\left(\frac{407}{384}+\frac{7387}{30240}\frac{d^2}{d\tau^2}
+\frac{143}{7560}\frac{d^4}{d\tau^4}
+\frac{1}{1890}\frac{d^6}{d\tau^6}\right)+
O\left(\delta^4\right)\right] \beta_2^\phi(\tau)\,.
\end{align}
For the second correction we used
\begin{align}\nonumber
\rho_i^{(1)}=&  \left[0+0+\frac{24}{15}\delta^2
\left(\frac{d^2}{d\tau^2}+\frac{d^4}{d\tau^4}\right)
+\delta^3\left(\frac{96}{5}\frac{d^2}{d\tau^2}+\frac{512}{15}\frac{d^4}{d\tau^4}+\frac{224}{15}\frac{d^6}{d\tau^6}\right)
+O\left(\delta^4\right)
\right]\beta_1^\rho(\tau) \nonumber
\\ \nonumber 
+ &
\delta^{1/2}\left[-4+\delta
\left(\frac{41}{5}+\frac{36}{5}\frac{d^2}{d\tau^2}\right)
+\delta^2\left(\frac{247}{120}+\frac{187}{15}\frac{d^2}{d\tau^2}+\frac{158}{15}\frac{d^4}{d\tau^4}\right)
\right. \\ &\phantom{\delta^{1/2}[} \left. 
+\delta^3\left(\frac{8419}{480}+\frac{26171}{360}\frac{d^2}{d\tau^2}+\frac{1295}{18}\frac{d^4}{d\tau^4}+\frac{754}{45}\frac{d^6}{d\tau^6}\right)
+O\left(\delta^4\right)\right]\beta_2^\rho(\tau)\,,\\
\nonumber
\phi^{(1)}=&
\left[0+0+0-\frac{512}{15}\delta^3
\left(\frac{d^2}{d\tau^2}+2\frac{d^4}{d\tau^4}+\frac{d^6}{d\tau^6}\right)
\right. \\  \nonumber &\phantom{[} \left.
+\frac{256}{15}\delta^4\log\delta\left(4\frac{d^2}{d\tau^2}+9\frac{d^4}{d\tau^4}+6\frac{d^6}{d\tau^6}+\frac{d^8}{d\tau^8}\right)+O\left(\delta^5\right)
\right]\beta_1^\phi(\tau)
\\ \nonumber +&
\delta^{3/2}\left[-\frac{84}{5}-
\delta\left(\frac{37}{5}-\frac{124}{5}\frac{d^2}{d\tau^2}\right)
-\delta^2\left(\frac{1223}{40}-\frac{713}{5}\frac{d^2}{d\tau^2}-\frac{242}{5}\frac{d^4}{d\tau^4}\right)
\right. \\   &\phantom{\delta^{3/2}[} \left.
+\delta^3\left(\frac{18399}{160}-\frac{55493}{120}\frac{d^2}{d\tau^2}
-\frac{10081}{30}\frac{d^4}{d\tau^4}
-\frac{766}{15}\frac{d^6}{d\tau^6}\right)+
O\left(\delta^4\right)\right] \beta_2^\phi(\tau)\,.
\end{align}
Notice the appearence of a $\delta^4\log\delta$ term. In general, an operator of order $2n$ in derivatives leads to $\delta^{2n+k}\log\delta$ terms, with $k=0,1,\ldots$, in the near boundary expansion.

\section{Bosonic action for a point particle with spin}\label{app_spin}

\subsection{Coadjoint orbit through spinors}

\hspace{5mm}
Let us consider a relativistic-point particle with spin $s$. We are interested in writing a worldline action which describes its spin. In the standard approach,  one introduces some suitably coupled Grassmanian variables on the worldline, whose Hilbert space describes the different spin states~\cite{Berezin:1976eg,Barducci:1976qu,Brink:1976uf}, see~\cite{Landry:2020obv} for a review. This approach relies on the introduction of supersymmetric gauge redundancies,  and the number of Grassmanian fields needed depends upon the size of the spin representation. 

A more straighftorward approach, which encompasses all spin representations, is the so-called coadjoint orbit technique \cite{Alekseev:1988tj,Wiegmann:1989hn}. Below, we review how to implement this construction in terms of spinors following \cite{Forte:2005ud}. In the next subsection we explain how this approach can be understood from the viewpoint of spontaneous symmetry breaking.

Let us first review the action for a non-relativistic spin in the $2s+1$-dimensional representation at a point in space.  We consider a bosonic $su(2)$ spinor $z=\{z_1,z_2\}$, subject to the constraint $\bar{z} z=2$. In this formulation, the action reads (see e.g. \cite{sachdev_2011})\footnote{The action~\eqref{eq_free_NRspin0} has been studied in several different contexts, see e.g. \cite{Lieb:1973vg,Rabinovici:1984mj,Clark_1997}. The constrained spinor $z$ is also equivalent to a charged particle moving on a sphere with a charge $s$ monopole at the center in the lowest Landau level  \cite{WU1976365,Dunne:1989hv,Hasebe:2010vp}. } 
\begin{equation}\label{eq_free_NRspin0}
S=i s\int d\tau\bz\dot z\,,\qquad
\bz z=2\, .
\end{equation}
The action~\eqref{eq_free_NRspin0} is invariant under $U(1)$ gauge transformations $z\rightarrow e^{i\alpha(\tau)}z$.  Invariance under large gauge-transformations requires
\begin{equation}
s\in\mathds{Z}/2\,.
\end{equation}
The target space is  $CP^1\simeq S^2$, physically parametrizing the orientation of the spin:
\begin{equation}
S^i=s\bz \frac{\sigma^i}{2}z\,.
\end{equation}
The canonical commutation relations $[z,\bz]=\mathds{1}/s$ imply that the spin satisfies the $su(2)$ algebra $[S^i,S^j]=i\varepsilon^{ijk} S^k$, while the constraint implies $S^i S^i=s(s+1)$.\footnote{To check this, it is important to use the correct ordering of the operators; we refer the reader to \cite{Cuomo:2022xgw} for a more detailed discussion of ordering issues in the path-integral formulation.}  Because of the $U(1)$ gauge-invariance and the constraint,  all the interactions with other fields on the worldline are straightforwardly written in terms of $S^i$.

To make contact with the standard literature on the subject \cite{Lieb:1973vg,Rabinovici:1984mj,sachdev_2011}, it is convenient to write $z$ in terms of polar and azimuthal angles $\theta_s$ and $\phi_s$:
\begin{equation}\label{eq_Bloch_par}
z=\sqrt{2}\left(\begin{array}{c}
e^{-i\phi_s/2}\cos\frac{\theta_s}{2}\\  
e^{i\phi_s/2}\sin\frac{\theta_s}{2}
\end{array}\right)
\end{equation}
Then the action~\eqref{eq_free_NRspin0} becomes
\begin{equation}\label{eq_S_free_NR}
S=s\int d\tau\cos\theta_s\dot\phi_s=-s\int_{\mathcal{M}} \epsilon_{ijk} \hat{n}^i d\hat{n}^j d\hat{n}^k\,,\qquad
\hat{n}^i=S^i/s\,,
\end{equation}
where in the second line we wrote the spin kinetic term as a Wess-Zumino integral, such that $\pd\mathcal{M}$ coincides with the worldline.

Let us now discuss the generalization to a relativistic point particle with worldline $X^\mu(\tau)$.\footnote{In the relativistic case, the coadjoint orbit technicque is equivalent to a certain nonlinear relalization of the coset $ISO(3,1)/\left(U(1)\times \mathds{R}\right)$, as we will see below. This coset spans an eight dimensional phase space, parametrized by the tree spatial positions $X^i$, the corresponding momenta and the two angles describing the spin variable \cite{Alekseev:1988tj,Wiegmann:1989hn,Forte:2005ud}.}  The motion of the particle is described via the standard action $-m\int d\tau\sqrt{\dot{X}^2}$. To account for the spin, we introduce a \emph{bosonic} Dirac spinor $\psi$ satisfying
\begin{equation}\label{eq_psi_constraints}
\bar{\psi}\psi=2\,,\qquad\slashed{u}\psi=\psi\,,
\end{equation}
where $u^\mu=\dot{X}^\mu/\sqrt{\dot{X}^2}$.  $\psi$ can be written explicitly in terms of a two-component spinor $z$ like before as (in Weyl basis)
\begin{equation}
\psi=\frac{1}{\sqrt{2}}\Lambda_u\cdot\{z,z\}\,,
\end{equation}
where $\Lambda_u$ is the Dirac boost matrix with velocity $u^\mu$. 
Then we include on the worldline an action analogous to eq.~\eqref{eq_free_NRspin0}:
\begin{equation}\label{eq_free_spin_app}
S_{spin}=s\int d\tau i\bar{\psi}\dot{\psi}\,.
\end{equation}
This action is manifestly Lorentz invariant and admits a $U(1)$ gauge-invariance similarly to eq.~\eqref{eq_free_NRspin0}, which fixes $s\in\mathds{Z}/2$.  

It is straightforward to check that in the nonrelativistic limit $u^\mu\simeq\delta^\mu_0$, the action~\eqref{eq_free_spin_app} reduces to~\eqref{eq_free_NRspin0}. To prove that the action~\eqref{eq_free_spin_app} describes a relativistic spin-$s$ particle in general,  it is convenient to quantize $\psi$ as an unconstrained variable, and impose the constraints~\eqref{eq_psi_constraints} on-shell.  One finds that the canonical commutation relations
\begin{equation}
[\psi,\bar{\psi}]=\frac{1}{s}\mathds{1}\,,
\end{equation}
imply that the spin part of the Lorentz generators, 
\begin{equation}
S_{\mu\nu}=\frac{i}{4}s\bar{\psi}[\gamma_\mu,\gamma_\nu]\psi\,,
\end{equation}
satisfies the Lorentz algebra:
\begin{equation}
[S_{\mu\nu},S_{\rho\sigma}]=-i\eta_{\mu\sigma}S_{\nu\rho}+\text{permutations}\,.
\end{equation}
One also finds that the on-shell constraint $\bar{\psi}\psi$ implies $S^\mu S_\mu|\text{phys}\rangle=-s(s+1)|\text{phys}\rangle$,\footnote{To prove this it is useful to write $S^\mu$ as \cite{Dreiner:2008tw}:
 \begin{equation}\label{eq_Smu}
S^\mu=\left(\vec{u}\cdot\vec{S}_{rest},\,
\vec{S}_{rest}+\frac{\vec{u}\cdot\vec{S}_{rest}}{1+u^0}\vec{u}
\right)\,,
\end{equation}
and use that $\vec{S}^{\,2}_{rest}|\text{phys}\rangle=s(s+1)|\text{phys}\rangle$. } where $S^\mu$ is the Pauli-Lubanski vector defined by
\begin{equation}
S_{\mu}=\frac{1}{2}\varepsilon^{\mu\nu\rho\sigma}u_\nu S_{\rho\sigma}\,.
\end{equation}

Finally, all interactions with other worldline fields are written in terms of $S_{\mu\nu}$. This is because of the $U(1)$ gauge-invariance and the following relation that follows from~\eqref{eq_psi_constraints}
\begin{equation}\label{eq_psi_psib}
\psi\bar{\psi}=P_u+P_u\gamma_5\gamma_\mu P_u \hat{n}^\mu
\,,
\end{equation}
where $\hat{n}^\mu=S^\mu/s$ and we introduced the following projector:
\begin{equation}
P_u=\frac{1}{2}\left(\mathds{1}+\slashed{u}\right)=P_u^2\,.
\end{equation}
Notice also that $S_{\mu\nu}u^\nu=0$ and $S_{\mu\nu}S^{\mu\nu}=-2S_\mu S^\mu=2s^2$ (at the classical level).

\subsection{Spin from Goldstones}

\hspace{5mm}
As emphasized in \cite{Delacretaz:2014oxa}, a point-particle, or more in general any relativistic object which appears point-like from large distances, defines a state which breaks \emph{spontaneously} the Poincar\'e group,  similarly to a string (as described in Section~\ref{sec_review}) or a membrane. A spinless particle in its rest frame, breaks the Poincar\'e group to time translations and transverse rotations $SO(3)$.  It is well known that we can think of the particle's spatial coordinates $X^i(\tau)$ as the Goldstone fields associated with such symmetry breaking pattern. Accordingly, the standard worldline action can be shown to be the most general action compatible with the nonlinear realization of the symmetry via the spacetime generalization of the Callan-Coleman-Wess-Zumino (CCWZ) construction \cite{Coleman:1969sm,Callan:1969sn}.  A particle with spin further breaks the transverse $SO(3)$ rotations to $SO(2)$ in the rest frame, due to the spin orientation.  Here we show that the action~\eqref{eq_free_spin_app} can also be derived insisting on the nonlinear realization of the symmetry via the CCWZ construction.\footnote{\cite{Landry:2020obv} proposed a generalization of the CCWZ construction to describe particles with spin in the Grassmanian formalism alluded before. The approach of \cite{Landry:2020obv} however relies on the introduction of supsersymmetric gauge redundancies, while the construction in this section instead does not require any extra ingredients compared to the usual CCWZ procedure for broken spacetime symmetries.}

It is again instructive to derive first from this perspective the non-relativistic action~\eqref{eq_free_NRspin0}. To this aim we consider the symmetry breaking pattern $SO(3)\rightarrow SO(2)$. Assuming that the spin points in the direction ``$1$", we consider the coset
\begin{equation}\label{eq_app_Omega_R}
\Omega_R=e^{-i\phi_s(\tau) J_3}e^{-i\left(\theta_s(\tau)-\frac{\pi}{2}\right) J_2}\,,
\end{equation}
where we denoted with $J_2$ and $J_3$ the broken generators, while $J_1$ is the unbroken one.  To construct $SO(3)$ invariants, we are instructed to consider the Maurer-Cartan one form
\begin{equation}\label{eq_app_MC_R}
\Omega^{-1}_R\dot{\Omega}_R=iD\Theta^a_s J_a+i A J_1\,,
\end{equation}
where $a=2,3$ and
\begin{equation}
D\Theta^2_s=-\dot{\theta}_s\,,\qquad D\Theta_s^3=-\sin\theta_s \dot{\phi}_s\,,\qquad
A=-\cos\theta_s \dot{\phi}_s\,.
\end{equation}
The $SO(3)$ group acts linearly via local $SO(2)$ rotations on the vector $D\Theta^a_s$, while $A$ transforms as a $U(1)$ connection, which may be used to define a covariant derivatives acting on the $D\Theta^a_s$ and matter fields.

We now have two choices for the kinetic term of the angular Goldstones compatibly with the $SU(2)$ symmetry. The simplest one is to consider a quadratic kinetic term
\begin{equation}\label{eq_app_rigid_rotor}
S=\frac{I}{2}\int d\tau D\Theta^a_s D\Theta^a_s=\frac{I}{2}\int d\tau\,\dot{\hat{n}}^i\dot{\hat{n}}^i\,,
\end{equation}
where $\hat{n}^i=S^i/s$ is a unit vector.
This choice gives the standard rigid rotor and, at the quantum level, describes an infinite tower of states arranged in all the representations of $SO(3)$. Eq.~\eqref{eq_S_free_NR} corresponds instead to a Chern-Simons like term,
\begin{equation}\label{eq_app_AspinNR}
S=-s\int d\tau A\,,
\end{equation}
which is first order in derivatives and it is allowed since $A$ shifts by a total derivative under $SO(3)$ transformations.  

From the viewpoint of symmetries, the choice between eq.~\eqref{eq_app_rigid_rotor} or~\eqref{eq_app_AspinNR} is dictated by the transformation properties of the angles under time reversal $T$ (assuming it is a symmetry). Indeed eq.~\eqref{eq_app_AspinNR} is allowed only if the action of $T$ is accompanied by $\{\theta_s,\,\phi_s\}\stackrel{T}{\rightarrow}\{\pi-\theta_s,\,\phi_s+\pi\}$ (up to $SO(3)$ rotations), that physically corresponds to the spin being a pseudovector.\footnote{Note that, within EFT, in the simplest option,  given the action~\eqref{eq_app_AspinNR}, terms with two or more derivatives, such as~\eqref{eq_app_rigid_rotor}, are treated perturbatively and do not yield new states when included.} 

Notice also that~\eqref{eq_app_AspinNR}, due to the transformation properties of $A$, is the only allowed term that cannot be written as a local integral in terms of the unit vector $\hat{n}^i$. To write $A$ in a convenient form, it is useful to evaluate $\Omega$ in a spin $j$ representation. Working in a basis that diagonalizes $J_1$ we have, in obvious notation,
\begin{equation}\label{eq_app_OmegaR_mat}
i A=\frac{1}{m}\langle j,m|\Omega^{-1}_R\dot{\Omega}_R|j,m\rangle\,.
\end{equation}
The spinor representation in eq. \eqref{eq_free_NRspin0} is the minimal choice, corresponding to $j=m=1/2$, so that the spinor~\eqref{eq_Bloch_par} is obtained from
\begin{equation}
z=\sqrt{2}\,\Omega |1/2,1/2\rangle=e^{-i\phi_s \frac{\sigma^3}{2}} e^{-i\left(\theta_s-\frac{\pi}{2}\right) \frac{\sigma^2}{2}}
\cdot\left(\begin{array}{c}
1\\
1
\end{array}\right)
\,.
\end{equation}

The generalization to the relativistic case is conceptually straightforward.  We refer the reader to \cite{ogievetsky1974nonlinear,Delacretaz:2014oxa} for an introduction to the CCWZ technology for nonlinearly realized spacetime symmetries and its application to the spinless relativistic point-particle. Here we focus directly on a particle with spin. The associated symmetry breaking pattern is
\begin{align}
\begin{split}
\text{unbroken}&=
\begin{cases}
P_0 
&\text{time translations}\,,\\
J_1
&\text{spatial rotation}\,,
\end{cases}
\\ 
\text{broken}&=
\begin{cases}
P_i &\text{spatial translations}\,, \\
K_{i}
& \text{boosts}\,, \\
J_a
& \text{spatial rotations}\,,\\
\end{cases}
\end{split}
\end{align}
where $i=1,2,3$ and $a=2,3$ as before. The coset can be parametrized as
\begin{equation}
\Omega=e^{iX^\mu(\tau)P_\mu}e^{i\eta^i(\tau) K_i}\Omega_R\,,
\end{equation}
where $\Omega_R$ was defined in eq.~\eqref{eq_app_Omega_R}. The Maurer-Cartan one-form is given by
\begin{equation}
\Omega^{-1}\dot{\Omega}=i E\left(P_0+\nabla\pi^i P_i+\nabla\eta^i K_i
+\nabla\Theta_s^aJ_a+A_P J_1\right)\,.
\end{equation}
Here $E$ transforms as a worldline vielbein,  while the coefficients of the broken generators can be thought as the covariant derivatives of the Goldstones and, under the action of any element of the full group, they transform as a linear (local) representation of the unbroken subgroup $SO(2)$. Finally $A_P$ transforms as $U(1)$ connection similarly to $A$ in eq.~\eqref{eq_app_MC_R}. We can write the covariant derivatives and the connection explicitly in terms of the following matrices
\begin{equation}
(e^{-i\eta^i K_i})^\mu_{\;\nu}=(\Lambda^{-1})^\mu_{\;\nu}=\Lambda_\nu^{\;\mu}\,,\qquad (\Omega_R)^\mu_{\;\nu}=R_\nu^{\;\mu}\,,\qquad
\bar{\Lambda}_\nu^{\;\mu}=
R_\nu^{\;\rho}\Lambda_\rho^{\;\mu}\,.
\end{equation}
We find
\begin{equation}
\begin{gathered}
E=\dot{X}^\nu\Lambda_\nu^{\;0}\,,\quad
\nabla\pi^i=E^{-1}\dot{X}^\nu\bar{\Lambda}_\nu^{\;i}\,,\quad
\nabla\eta^i=-E^{-1}\bar{\Lambda}_\nu^{\;0}\dot{\bar{\Lambda}}^{\nu i}\,,\\
\nabla\Theta^a_s=-\frac{1}{2}\varepsilon_{aij}
E^{-1}\bar{\Lambda}_\nu^{\;i}\dot{\bar{\Lambda}}^{\nu j}\,,\qquad
A_P=-E^{-1}\bar{\Lambda}_\nu^{\;2}\dot{\bar{\Lambda}}^{\nu 3}\,.
\end{gathered}
\end{equation}
As in the spinless case, the boost Goldstones $\eta^i$ are redundant \cite{Low:2001bw}. This means that we can eliminate them imposing an \emph{Inverse Higgs Constraint} \cite{IvanovIHC}, namely a covariant condition which allows to express the redundant Goldstones in terms of other Goldstone fields. For the case at hand the relevant condition is 
\begin{equation}\label{eq_app_IHC}
\nabla\pi^i=0\quad\implies\quad
 \quad
\frac{\eta^i}{\eta}\tanh{\eta}=\frac{\dot{X}^i}{\dot{X}^0}\,.
\end{equation}
As explained in \cite{Delacretaz:2014oxa}, eq.~\eqref{eq_app_IHC} implies $\Lambda_\mu^{\;0}=u_\mu$, so that the vectors $\{\Lambda^\mu_{\;i}\}$ form an orthonormal basis for the normal vectors on the worldline. 

We may finally write the action for the point-particle. To lowest order in derivatives the kinetic term for its coordinates is simply given by the vielbein
\begin{equation}
S=-m\int d\tau E=-m\int d\tau \sqrt{\dot{X}^2}\,,
\end{equation}
where we used eq.~\eqref{eq_app_IHC}. It is possible to check that, upon using eq.~\eqref{eq_app_IHC}, the $\nabla\eta^i$ become proportional to the particle's acceleration and therefore are of higher order in derivative expansion. As in the non-relativistic case, we have two choices for the leading order action for the rotational Goldstones $(\theta_s,\phi_s)$ analogously to the non-relativistic case. If we assume that they do not transform under $T$, we consider a rigid-rotor like kinetic term:
\begin{equation}\label{eq_app_rigid_rotor_rel}
\frac{I}{2}\int d\tau E\, \nabla\Theta^a_s \nabla\Theta^a_s\,.
\end{equation}
Eq.~\eqref{eq_app_rigid_rotor_rel} has physical applications in the description of astrophysical objects \cite{Delacretaz:2014oxa}. Unsurprisingly, the spin action~\eqref{eq_free_spin_app} is given instead by 
\begin{equation}\label{eq_app_AspinRel}
S_{spin}=-s\int d\tau E\, A_P\,,
\end{equation}
which as eq.~\eqref{eq_app_AspinNR} is compatible with time reversal only if the spin flips sign under $T$.

Finally, we remark that~\eqref{eq_app_AspinRel} is the only allowed term that cannot be written as a local integral in terms of the vector $\hat{n}^\mu=S^\mu/s$, which satisfies $\hat{n}^\mu\hat{n}_{\mu}=-1$ and $\hat{n}^\mu u_\mu=0$. For instance, we have
\begin{equation}
E \nabla \eta^1=-\dot{u}^\mu\hat{n}_\mu\,,\quad
E^2\nabla\eta^a\nabla\eta^a=\dot{u}^\mu \hat{S}_{\mu\nu}\hat{S}^{\nu}_{\;\rho}\dot{u}^\rho\,,\quad
E^2\nabla\Theta^a_s\nabla\Theta_s^a=\dot{\hat{S}}^{\mu\nu}\dot{\hat{S}}_{\mu\nu}+
\dot{u}^\mu \hat{S}_{\mu\nu}\hat{S}^{\nu}_{\;\rho}\dot{u}^\rho\,,
\end{equation}
where $\hat{S}_{\rho\sigma}=S_{\rho\sigma}/s=\varepsilon_{\mu\nu\rho\sigma}\hat{n}^\mu u^\nu$. To write the connection $A_P$ we instead need to evaluate the coset matrix in a specific representation, as in eq.~\eqref{eq_app_OmegaR_mat}. The Dirac spinor in eq.~\eqref{eq_free_spin_app} is the minimal choice, that allows writing $E\,A_P$ without picking additional terms (e.g. a Weyl spinor always picks up a term proportional to $E\,\nabla\eta^1$).

\section{Worldsheet axion energy levels from the Mathieu equation}\label{app_Mathieu}

\hspace{5mm}
Eq.~\eqref{eq_a_m_EOM} can be written as a Mathieu equation in the form
\begin{equation}\label{eq_Mathieu}
a''+\left[R-2Q\cos(2s)\right]a=0\,,\quad
R=\epsilon ^2-\frac{m^2_a}{2 \omega ^2}\,,\quad
Q=\frac{m^2_a}{4 \omega ^2}\,.
\end{equation}
The most general solution is a linear combination of the two Mathieu functions. For fixed $Q$, only a series of discrete values of $R$ labeled by $k\in\mathds{N}$ yields solutions satisfying Dirichlet or Neumann boundary conditions. These are
\begin{equation}
\text{Dirichlet}\quad\implies\quad
R= D_k(Q)\equiv \begin{cases}
B_k(Q)  & k=\text{even}>0\\
A_k(Q) & k=\text{odd}>0 \,,
\end{cases}
\end{equation}
\begin{equation}
\text{Neumann}\quad\implies\quad
R= N_k(Q)\equiv\begin{cases}
A_k(Q)  & n=\text{even}\geq 0\\
B_k(Q) & n=\text{odd}>0 \,,
\end{cases}
\end{equation}
where $A_k$ and $B_k$ are the so called Mathieu characteristics (\texttt{MathieuCharacteristicA} and \texttt{MathieuCharacteristicB} in \texttt{Mathematica}), and we defined $D_k$ and $N_k$ for future reference.  (Notice that $k=0$ is allowed only for Neumann). For $Q=0$ we have $A_k(0)=B_k(0)=k^2$, and we recover thus the massless solution discussed in Section~\ref{subsec_toy_bc}.

As explained in Section~\ref{subsec_pseudoaxion}, we are interested in the large $Q\sim m^2_a/\omega^2$ limit. In this limit the Mathieu Characteristics admit the following expansion \cite{NIST:DLMF}
\begin{equation}
A_k(Q)\sim B_{k+1}(Q)=-2 Q+2(2k+1)\sqrt{Q}+O(Q^0)\,.
\end{equation}
Additionally, the difference $B_{k+1}(Q)-A_k(Q)\sim e^{-4\sqrt{Q}}$ is nonperturbative in $Q$.  This is because the Mathieu equation has the structure of a Schr\"odinger equation in a double-well potential, as explained below eq.~\eqref{eq_a_m_EOM_rewritten}.

In conclusion, depending on the boundary conditions, the single-particle energy levels are given by 
\begin{equation}\label{eq_e_pseudoaxion_app_Dir}
\epsilon_n^2=\frac{m^2_a}{2\omega^2}+D_{n+1}\left(\frac{m^2_a}{4\omega^2}\right)\simeq \begin{cases}
(2n+3)\frac{m_a}{\omega} & n\text{ even}\\
(2n+1)\frac{m_a}{\omega} & n\text{ odd}\,,
\end{cases}
\end{equation}
for Dirichlet, and by
\begin{equation}\label{eq_e_pseudoaxion_app_Neu}
\epsilon_n^2=\frac{m^2_a}{2\omega^2}+N_{n}\left(\frac{m^2_a}{4\omega^2}\right)\simeq \begin{cases}
(2n+1)\frac{m_a}{\omega} & n\text{ even}\\
(2n-1)\frac{m_a}{\omega} & n\text{ odd}\,,
\end{cases}
\end{equation}
for Neumann, where $n\in \mathds{N}$ and we ordered such that $\epsilon_n<\epsilon_{n+1}$.
Eq.~\eqref{eq_e_pseudoaxion_app_Dir} and~\eqref{eq_e_pseudoaxion_app_Neu} imply a double-degeneracy to all orders in $1/Q\sim \omega/m_a$.  Notice that this result coincides with eq.~\eqref{eq_e_pseudoaxion} to leading order in $\omega/m_a$. Under $\sigma\rightarrow-\sigma$, the wave-functions are even (\texttt{MathieuC} in \texttt{Mathematica}) for $n=$even, and odd (\texttt{MathieuS} in \texttt{Mathematica}) for $n=$odd.

\section{Daughter Regge trajectories for heavy quarks}\label{app_heavy_excitations}

\hspace{5mm}
In this appendix we briefly analyze the spectrum of fluctuations for heavy quarks. We limit ourselves to the leading order in derivatives, therefore neglecting the contribution proportional to $\bar{b}_1=0$ in eq.~\eqref{eq_heavy_Sbdry}. The analysis is largely analogous to that of Section~\ref{sec_bdry_action}. The main difference is that the boundary mode $\alpha$ cannot be thought of as an auxiliary field anymore. This is because the expansion of the boundary action~\eqref{eq_heavy_Sbdry} results in the following contribution to the quadratic action
\begin{equation}\label{eq_heavy_Sbdry2}
\begin{split}
\sum_{\sigma=\pm 1}
\frac{1/(\alpha_{f,0}\omega^2)}{2 \ell_s^2 \sqrt{1-\alpha_{f,0}^2} }\int d\tau
\left[\alpha_{f,0}^2 \dot\phi^2+
\left(1-\alpha_{f,0}^2\right) \dot{\alpha}^2+\alpha^2\pm 2 \alpha_{f,0} \left(2-\alpha_{f,0}^2\right) \alpha \dot{\phi}+(1-\alpha_{f,0}^2 )\dot{\rho}_i^2\right]\,,
\end{split}
\end{equation}
where we wrote $\tilde{m}$ in terms of $\alpha_{f,0}$ to make manifest that this contribution is of the same order as the NG action. Therefore we cannot integrate out $\alpha$ as for light quarks. Note that $\alpha$ mixes nontrivially with the bulk field $\phi$, and therefore it does not correspond to a special boundary mode at generic values of $\alpha_{f,0}$

The spectrum of fluctuations is then extracted from the solutions of the EOMs that follow from the action~\eqref{eq_heavy_Sbdry2}$+$\eqref{eq_S2}. The analysis is straightforward and we only report the results. We refer the reader to \cite{Sonnenschein:2018aqf} for details.

For the transverse modes $\rho_i$, in units of $\omega$ the frequencies are the solutions of the following equation
\begin{equation}\label{eq_heavy_rho_eps_condition}
2 \epsilon\,\alpha_{f,0} \sqrt{1-\alpha_{f,0}^2}   \cos \left(2 \epsilon  \arcsin(\alpha_{f,0})\right)+ \left[\alpha_{f,0}^2+
 \epsilon^2\left(1-\alpha_{f,0}^2\right)\right] \sin \left(2 \epsilon  \arcsin(\alpha_{f,0})\right)=0\,.
\end{equation}
Eq.~\eqref{eq_heavy_rho_eps_condition} is solved by $\epsilon =0,1$ for all values of $\alpha_{f,0}$, as expected due to the protected nature of these modes. In the relativistic regime $\alpha_{f,0}\simeq 1$ the other modes have frequencies given by
\begin{equation}
    \epsilon_{n}=n+
    \frac{2 \left(m\omega \ell_s^2\right)^{3/2} }{3 \pi } n \left(n^2-1\right)+\ldots\qquad
    n=2,3,\ldots\,,
\end{equation}
in agreement with eq.~\eqref{eq_rho_spectrum_NNLO}. In the NR regime $\alpha_{f,0}\ll 1$ there is a parametric separation between the first two protected modes $\epsilon=0,1$ and the other string excitations, whose gap scales as $1/\alpha_{f,0}\sim m\omega \ell_s^2\gg 1$ and is given by
\begin{equation}\label{eq_heavy_rho_spectrum}
\begin{split}
    \epsilon_{n}&\simeq\frac{\pi (n-1)}{2\alpha_{f,0}}-
    \alpha_{f,0}\frac{\pi ^2 (n-1)^2/4-6}{3 \pi  (n-1)}+\ldots \\
    &=\frac{\pi   (n-1) m\omega }{2\ell_s^{-2}}+
    \frac{\left[5 \pi ^2 (n-1)^2/4+6\right]\ell_s^{-2}}{3 \pi   (n-1) m\omega }+\ldots
    \qquad
    n=2,3,\ldots\,.
    \end{split}
\end{equation}
The leading order term in eq.~\eqref{eq_heavy_rho_spectrum} coincides with the answer for the energy levels on a static string of length $2\alpha_{f,0}/\omega$ and Dirichlet boundary conditions.

The eigenfrequencies of the $\alpha/\phi$ system are the solutions of the following equation:
\begin{multline}\label{eq_heavy_phi_eps_condition}
\left[\left(1+\alpha_{f,0}^2\right)^2-2 \left(1
+2 \alpha_{f,0}^2-3 \alpha_{f,0}^4\right) \epsilon ^2+
\left(1-\alpha_{f,0}^2\right)^2 \epsilon ^4
\right] \sin \left(2 \epsilon  \arcsin(\alpha_{f,0})\right)\\
+4\epsilon\, \alpha_{f,0} \sqrt{1-\alpha_{f,0}^2}   \left[1+\alpha_{f,0}^2-\left(1-\alpha_{f,0}^2\right) \epsilon ^2\right] \cos \left(2 \epsilon  \arcsin(\alpha_{f,0})\right)=0\,.
\end{multline}
As before, the lowest modes are protected and have $\epsilon=0,1$, that solve eq.~\eqref{eq_heavy_phi_eps_condition} for any value of $\alpha_{f,0}$. In the relativistic regime the other modes have frequency given by
\begin{equation}
    \epsilon_{n}=n
    -\frac{1}{3 \pi } \left(m\omega\ell_s^2\right)^{3/2}n \left(n^2-1\right)+\ldots\qquad
    n=2,3,\ldots\,,
\end{equation}
in agreement with the result for light quarks eq.~\eqref{eq_phi_spectrum_NNLO}. In the NR regime the first unprotected mode has frequency given by
\begin{equation}\label{eq_heavy_eps2}
  \epsilon_2=\sqrt{3}  +\frac{\alpha_{f,0}^2}{3 \sqrt{3}}+\ldots
 =\sqrt{3} +\frac{1}{3\sqrt{3}m^2\omega^2\ell_s^4}+\ldots\,.
\end{equation}
The other modes are parametrically heavier and, similarly to eq.~\eqref{eq_heavy_rho_spectrum}, their gap at leading order is as for a static string of length $2\alpha_{f,0}/\omega$ and Dirichlet boundary conditions:
\begin{equation}
\begin{split}
 \epsilon_n& =   \frac{\pi  (n-2)}{2 \alpha_{f,0}}-\frac{\alpha_{f,0} \left[\pi ^2 (n-2)^2-48\right]}{12 \pi  (n-2)} +\ldots\\
 &\simeq \frac{\pi   (n-2)m \omega }{2 \ell_s^{-2}}+
 \frac{\left[5 \pi ^2 (n-2)^2+48\right] \ell_s^{-2}}{12 \pi  (n-2) m\omega }+\ldots\qquad
 n=3,4,\ldots\,.
 \end{split}
\end{equation}
Note that the $n$th mode of $\rho^i$ and the $(n+1)$th $\phi/\alpha$ mode have the same energy at leading order in the NR limit.

A similar analysis can be performed for the axion. In the relativistic regime $m\omega\ell_s^2\ll 1$ we recover the results of Section~\ref{subsec_pseudoaxion}, while for a NR rotating string its energy levels approximately coincide with those of a massive particle on a static string:
\begin{equation}
    \epsilon_n\simeq\begin{cases}
\sqrt{\frac{m_a^2}{\omega^2}+\frac{\pi ^2 (n+1)^2 }{4 \alpha_{f,0}^2}}=\sqrt{\frac{m_a^2}{\omega^2}+\frac{\pi ^2  }{4 }(n+1)^2 m^2\omega ^2\ell_s^4} &\text{for Dirichlet}\\[0.6em]
\sqrt{\frac{m_a^2}{\omega^2}+\frac{\pi ^2 n^2 }{4 \alpha_{f,0}^2}}=\sqrt{\frac{m_a^2}{\omega^2}+\frac{\pi ^2  }{4 }n^2 m^2\omega ^2\ell_s^4} &\text{for Neumann}
    \end{cases}
    \qquad
    n=0,1,\ldots\,,
\end{equation}
where we reported the result both for Dirichlet and Neumann boundary conditions according to the discussion in Section~\ref{subsec_pseudoaxion}.

In summary, in the relativistic limit, we recover results analogous to those for light quarks, while in the NR regime the spectrum coincides with that of a static string with Dirichlet boundary conditions, but for the protected modes and the excitation~\eqref{eq_heavy_eps2}, that creates the only parametrically light states. As explained in the main text in Section~\ref{subsec_NR_limit}, these light states admit a simple interpretation in terms of a quark model with linear potential.

\section{Tables of mesons}\label{meson_table}

\hspace{5mm}
This appendix reports the data that we used for mesons and lists the potential $stringy$ interpretation of each particle purely based on its quantum numbers.  Fig. \ref{fig:mesons_all} summarizes our predictions for the lightest excited states on top of the leading Regge trajectory, and tables \ref{tableI=1} to \ref{tableI=1/2} list and compare the mesons from PDG with these predictions.

At each spin $J$, we take the lightest $\rho/a$ meson for $I=1$ and the lightest $\omega/f$ meson for $I=0$ to be the string gorund-state. Fig.~\ref{fig:leading_Regge} and fig.~\ref{fig:mesons_all} show that the parity assignment of these states follow the same behavior expressed in eq.~\eqref{general_parity}. As discussed in the first part of this paper, we expect to have two string excitations at the second energy level on top of each state on the Regge trajectory. Looking at PDG, we see candidates for this prediction with parity assignments that match with table \ref{table_PC}. Due to the subleading order corrections, these two degenerate excited states would have an energy splitting that results in the angular fluctuation to be lighter than the transverse one. The first plot in fig. \ref{fig:candidates} represents a subset of particles with $I=1$ that approximately follow this mass hierarchy. 

Based on the discussion on parity assignments for spin states explained in Section~\eqref{subsec_PC}, for each string ground-state, we might have up to three associated spin states in the vicinity of leading Regge trajectory.  The second plot in fig. \ref{fig:candidates} shows the $I=1$ PDG candidates for these spin states. Following the parity assignment for pseudoscalar in table \ref{table_PC}, we also expect to see up to three axionic excitations among the particles around the first two daugther Regge trajectories; we included three because the gap of the axion is not determined within EFT, and it is a priori possible that the resulting modes are parametrically light. In fig. \ref{fig:candidates}, the third plot shows the PDG $I=1$ candidates for these axionic excitations.

\begin{figure}[t!]
    \centering
    \includegraphics[width=0.8\textwidth]{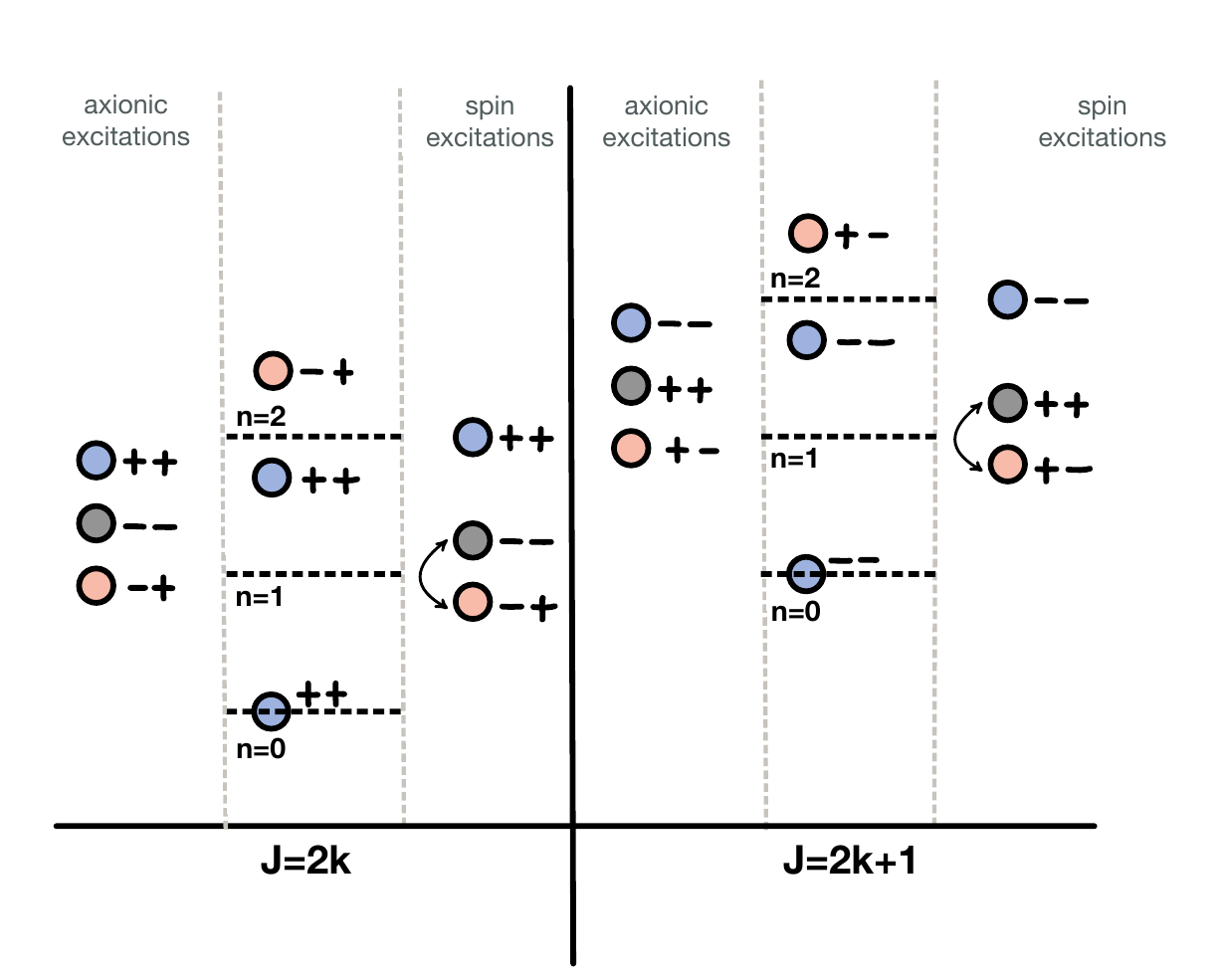}
    \caption{\small \textbf{Qualitative presentation of the predictions for the lightest daughter Regge trajectories:}  For each spin $J$, the states are classified into three categories, separated by gray dashed lines; we know the energy hierarchy in each category, but we don't know the hierarchy between different categories. The first column shows the axionic excitations ($n=0$, $n=1$, and two $n=0$). The second one shows the string excitations around second daughter Regge trajectory. The last column shows the spin excitations. The double-side arrows are included because the hierarchy between those two states is not determined from EFT. }
    \label{fig:predictions}
\end{figure}

\begin{figure}[ht!]
  \centering
  \subfloat[PDG candidates for string excitations]{\includegraphics[width=0.43\textwidth]{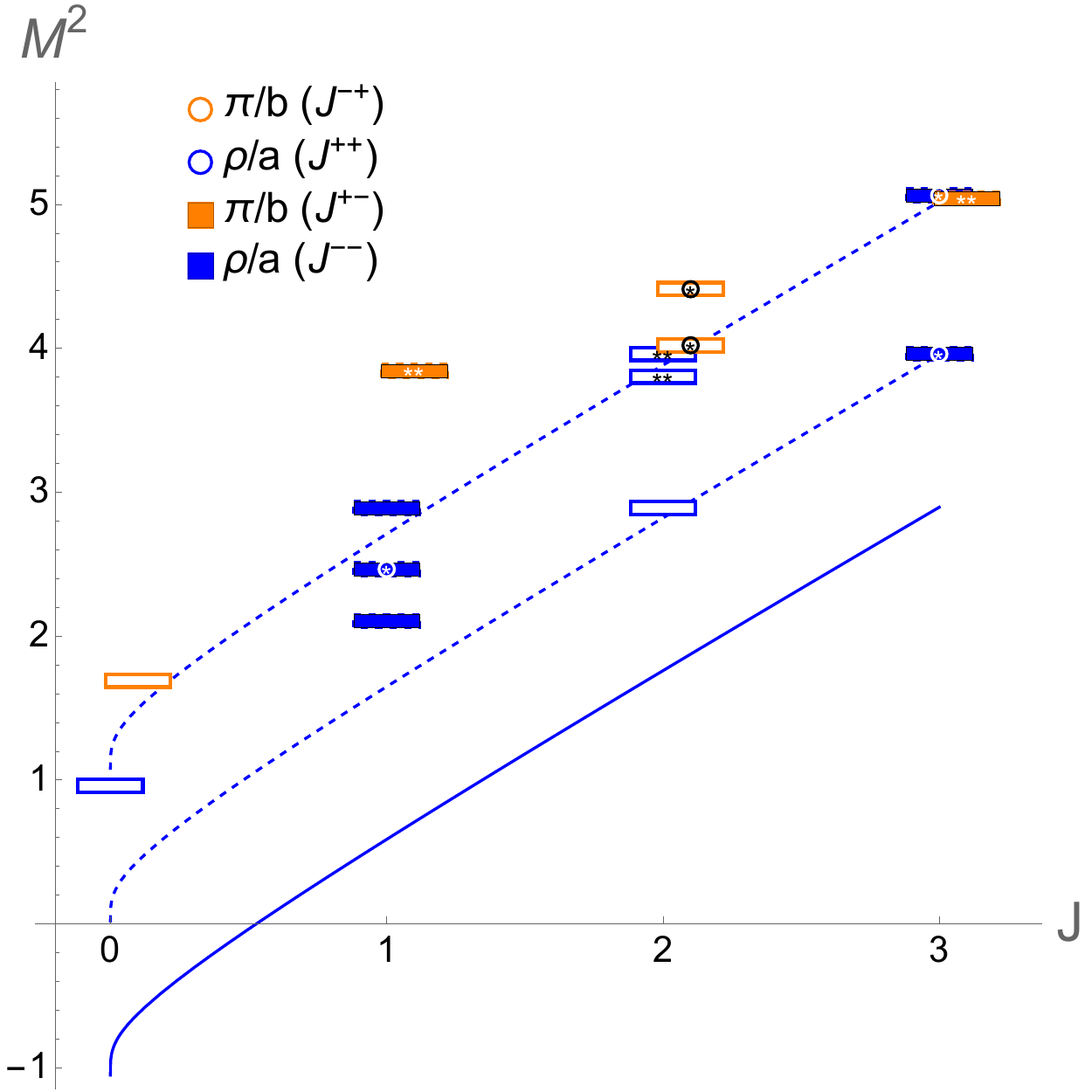}}\\
   \subfloat[PDG candidates for spin states]{\includegraphics[width=0.43\textwidth]{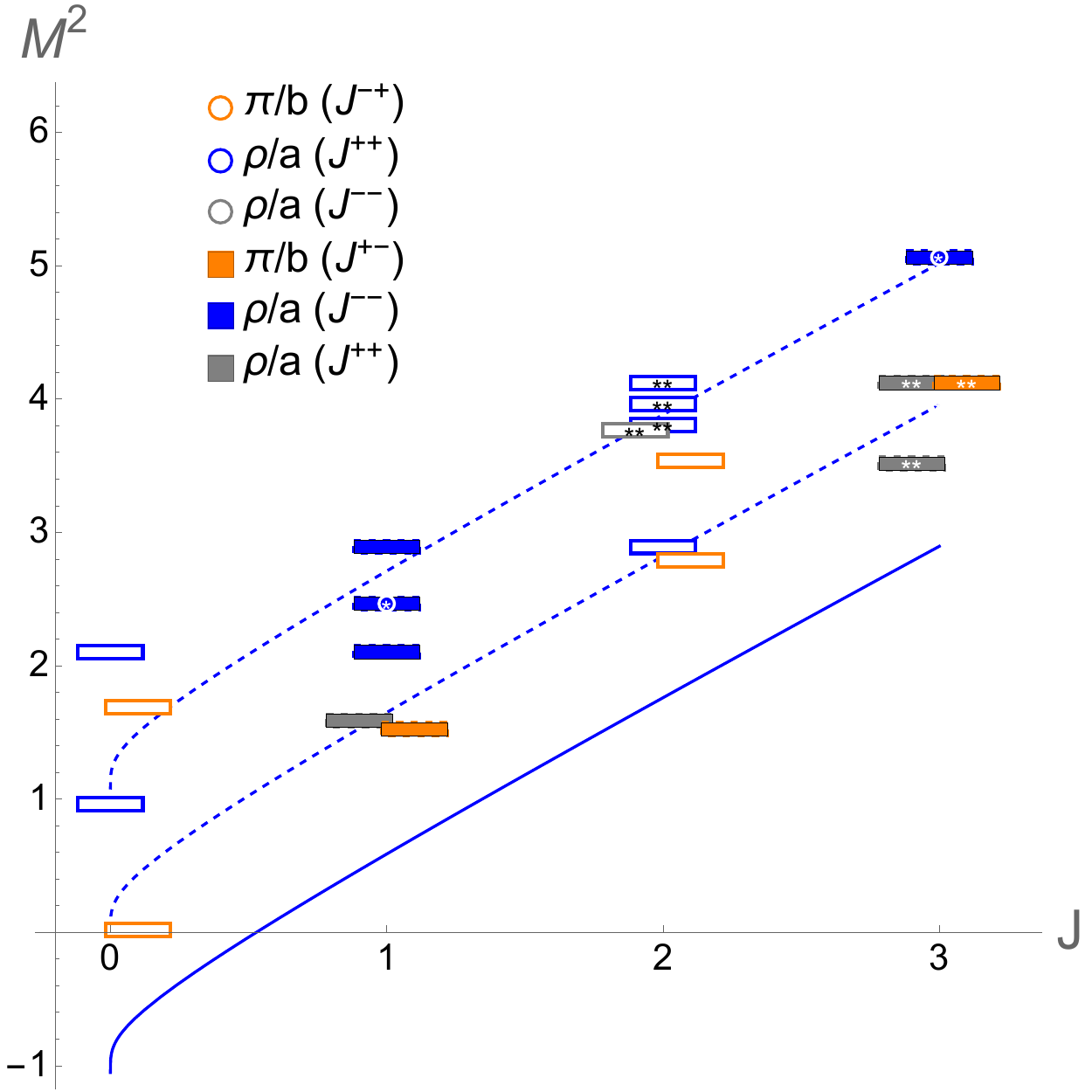}}\hfill
    \subfloat[PDG candidates for axionic excitations]{\includegraphics[width=0.43\textwidth]{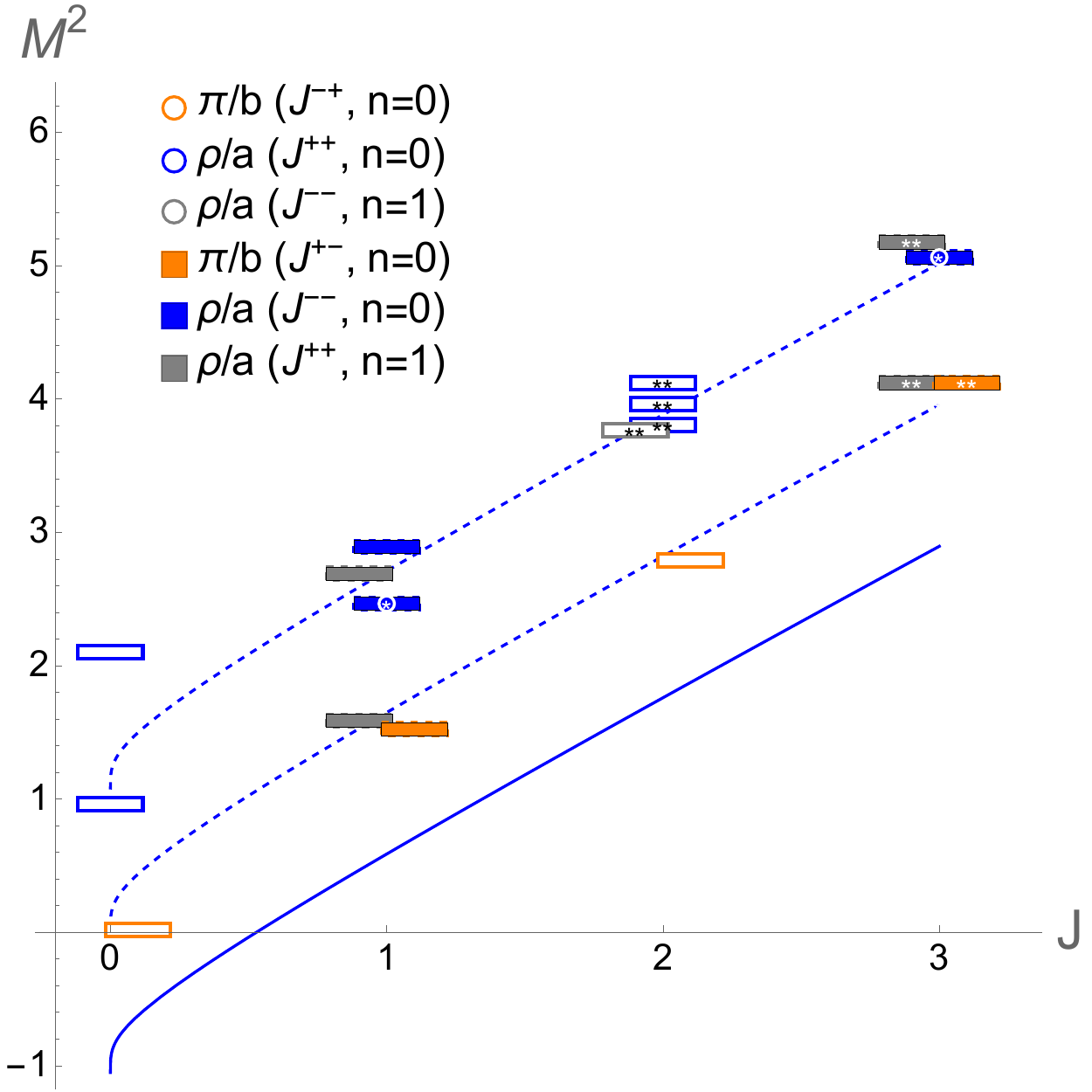}}
  \caption{\small \textbf{I=1 PDG candidates for our predictions:} The solid line shows the leading Regge trajectory, which is associated with $\rho/a$ meson. The dashed lines show the first two daughter Regge trajectories: $M^2_n(J)=M^2(J)_{leading}+n/\ellb_s^2$ with $n=1,2$. In (a),  the blue markers represent the candidates for the angular excitation, while particles in orange are candidates for the transverse excitation.  In (b), the orange and gray markers stand for candidates for the first two spin states in plot~\ref{fig:predictions}. The last, expectedly heavier, spin state with the same quantum numbers as the ground-state,  is shown in blue. In (c), the candidates for the lightest axionic excitation, associated with level $n=0$, are shown in orange. The particles in blue are candidates for double-axion excitation with $n=0$. An axionic excitation with $n=1$, heavier than the first $n=0$ excitation, could appear either above or below the second $n=0$ excitation in gray.}
  \label{fig:candidates}
\end{figure}

\begin{figure}[ht!]
  \centering
  \subfloat[PDG candidates for axionic excitations of spin states]{\includegraphics[width=0.43\textwidth]{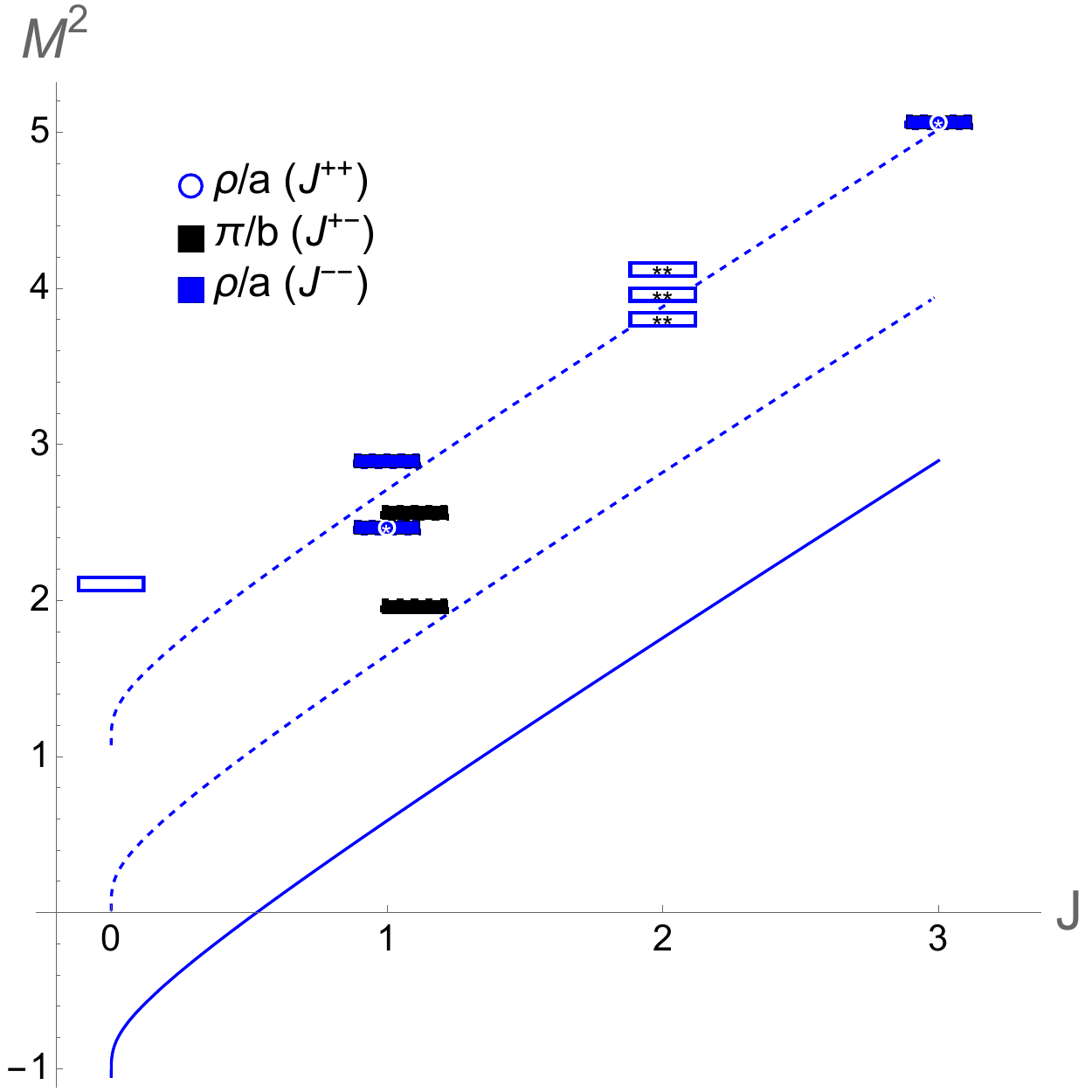}}\hfill
   \subfloat[PDG candidates for angular string fluctuations of spin states]{\includegraphics[width=0.43\textwidth]{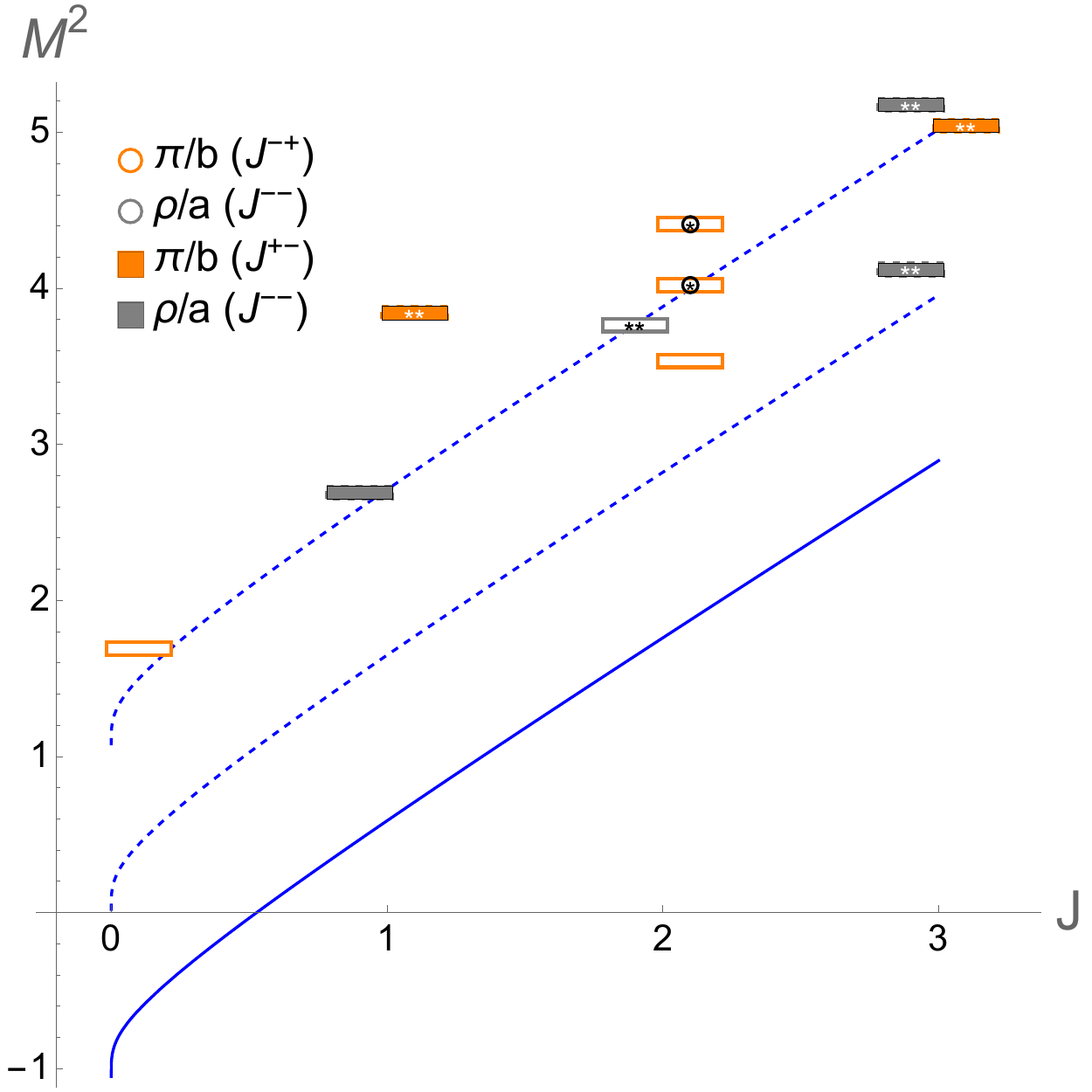}}
  \caption{\small \textbf{I=1 PDG candidates for mixed excitations:} In (a), we represent the candidates for $n=0$ and $n=1$ axionic excitations of spin states, which could be light enough to be observable in the range of our study. Noting that the angular string excitation is lighter than transverse string excitation, we only show the candidates for angular fluctuations of spin states in (b).}
  \label{fig:mix}
\end{figure}

Other than the single excitations and the double-axion of first-level (presented in fig.~\ref{fig:predictions}), there are some multiple excitations that could be detectable in the range of our study. Potential PDG candidates for such states are marked on the last two columns of tables \ref{tableI=1} to \ref{tableI=1/2}. The $I=1$ candidates for such excitations are demonstrated in fig.~\ref{fig:mix}. They all have overlaps with the particles shown in fig.~\ref{fig:candidates} except for $\pi_1(1400)$ and $\pi_1(1600)$\footnote{In fig. \ref{fig:mix}, they are represented in black.}, known as isovector exotic mesons. As mentioned in the main text, in our framework, their quantum numbers could only match with an axionic excitation of a spin state.

\definecolor{dark_blue}{rgb}{0,0.2,1}
\definecolor{Blue}{rgb}{0,0.6,1}

\begin{longtable}[ht!]{ |P{0.5cm}|P{1.7cm}|P{1cm}|P{0.5cm}|P{0.5cm}|P{0.5cm}|P{0.5cm}|P{0.5cm}|P{0.5cm}|P{0.8cm}|}
\caption{\small \textbf{Flavourless light mesons with isospin one, $I=1$:} Numbers in parenthesis are particle masses in MeV. In the first column, ``$?$" means that the particle is on Meson Summary Table but has yet to be established; the particles that PDG refers to as Further States are marked by ``$??$". The particles are first ordered by spin and then, by mass. The spin and discrete quantum numbers are given in the third column. In the remaining columns, 
we put a check symbol if the particle has the right quantum numbers to be interpreted as respectively a string ground-state, a transverse string excitation, an angular string excitation, a spin state, an axionic excitation of the respective ground-state, an axionic excitation of a spin state, or the angular fluctuation of a spin state. For the axionic states, the level of axion excitation $n$ is also determined. The symbol ``$*$" stands double-axion excitations. The color codes are the same as those used in our plots.}\\
 \hline
 \centering  & & \centering $J^{PC}$ & GS & $\rho$ & $\phi$ & $S$ & $n_a$ & $S,a$ & $S,\phi$ \\
 \hline
 \hline
\endfirsthead
 \hline
 \centering  & & \centering $J^{PC}$ & GS & $\rho$ & $\phi$ & $S$ & $n_a$ & $S,a$ & $S,\phi$ \\
 \hline
 \hline
\endhead
 \centering
 & \colorbox{orange!35}{$\pi^0(135)$}  & \centering $0^{-+}$ & & & & \checkmark & 0 & &\\
 \hline
  \centering
 & \colorbox{dark_blue!35}{$a_0(980)$}  & \centering $0^{++}$ & & & \checkmark & \checkmark & $0^*$ & &\\
 \hline
   \centering
 & \colorbox{orange!35}{$\pi(1300)$}  & \centering $0^{-+}$ & & \checkmark & & \checkmark & & & \checkmark \\
 \hline
   \centering
 & \colorbox{dark_blue!35}{$a_0(1450)$}  & \centering $0^{++}$ & & & & \checkmark & $0^*$ & \checkmark & \\
 \hline
 \hline
 \centering
 &\colorbox{dark_blue!35}{$\rho(770)$}  & \centering $1^{--}$ & \checkmark & & & & & & \\
 \hline
  \centering
 & \colorbox{orange!35}{$b_1(1235)$}  & \centering $1^{+-}$ & & & & \checkmark & $0$ & & \\
  \hline
        \centering
 & \colorbox{gray!35}{$a_1(1260)$}  & \centering $1^{++}$ & & & & \checkmark & $1$ & & \\
 \hline
        \centering
 & \colorbox{white!35}{$\pi_1(1400)$}  & \centering $1^{-+}$ & & & & & & \checkmark & \\
 \hline
        \centering
 & \colorbox{dark_blue!35}{$\rho_1(1450)$}  & \centering $1^{--}$ & & & \checkmark &\checkmark & & & \\
 \hline
        \centering
 ? & \colorbox{dark_blue!35}{$\rho_1(1570)$}  & \centering $1^{--}$ & & & \checkmark &\checkmark & $0^*$ & \checkmark & \\
 \hline
        \centering
 & \colorbox{white!35}{$\pi_1(1600)$}  & \centering $1^{-+}$ & & & & & & \checkmark & \\
 \hline
        \centering
 & \colorbox{gray!35}{$a_1(1640)$}  & \centering $1^{++}$ & & & & & $1$ & & \checkmark \\
 \hline
        \centering
 & \colorbox{dark_blue!35}{$\rho(1700)$}  & \centering $1^{--}$ & & & \checkmark & \checkmark & $0^*$ & \checkmark & \\
 \hline
          \centering
 ?? & \colorbox{orange!35}{$\pi_1(1960)$}  & \centering $1^{+-}$ & & \checkmark & & & & & \checkmark \\
 \hline
 \hline
        \centering
 & \colorbox{dark_blue!35}{$a_2(1320)$}  & \centering $2^{++}$ & \checkmark & &  & & & & \\
 \hline
        \centering
 & \colorbox{orange!35}{$\pi_2(1670)$}  & \centering $2^{-+}$ & & & & \checkmark & 0 & & \\
 \hline
        \centering
 & \colorbox{dark_blue!35}{$a_2(1700)$}  & \centering $2^{++}$ & & & \checkmark & \checkmark & & & \\
 \hline 
         \centering
 & \colorbox{orange!35}{$\pi_2(1880)$}  & \centering $2^{-+}$ & & & & \checkmark & & & \checkmark \\
 \hline
?? & \colorbox{gray!35}{$\rho_2(1940)$}  & \centering $2^{--}$ & & & & \checkmark & 1 & & \checkmark \\
  \hline
          \centering
 ?? & \colorbox{dark_blue!35}{$a_2(1950)$}  & \centering $2^{++}$ & & & \checkmark & \checkmark & $0^*$ & \checkmark & \\
 \hline
           \centering
 ?? & \colorbox{dark_blue!35}{$a_2(1990)$}  & \centering $2^{++}$ & & & \checkmark & \checkmark & $0^*$ & \checkmark & \\
 \hline
         \centering
 ? & \colorbox{orange!35}{$\pi_2(2005)$}  & \centering $2^{-+}$ & & \checkmark & & & & & \checkmark \\
 \hline
            \centering
 ?? & \colorbox{dark_blue!35}{$a_2(2030)$}  & \centering $2^{++}$ & & & & \checkmark & $0^*$ & \checkmark & \\
 \hline
          \centering
 ? & \colorbox{orange!35}{$\pi_2(2100)$}  & \centering $2^{-+}$ & & \checkmark & & & & & \checkmark \\
 \hline
 \hline
          \centering
 & \colorbox{dark_blue!35}{$\rho_3(1690)$}  & \centering $3^{--}$ & \checkmark & & & & & & \\
 \hline
            \centering
 ?? & \colorbox{gray!35}{$a_3(1875)$}  & \centering $3^{++}$ & & & & \checkmark & & & \\
 \hline 
           \centering
 ? & \colorbox{dark_blue!35}{$\rho_3(1990)$}  & \centering $3^{--}$ & & & \checkmark & & & & \\
 \hline
             \centering
 ?? & \colorbox{orange!35}{$b_3(2030)$}  & \centering $3^{+-}$ & & & & \checkmark & 0 & & \\
 \hline 
 \centering
  ?? & \colorbox{gray!35}{$a_3(2030)$}  & \centering $3^{++}$ & & & & \checkmark & 1 & & \checkmark \\
 \hline
   \centering
  ?? & \colorbox{orange!35}{$b_3(2245)$}  & \centering $3^{+-}$ & & \checkmark & & & & & \checkmark \\
 \hline
 \centering
  ? &\colorbox{dark_blue!35}{$\rho_3(2250)$} & \centering $3^{--}$ & & & \checkmark & \checkmark & $0^*$ & \checkmark & \\
 \hline  
 \centering
  ?? & \colorbox{gray!35}{$a_3(2275)$}  & \centering $3^{++}$ & & & & & 1 & & \checkmark \\
 \hline
 \hline
           \centering
 & \colorbox{dark_blue!35}{$a_4(1970)$}  & \centering $4^{++}$ & \checkmark & & & & & & \\
 \hline
  \hline
           \centering
 ? & \colorbox{dark_blue!35}{$\rho_5(2350)$}  & \centering $5^{--}$ & \checkmark & & & & & & \\
 \hline
  \hline
           \centering
 ?? & \colorbox{dark_blue!35}{$a_6(2450)$}  & \centering $6^{++}$ & \checkmark & & & & & & \\
 \hline
 \caption*{\ }
 \label{tableI=1}
\end{longtable}

\begin{longtable}[ht!]{ |P{0.5cm}|P{1.7cm}|P{1cm}|P{0.5cm}|P{0.5cm}|P{0.5cm}|P{0.5cm}|P{0.5cm}|P{0.5cm}|P{0.8cm}|}
\caption{\small \textbf{Flavourless light mesons with isospin zero, $I=0$}}\\
 \hline
 \centering  & & \centering $J^{PC}$ & GS & $\rho$ & $\phi$ & $S$ & $n_a$ & $S,a$ & $S,\phi$ \\
 \hline
 \hline
\endfirsthead
 \hline
 \centering  & & \centering $J^{PC}$ & GS & $\rho$ & $\phi$ & $S$ & $n_a$ & $S,a$ & $S,\phi$ \\
 \hline
 \hline
\endhead
 \centering
 & \colorbox{dark_blue!35}{$f_0(500)$}  & \centering $0^{++}$ & & & \checkmark & & & & \\
 \hline
 \centering
 & \colorbox{orange!35}{$\eta(548)$}  & \centering $0^{-+}$ & & & & \checkmark & 0 & & \\
 \hline
 \centering
 & \colorbox{orange!35}{$\eta(958)$}  & \centering $0^{-+}$ & & & & \checkmark & 0 & & \checkmark \\
 \hline
  \centering
 & \colorbox{dark_blue!35}{$f_0(980)$}  & \centering $0^{++}$ & & & \checkmark & \checkmark & $0^*$ & & \\
 \hline
   \centering
 & \colorbox{orange!35}{$\eta(1295)$}  & \centering $0^{-+}$ & & \checkmark & & \checkmark & & & \checkmark \\
 \hline
   \centering
 & \colorbox{dark_blue!35}{$f_0(1370)$}  & \centering $0^{++}$ & & & & \checkmark & $0^*$ & \checkmark & \\
 \hline
 \centering
 & \colorbox{orange!35}{$\eta(1405)$}  & \centering $0^{-+}$ & & \checkmark & & & & & \\
 \hline
  \centering
 & \colorbox{orange!35}{$\eta(1475)$}  & \centering $0^{-+}$ & & \checkmark & & & & & \\
 \hline
    \centering
 & \colorbox{dark_blue!35}{$f_0(1500)$}  & \centering $0^{++}$ & & & & \checkmark & $0^*$ & \checkmark & \\
 \hline
 \hline
    \centering
 & \colorbox{dark_blue!35}{$\omega(782)$}  & \centering $1^{--}$ & \checkmark & & & & & & \\
 \hline
     \centering
 & \colorbox{dark_blue!35}{$\phi(1020)$}  & \centering $1^{--}$ & & & \checkmark & & & & \\
 \hline
   \centering
 & \colorbox{orange!35}{$\eta(1170)$}  & \centering $1^{+-}$ & & & & \checkmark & 0 & & \\
 \hline
     \centering
 & \colorbox{gray!35}{$f_1(1285)$}  & \centering $1^{++}$ & & & & \checkmark & 1 & & \\
 \hline
    \centering
 & \colorbox{orange!35}{$h_1(1415)$}  & \centering $1^{+-}$ & & & & \checkmark & 0 & & \checkmark \\
 \hline
      \centering
 & \colorbox{gray!35}{$f_1(1420)$}  & \centering $1^{++}$ & & & & \checkmark & 1 & & \checkmark \\
 \hline
       \centering
 & \colorbox{dark_blue!35}{$\omega(1420)$}  & \centering $1^{--}$ & & & \checkmark & \checkmark & $0^*$ & & \\
 \hline
       \centering
 ? & \colorbox{gray!35}{$f_1(1510)$}  & \centering $1^{++}$ & & & & \checkmark & 1 & & \checkmark \\
 \hline
     \centering
 ? & \colorbox{orange!35}{$h_1(1595)$}  & \centering $1^{+-}$ & & & & \checkmark & & & \checkmark \\
 \hline
        \centering
 & \colorbox{dark_blue!35}{$\omega(1650)$}  & \centering $1^{--}$ & & & \checkmark & \checkmark & $0^*$ & \checkmark & \\
 \hline
       \centering
 & \colorbox{dark_blue!35}{$\phi(1680)$}  & \centering $1^{--}$ & & & & \checkmark & $0^*$ & \checkmark & \\
 \hline
         \centering
 ? & \colorbox{white!35}{$\eta_1(1855)$}  & \centering $1^{-+}$ & & & & & & \checkmark & \\
 \hline
 \hline
        \centering
 & \colorbox{dark_blue!35}{$f_2(1270)$}  & \centering $2^{++}$ & \checkmark & & & & & & \\
 \hline
         \centering
 ? & \colorbox{dark_blue!35}{$f_2(1430)$}  & \centering $2^{++}$ & & & \checkmark & & & & \\
 \hline
           \centering
 & \colorbox{dark_blue!35}{$f'_2(1525)$}  & \centering $2^{++}$ & & & \checkmark & & & & \\
 \hline
          \centering
 ? & \colorbox{dark_blue!35}{$f_2(1565)$}  & \centering $2^{++}$ & & & \checkmark & & & & \\
 \hline
        \centering
  ? & \colorbox{dark_blue!35}{$f_2(1640)$}  & \centering $2^{++}$ & & & \checkmark & & & & \\
  \hline
       \centering
 & \colorbox{orange!35}{$\eta_2(1645)$}  & \centering $2^{-+}$ & & & & \checkmark & 0 & & \\
 \hline
          \centering
  ?? & \colorbox{dark_blue!35}{$f_2(1750)$}  & \centering $2^{++}$ & & & \checkmark & \checkmark & & & \\
  \hline
          \centering
  ? & \colorbox{dark_blue!35}{$f_2(1810)$}  & \centering $2^{++}$ & & & \checkmark & \checkmark & & & \\
  \hline
         \centering
 & \colorbox{orange!35}{$\eta_2(1870)$}  & \centering $2^{-+}$ & & & & \checkmark & & & \checkmark \\
 \hline
          \centering
  ? & \colorbox{dark_blue!35}{$f_2(1910)$}  & \centering $2^{++}$ & & & \checkmark & \checkmark & $0^*$ & & \\
  \hline
          \centering
  & \colorbox{dark_blue!35}{$f_2(1950)$}  & \centering $2^{++}$ & & & \checkmark & \checkmark & $0^*$ & \checkmark & \\
  \hline
              \centering
  ?? & \colorbox{gray!35}{$\omega_2(1975)$}  & \centering $2^{--}$ & & & & \checkmark & 1 & & \checkmark \\
  \hline
          \centering
  ?? & \colorbox{dark_blue!35}{$f_2(2000)$}  & \centering $2^{++}$ & & & \checkmark & \checkmark & $0^*$ & \checkmark &  \\
  \hline
          \centering
  & \colorbox{dark_blue!35}{$f_2(2010)$}  & \centering $2^{++}$ & & & \checkmark & \checkmark & $0^*$ & \checkmark & \\
  \hline
           \centering
 ?? & \colorbox{orange!35}{$\eta_2(2030)$}  & \centering $2^{-+}$ & & \checkmark & & & & & \\
 \hline
 \hline
           \centering
  & \colorbox{dark_blue!35}{$\omega_3(1670)$}  & \centering $3^{--}$ & \checkmark & & & & & & \\
  \hline
             \centering
  & \colorbox{dark_blue!35}{$\phi_3(1850)$}  & \centering $3^{--}$ & & & \checkmark & & & & \\
  \hline
             \centering
  ?? & \colorbox{dark_blue!35}{$\omega_3(1945)$}  & \centering $3^{--}$ & & & \checkmark & & & & \\
  \hline
               \centering
 ?? & \colorbox{orange!35}{$h_3(2025)$}  & \centering $3^{+-}$ & & & & \checkmark & 0 & & \\
 \hline
              \centering
  ?? & \colorbox{gray!35}{$f_3(2050)$}  & \centering $3^{++}$ & & & & \checkmark & 1 & & \\
  \hline
             \centering
  ?? & \colorbox{dark_blue!35}{$\omega_3(2255)$}  & \centering $3^{--}$ & & & \checkmark & \checkmark & $0^*$ & \checkmark & \\
  \hline
                 \centering
 ?? & \colorbox{orange!35}{$h_3(2275)$}  & \centering $3^{+-}$ & & \checkmark & & & & & \checkmark \\
 \hline
 \hline
           \centering
 & \colorbox{dark_blue!35}{$f_4(2050)$}  & \centering $4^{++}$ & \checkmark & & & & & & \\
 \hline
  \hline
           \centering
 ?? & \colorbox{dark_blue!35}{$\omega_5(2250)$}  & \centering $5^{--}$ & \checkmark & & & & & & \\
 \hline
  \hline
           \centering
 ? & \colorbox{dark_blue!35}{$f_6(2510)$}  & \centering $6^{++}$ & \checkmark & & & & & & \\
 \hline
 \caption*{\ }
 \label{tableI=0}
\end{longtable}

\begin{longtable}[ht!]{ |P{0.5cm}|P{1.7cm}|P{1cm}|P{0.5cm}|P{0.5cm}|P{0.5cm}|P{0.5cm}|P{0.5cm}|P{0.5cm}|P{0.8cm}|}
\caption{\small \textbf{Strange mesons with isospin half, $I=1/2$}}\\
 \hline
 \centering  & & \centering $J^{PC}$ & GS & $\rho$ & $\phi$ & $S$ & $n_a$ & $S,a$ & $S,\phi$ \\
 \hline
 \hline
\endfirsthead
 \hline
 \centering  & & \centering $J^{P}$ & GS & $\rho$ & $\phi$ & $S$ & $n_a$ & $S,a$ & $S,\phi$ \\
 \hline
 \hline
\endhead
 \centering
 & \colorbox{orange!35}{$K^0(498)$}  & \centering $0^{-}$ & & & & \checkmark & 0 & & \\
 \hline
 \centering
 & \colorbox{dark_blue!35}{$K_0^*(700)$}  & \centering $0^{+}$ & & & \checkmark & & & & \\
 \hline
  \centering
 & \colorbox{dark_blue!35}{$K_0^*(1430)$}  & \centering $0^{+}$ & & & & \checkmark & $0^*$ & \checkmark & \\
 \hline
  \centering
 & \colorbox{orange!35}{$K(1460)$}  & \centering $0^{-}$ & & \checkmark & & \checkmark & 1 & & \\
 \hline
 \hline
   \centering
 & \colorbox{dark_blue!35}{$K^*(892)$}  & \centering $1^{-}$ & \checkmark & & & & & & \\
 \hline
   \centering
 & \colorbox{orange!35}{$K_1(1270)$}  & \centering $1^{+}$ & & & & \checkmark & 0 & & \\
 \hline
    \centering
 & \colorbox{orange!35}{$K_1(1400)$}  & \centering $1^{+}$ & & & & \checkmark & 0/1 & & \\
 \hline
    \centering
 & \colorbox{dark_blue!35}{$K^*(1410)$}  & \centering $1^{-}$ & & & \checkmark & \checkmark & & & \\
 \hline
     \centering
 & \colorbox{orange!35}{$K_1(1650)$}  & \centering $1^{+}$ & & & & \checkmark & 1 & & \checkmark \\
 \hline
     \centering
 & \colorbox{dark_blue!35}{$K^*(1680)$}  & \centering $1^{-}$ & & & \checkmark & \checkmark & $0^*$ & \checkmark & \\
 \hline
 \hline
      \centering
 & \colorbox{dark_blue!35}{$K_2^*(1430)$}  & \centering $2^{+}$ & \checkmark & & & & & & \\
 \hline
      \centering
? & \colorbox{orange!35}{$K_2(1580)$}  & \centering $2^{-}$ & & & & & 0 & & \\
 \hline
       \centering
 & \colorbox{orange!35}{$K_2(1770)$}  & \centering $2^{-}$ & & & & \checkmark & 0/1 & & \\
 \hline
       \centering
 & \colorbox{orange!35}{$K_2(1820)$}  & \centering $2^{-}$ & & & & \checkmark & 0/1 & & \\
 \hline
       \centering
 & \colorbox{dark_blue!35}{$K_2^*(1980)$}  & \centering $2^{+}$ & & & \checkmark & \checkmark & $0^*$ & \checkmark & \\
 \hline
 \hline
        \centering
 & \colorbox{dark_blue!35}{$K_3^*(1780)$}  & \centering $3^{-}$ & \checkmark & & & & & & \\
 \hline
        \centering
 ? & \colorbox{orange!35}{$K_3(2320)$}  & \centering $3^{+}$ & & \checkmark & & \checkmark & 1 & & \\
 \hline
 \hline
         \centering
 & \colorbox{dark_blue!35}{$K_4^*(2045)$}  & \centering $4^{+}$ & \checkmark & & & & & & \\
 \hline
 \hline
          \centering
 ? & \colorbox{dark_blue!35}{$K_5^*(2380)$}  & \centering $5^{-}$ & \checkmark & & & & & & \\
 \hline
\caption*{\ }
 \label{tableI=1/2}
\end{longtable}

 \newpage
 \clearpage
\bibliography{Biblio}
	\bibliographystyle{JHEP.bst}

\end{document}